\documentclass[a4paper, 12pt]{article}

\usepackage[utf8]{inputenc}
\usepackage{amsfonts,amsmath,amssymb}
\usepackage{bbold}
\usepackage{siunitx}
\usepackage{xcolor}
\usepackage{hyperref}
\usepackage{cite}
\usepackage{graphicx}
\usepackage{empheq}
\usepackage{array}
\usepackage{caption}
\usepackage{subcaption}
\usepackage[a4paper,bottom=3cm,top=2.5cm,head=0mm,width=17cm,dvipdfm]{geometry}
\usepackage[normalem]{ulem}
\usepackage{authblk}

\newcommand{\op}[1]{\mathcal{O} #1}
\def\gcw{{\overline g_{2}}}
\def\gcZ{{\overline g_{Z}}}

\def\sc{{\overline s}}

\usepackage{array,multirow}
\usepackage{booktabs}
\usepackage{colortbl}
\usepackage{float}

\title{DUNE potential as a New Physics probe}

\author[1,2]{Adriano Cherchiglia}
\author[2]{José Santiago}
\affil[1]{Instituto de F\'isica Gleb Wataghin, Universidade Estadual de Campinas,\hspace{0.5cm}Rua Sérgio Buarque de Holanda, 777, 13083-859, Campinas, SP, Brazil}
\affil[2]{Departamento de Física Teórica y del Cosmos, Universidad de Granada,
Campus de Fuentenueva, E-18071 Granada, Spain}
\date{}
\begin{document}

\maketitle

\begin{abstract}
\noindent
Neutrino experiments, in the next years, aim to determine with precision all the six parameters of the three-neutrino standard paradigm. The complete success of the experimental program is, nevertheless, attached to the non-existence (or at least smallness) of Non-Standard Interactions (NSI). In this work, anticipating the data taken from long-baseline neutrino experiments, we map all the weakly coupled theories that could induce sizable NSI, with the potential to be determined in these experiments, in particular DUNE. Once present constraints from other experiments are taken into account, in particular charged-lepton flavor violation, we find that only models containing leptoquarks (scalar or vector) and/or neutral isosinglet vector bosons are viable. We provide the explicit matching formulas connecting weakly coupled models and NSI, both in propagation and production. Departing from the weakly coupled completion with masses at TeV scale, we also provide a global fit on all NSI for DUNE, finding that NSI smaller than $10^{-2}$ cannot be probed even in the best-case scenario.
\end{abstract}

\section{Introduction}

Neutrino experiments are entering a high precision era. Within the next decade we expect to have new long-baseline experiments, such as DUNE and T2HK, whose main goal is to measure with high accuracy two of the parameters of the three-neutrino paradigm, namely $\theta_{23}$ and $\delta_{CP}$. These measurements, however, can be sensitive to the presence of Non-Standard Interactions (NSI), which represent new sources for neutral (charged) currents apart from the usual Z (W) mediated processes, see~\cite{Farzan:2017xzy,Proceedings:2019qno,Arguelles:2022tki} for reviews. Given their nature, NSI can affect neutrino production, detection and/or propagation through matter~\cite{Wolfenstein:1977ue,Mikheyev:1985zog}. Their impact, nonetheless, will only be known in the 2030s or later, after the data taken from these planned experiments has been analyzed. 
\\

\noindent
In the meantime, there are currently significant efforts towards the search for new physics at colliders~\cite{Gourlay:2022odf} and at low-energy experiments~\cite{Artuso:2022ouk}. Even though there are some hints of anomalies in experimental data, no clear picture can be drawn yet. In view of this scenario, it is desirable to merge the present knowledge in a unified front, such that any theoretical model can be scrutinized by as many experimental data as currently available. This is especially relevant if an anomaly is observed in a particular experiment, as one can immediately rule out or confirm, based on other experiments, if a particular model can be the origin of the anomaly, also making predictions of unmeasured observables that could further confirm the nature of the anomaly.
\\

\noindent
In this work, anticipating the results from planned long-baseline experiments, in particular DUNE, we give a comprehensive list of which weakly coupled models, based only on the Standard Model (SM) gauge group, could be responsible for any sizeable anomaly seen by the experiment. This can be achieved under the assumption that new physics is heavy, as compared with the electroweak scale, and weakly coupled, making the loop expansion meaningful. Under these circumstances we can use the tree-level SM effective field theory (SMEFT) dictionary~\cite{deBlas:2017xtg} that includes the most general model with heavy particles of spin 1 or less that can contribute to the SMEFT Lagrangian at tree level and mass dimension 6~\footnote{A similar analysis, taking into account operators of dimension 6 and 8 containing only leptons was performed in~\cite{Gavela:2008ra}. The complete tree-level dictionary has been recently extended up to dimension 7 in~\cite{Li:2023cwy} and the calculation of the dimension 6, one-loop dictionary is currently underway~\cite{Guedes:2023azv}.}. Given that the new particles are heavier than the electroweak scale, we will use an effective approach based on the SMEFT. As first noticed in~\cite{Bergmann:1999pk}, an immediate problem to be faced by any weakly coupled model with sizable non-diagonal NSI is to evade the strong bounds imposed by the null observation of charged lepton flavor violation (CLFV) processes. In references~\cite{Gavela:2008ra,Bergmann:1999pk}, it is argued that, by a suitable cancellation of the Wilson coefficients (WC), it is in principle possible to generate large non-diagonal NSI via dimension six-operators at the same time that CLFV processes are under control. In particular, the condition $\mathcal{O}^1_{psrt}= (\bar{\ell}_p\gamma_{\mu}\ell_s)(\bar{\ell}_r\gamma^{\mu}\ell_t)=-\mathcal{O}^1_{ptrs}=-\mathcal{O}^1_{rspt}$ must be fulfilled, which happens for a model containing a charged scalar singlet (${\cal S}_1$ in the notation of~\cite{deBlas:2017xtg}). However, as shown in~\cite{Antusch:2008tz,Gavela:2008ra}, bounds from lepton universality still apply, restricting the non-diagonal NSI to be of $\mathcal{O}(10^{-3})$. This is beyond the reach of present (and planned) long-baseline neutrino experiments. Therefore, it is not possible to obtain large non-diagonal NSI by considering only four-lepton operators. By considering operators containing two leptons and two quarks, the situation is even more dramatic given the very strong bounds coming from $\mu-e$ conversion in nuclei, for instance\footnote{By considering dimension-eight operators, it may be possible to enforce cancellations in CLFV processes, allowing non-diagonal NSI to be larger\cite{Gavela:2008ra,Antusch:2008tz,Davidson:2019iqh}. However, since the tree-level SMEFT dictionary at dimension 8 is not available, a systematic study is beyond the scope of our work.}.
\\

\noindent
Although the previous discussion was mainly related to NSI coming from propagation, a similar reasoning can be applied for those coming from production/detection. The main question is: how sensitive will future long-baseline experiments (such as DUNE) be to non-diagonal NSI in comparison with bounds from CLFV processes. By performing a global fit, we will obtain that DUNE can be sensitive to NSI at most of order $10^{-2}$. However, CLFV constraints from processes involving muon and electron are of order $10^{-6}$ (or lower) while those involving taus are of order $10^{-4}$ (or lower)~\cite{Bischer:2019ttk,Falkowski:2021bkq}. Therefore, it is clear that, if DUNE sees any anomaly related to non-diagonal NSI, it will be very challenging to invoke a weakly coupled model at the TeV scale as an explanation. Other possibilities such as considering lighter mediators and/or enlarging the gauge group could be envisaged (see~\cite{Farzan:2017xzy,Arguelles:2022tki} and references therein).
\\

\noindent
Analyses discussing the connection between NSI and EFT previously to the definition of the Warsaw basis were presented, for instance, in \cite{Bergmann:1999pk,Antusch:2008tz,Gavela:2008ra,Meloni:2009cg} while more recent analyses can be found, for instance, in~\cite{Altmannshofer:2018xyo,Falkowski:2019kfn,Falkowski:2019xoe, Babu:2019mfe,Bischer:2019ttk,Davidson:2019iqh,Terol-Calvo:2019vck,Babu:2020nna,Du:2020dwr,Falkowski:2021bkq,Du:2021rdg,Breso-Pla:2023tnz}. In \cite{Babu:2019mfe}, a particular set of models is discussed focusing on neutrino mass mechanisms. It is shown that, for some portions of the parameter space, it is possible to generate sizeable diagonal NSI, evading the strong bounds from charged lepton flavor violating processes. It is not clear, however, if there are other weakly coupled models that could still be viable. This work intends to fill this gap. In summary, this work goes beyond the current results in the literature in the following.
\begin{itemize}
    \item We provide a complete list of tree-level weakly coupled models relevant for long-baseline experiments, as well as their matching to the NSI.
    \item We adopt the quantum field theory (QFT) formalism to treat NSI at production, including indirect effects recently pointed out in~\cite{Breso-Pla:2023tnz}. We show that, if the non-diagonal NSI in production are suppressed, the diagonal ones cannot be probed. This implies that production can be treated as SM-like, allowing the bounds extracted in~\cite{Coloma:2023ixt} to be applied, even though they were derived under the hypothesis of SM fluxes.
    \item We go beyond the weakly coupled theories paradigm by performing a global fit considering arbitrary NSI at production and propagation, in order to asses the best-case scenario sensitivity of DUNE on such parameters. 
\end{itemize}

\noindent
The work is organized as follows. In Section~\ref{sec:rate} we review the connection between NSI and the neutrino transition rate in the QFT formalism, while in Section~\ref{sec:nsi} we present the matching at tree-level between NSI and SMEFT WC. In Section~\ref{sec:analitcal_dune} we identify and analyze the weakly coupled models, with new particles heavier than the electroweak scale, that can potentially generate sizeable NSI to be probed by DUNE. In Section~\ref{sec:numerical} we perform the numerical analysis, and present our results. Section~\ref{sec:conclusion} is devoted to the summary of our work while we briefly discuss the impact of NSI at detection in Appendix~\ref{ap:detection}, collecting relevant notation and the explicit matching result in Appendices~\ref{ap:notation} and \ref{ap:matching}, respectively.

\section{Neutrino transition rate}
\label{sec:rate}

In this work we will adopt the more general QFT framework to describe neutrino oscillations in long-baseline experiments, following~\cite{Falkowski:2019kfn} (see also~\cite{Giunti:1993se,Akhmedov:2010ms,Kobach:2017osm} for previous treatment in QFT not including NSI). The main point is to consider the entire process (from neutrino production to its detection) as an unique process in QFT. Thus, the differential event rate for neutrinos of flavor $\beta$ with energy $E_\nu$ to be detected at a distance $L$ from the source $S$, where they were produced with flavor $\alpha$, is given by 
\begin{equation}
\label{eq:rate}
    R_{\alpha\beta}^S = \frac{N_T}{32 \pi L^{2} m_S m_{T} E_{\nu}} 
    \sum_{k,l}
    e^{-i\frac{L \Delta m_{kl}^{2}}{2E_{\nu}}} \int d \Pi_{P^{\prime}} \mathcal{M}^{P}_{\alpha k} \overline{\mathcal{M}}^{P}_{\alpha l} \int d \Pi_{D} \mathcal{M}^{D}_{\beta k} \overline{\mathcal{M}}^{D}_{\beta l} ~,
\end{equation}
where $N_{T}$ is the number of target particles, $m_{S,T}$ are the masses of the source and target particles respectively, and we denote complex conjugation with a bar. As usual, $\Delta m_{kl}^{2}\equiv m_k^2 - m_l^2$ denotes the difference of neutrino masses squared, while the phase-space integrals are given by {$d \Pi \equiv \frac{d^3 k_1} { (2 \pi)^3 2 E_1} \dots  \frac{d^3 k_n}{(2 \pi)^3 2 E_n} (2\pi)^4 \delta^4(\sum p_n - \sum k_i )$}. Here, we denote by $k_i$ and $E_i$ the four-momenta and energies of the final states, while the sum $\sum p_n$ amounts to the total four-momentum of the initial state. Since we are interested in the differential number of events per incident neutrino energy $E_{\nu}$, we actually define $d \Pi_{P}\equiv d \Pi_{P^{\prime}} dE_{\nu} $, as can be seen in eq.\eqref{eq:rate}. Notice also that we are considering oscillation in vacuum only. Matter effects will be taken into account below.
\\

\noindent
As already stated, we will focus on long-baseline experiments, in which neutrinos are mainly produced by pion decay. Moreover, we will only consider beyond the SM (BSM) possible effects at neutrino production and/or propagation (by interaction with the medium). This is justified by the complicated nature of detection in long-baseline experiments. To properly include BSM effects, it would require first a theoretical description of standard neutrino interaction with nucleus, which is still not satisfactory \cite{NuSTEC:2017hzk}(see appendix~\ref{ap:detection} a brief description of the present status). Thus, regarding the detection amplitude, $\mathcal{M}^{D}_{\beta k}$, it will be given by
\begin{equation}
    \mathcal{M}^{D}_{\beta k} = U_{\beta k} A^{D}\,,
\end{equation}
where $A^{D}$ is a reduced matrix element whose explicit form is not relevant, and $U_{ij}$ is the PMNS matrix.
\\

\noindent
As recently pointed out \cite{Breso-Pla:2023tnz}, in the presence of BSM effects care must be exercised when calculating the phase-space integral related to the production amplitude. The amplitude concerning pion decay, $\pi^+ \to \ell_\alpha^+ \nu_k$,  is given by \cite{Breso-Pla:2023tnz}
\begin{align}\label{eq:m_Pion}
{\cal M}^{P,\pi}_{\alpha k} 
\equiv
\mathcal{M} (\pi^+ \to \ell_\alpha^+ \nu_k)  
= 
 - i \, m_\mu   f_{\pi^\pm}  \frac{V_{ud}}{v^2} [ {\cal P} U]_{\alpha k}^*(\bar{u}_{\nu_k} P_L v_{\ell_\alpha})~,
\end{align}
where $f_{\pi^\pm}$ is the pion decay constant, $v_{\ell_\alpha}$, $\bar{u}_{\nu_k}$ are the Dirac spinor wave functions of the charged lepton and the neutrino, respectively, and we are using the notation 
\begin{eqnarray}
 [ {\cal P}]_{\alpha \beta} 
 &\equiv& \delta_{\alpha \sigma}  +   \bar{\epsilon}_{\alpha \sigma},    
\quad\mbox{where}\quad \bar{\epsilon}_{\alpha\sigma} = (\epsilon_L)_{\alpha \sigma} -  (\epsilon_R)_{\alpha \sigma} 
-    \frac{ m_{\pi^\pm}^2}{   m_{\ell_\alpha}  (m_u + m_d) } (\epsilon_P)_{\alpha \sigma}\, .   
\end{eqnarray}
The NSI $\epsilon_{L},\epsilon_{R},\epsilon_{P}$ are defined in eq.\eqref{eq:Lag_Pion}. As can be seen, $\bar{\epsilon}_{\alpha \sigma}$ actually encodes the deviation from the PMNS matrix $U$ due to the presence of NSI.
\\

\noindent
At first sight, one could insert eq.\eqref{eq:m_Pion} into eq.\eqref{eq:rate}, perform the phase-space integration, and obtain that BSM effects appear only in the numerator of eq.\eqref{eq:rate}. However, this reasoning is incorrect, since we are assuming the BSM presence in all cases that the pion decay is involved, related (or not) to experiments that aim to probe neutrinos properties. Therefore, we should re-express eq.\eqref{eq:m_Pion} in terms of the experimental measured decay width of the pion, which was already affected by BSM effects as well. Proceeding this way, we obtain \cite{Breso-Pla:2023tnz}

\begin{align}
\int  d \Pi_{P^{\prime}}\mathcal{M}_{\alpha k}^{P,\pi} \bar{\mathcal{M}}_{\alpha l}^{P,\pi} 
= 
  2\,m_{\pi^\pm} \Gamma_{\pi\to\ell_{\alpha}\nu} \frac{[ {\cal P} U ]_{\alpha l}   [U^\dagger  {\cal P}^\dagger]_{k \alpha}} {[{\cal P}{\cal P}^\dagger]_{\alpha\alpha}}
  \,\delta \bigg ( E_\nu - E_{\nu,\pi} \bigg )~,
\end{align}
where $E_{\nu,\pi} = (m^2_{\pi^{\pm}}-m_{\alpha}^2)/(2m_{\pi^{\pm}})$ stands for the energy of the neutrino emitted. Finally, we can write eq.\eqref{eq:rate} in the form

\begin{equation}\label{eq:rate:final}
    R_{\alpha\beta}^S = N_T\sigma_\beta^{\text{SM}}(E_{\nu})\Phi_\alpha^{\text{SM}}(E_{\nu})
    \sum_{k,l}
    e^{-i\frac{L \Delta m_{kl}^{2}}{2E_{\nu}}} \frac{[ {\cal P} U ]_{\alpha l}   [U^\dagger  {\cal P}^\dagger]_{k \alpha}} {[{\cal P}{\cal P}^\dagger]_{\alpha\alpha}} U_{\beta k}U_{\beta l}^{*}~,
\end{equation}
where $\Phi_\alpha^{\text{SM}}$ is the SM flux (with the decay width of the pion, $\Gamma_{\pi\to\ell_{\alpha}\nu}$, as input), while $\sigma_\beta^{\text{SM}}$ is the SM cross-section. Notice that, by including the indirect effects (the denominator in $R_{\alpha\beta}^{S}$), there is no sensitivity to diagonal-only NSI.
\\

\noindent
At this stage we would like to include matter effects. As is well-known, the vector part of the neutral current (NC) NSI interactions introduced in eq.\eqref{eq:LagNC} will introduce flavor dependent matter effects, which can be parametrized in the Hamiltonian
\begin{equation}
    H = \frac{1}{2E}\left[ U\left(\begin{array}{ccc}
     0    &  & \\
         & \Delta m_{21}^{2} & \\
         & & \Delta m_{31}^{2} 
    \end{array}\right)U^{\dagger} + a \left(\begin{array}{ccc}
     1 + \epsilon_{ee}  &  \epsilon_{e\mu} & \epsilon_{e\tau}  \\
     \epsilon_{e\mu}^{*}    & \epsilon_{\mu\mu} & \epsilon_{\mu\tau} \\
     \epsilon_{e\tau}^{*} & \epsilon_{\mu\tau}^{*} & \epsilon_{\tau\tau}
    \end{array}\right)\right]\,.
\end{equation}
We introduced the parameter $a\equiv 2\sqrt{2} G_{F} N_{e} E$ which is known as the Wolfenstein matter potential, $N_{e}$ is the electron number density, and $E$ is the neutrino energy. Notice that, in the absence of NSI effects, only the first element of the matter matrix survives, which stands for the standard charged current (CC) mediated by the W-boson. In the equation above we are using the NSI notation in the Hamiltonian framework. The connection to the parameters introduced at Lagrangian level, see eq.\eqref{eq:LagNC}, is straightforward~\cite{Farzan:2017xzy}
\begin{equation}
    \epsilon_{\alpha\beta} = \sum_{f\in\left\{e,u,d\right\}}\left\langle\frac{N_{f}(x)}{N_{e}(x)}\right\rangle\left(\epsilon^{ff}_{V}\right)_{\alpha\beta}\,,
\end{equation}
where $N_{f}(x)$ is the number density of fermion of type $f$ at position $x$. In Earth, we generally have $N_{u}/N_{e}\sim N_{d}/N_{e}\sim 3$~\cite{Lisi:1997yc}. Moreover, in oscillation experiments one can, without loss of generality, remove a free parameter from the diagonal. In general, $\epsilon_{\mu\mu}$ is removed, which we will also adopt.

\section{Matching NSI to the SMEFT}
\label{sec:nsi}

The connection between NSI and the dimension-six operators in the SMEFT was, to the best of our knowledge, discussed in general first in~\cite{Bischer:2019ttk}. Here we review the formalism, and present a concise formula relating both quantities at the end of each subsection.

\subsection{Charged current: neutrino production}

The Lagrangian relevant for neutrino production (mainly through pion decay) is given by \cite{Cirigliano:2009wk,Jenkins:2017jig}
\begin{align}
\label{eq:Lag_Pion}
    \mathcal{L}_{\rm WEFT} \supset - \frac{2\, V_{jk}}{v^2} \Big\{
    & [1+\epsilon^{jk}_L]_{\alpha \beta} \left( \bar{u}^{j}\gamma^{\mu}P_L d^{k}\right)\left(\bar{\ell}_{\alpha}\gamma_{\mu}P_L \nu_{\beta} \right) + [\epsilon^{jk}_R]_{\alpha \beta} \left( \bar{u}^{j}\gamma^{\mu}P_R d^{k}\right)\left(\bar{\ell}_{\alpha}\gamma_{\mu}P_L \nu_{\beta} \right) 
    \nonumber\\
    & \left. +\frac{1}{2}\left[ \epsilon_S^{jk}\right]_{\alpha \beta} \left(\bar{u}^{j}d^{k} \right)\left(\bar{\ell}_{\alpha}P_L \nu_{\beta} \right)
    -\frac{1}{2}\left[ \epsilon_P^{jk}\right]_{\alpha \beta} \left(\bar{u}^{j} \gamma^5 d^{k} \right)\left(\bar{\ell}_{\alpha}P_L \nu_{\beta} \right)\right. 
    \nonumber\\
    & +  \frac{1}{4}\left[ \epsilon_T^{jk}\right]_{\alpha \beta} \left(\bar{u}^{j} \sigma^{\mu \nu} P_L d^{k} \right)\left(\bar{\ell}_{\alpha} \sigma_{\mu \nu}P_L \nu_{\beta} \right) +\mathrm{h.c.} \Big\}~,
\end{align}
We are considering the quark fields as well as the charged leptonic fields in the mass eigenstate basis, $V$ is the CKM matrix, and $v\approx 246$ GeV is the Higgs vacuum expectation value. This Lagrangian can be directly translated to the Low Energy EFT (LEFT) one, which reads, following the notation in~\cite{Jenkins:2017jig} 
\begin{align}
    &L_{r}^{V,LL} = - \frac{2\, V_{kj}^{*}}{v^2} \left[\delta_{\beta\alpha}+(\epsilon_{L}^{kj})_{\beta\alpha}\right]^{*}\,,\qquad
    L_{r}^{V,LR} = - \frac{2\, V_{kj}^{*}}{v^2} (\epsilon_{R}^{kj})_{\beta\alpha}^{*},  \\    &L_{r}^{S,RR}+L_{r}^{S,RL} = - \frac{2V_{kj}^{*}}{v^2} (\epsilon_{S}^{kj})_{\beta\alpha}^{*}\,,\qquad 
    L_{r}^{S,RL}-L_{r}^{S,RR} = - \frac{2V_{kj}^{*}}{v^2} (\epsilon_{P}^{kj})_{\beta\alpha}^{*},\\
  & L_{r}^{T,RR} = -\frac{V_{k j}^{*}}{2v^{2}} (\epsilon_{T}^{kj})_{\beta\alpha}^{*},
\end{align}
where $r = \substack{\nu e d u\\\alpha\beta j k} $ and we are using an asterisk to denote complex conjugation here. The matching to the SMEFT is also given in \cite{Jenkins:2017jig}, which we reproduce below 
\begin{equation}\label{eq:LEFT_VLL}
L_{\substack{\nu e d u\\prst}}^{V,LL} = \sum_{x}V^{*}_{xs}L_{\substack{\nu e d u\\prxt}}^{V,LL} = \sum_{x}V^{*}_{xs}\left(2 C^{(3)}_{\substack{ lq \\ prxt}} -\frac{\gcw^2}{2 M_W^2} 
\left[W_l\right]_{pr} \left[W_q\right]_{tx}^*\right),
\end{equation}
\begin{equation}
   L_{\substack{\nu e d u\\prst}}^{V,LR} = -\frac{\gcw^2}{2 M_W^2} 
\left[W_l\right]_{pr} \left[W_R\right]_{ts}^*,
\end{equation}
\begin{align}
    L_{\substack{\nu e d u\\prst}}^{S,RR} = \sum_{x}V_{xs}^{*}C^{(1)}_{\substack {lequ \\  prxt}},\quad
    L_{\substack{\nu e d u\\prst}}^{S,RL} = C_{\substack {ledq \\  prst}},\quad
    L_{\substack{\nu e d u\\prst}}^{T,RR} = \sum_{x}V_{xs}^{*}C^{(3)}_{\substack {lequ \\  prxt}},
\end{align}
where
\begin{align}
[W_l]_{pr} = \left[\delta_{pr} + v_T^2  C^{(3)}_{\substack {Hl \\  pr}} \right], \quad
[W_q]_{pr} = \left[\delta_{pr} +  v_T^2  C^{(3)}_{\substack {Hq \\  pr}} \right], \quad
[W_R]_{pr} = \left[ \frac12 v_T^2  C_{\substack {Hud \\  pr}} \right].
\end{align}
We are adopting the up basis where the up quark and charged lepton Yukawa matrices are diagonal. In this case, flavor and mass eingenstates of the left-handed down-quarks are connected by the CKM matrix.
\\

\noindent
Once all the pieces are known, it is straightforward to express the NSI parameter in production in terms of the WC of the SMEFT as below
\begin{align}\label{eq:NSI_prod}
 \bar{\epsilon}_{\beta \alpha} = \frac{v^{2}}{2}\left[2C^{(3)}_{\substack {Hl \\  \alpha\beta}}+2\sum_{x}V_{x1}^{*}\left(\delta_{\alpha\beta}C^{(3)*}_{\substack {Hq \\  1x}}-C^{(3)}_{\substack {lq \\  \alpha\beta 1x}}\right)+\delta_{\alpha\beta}C^{*}_{\substack {Hud \\  11}}+p_{\beta}\left(C_{\substack {ledq \\  \alpha\beta11}-}-\sum_{x}V_{x1}^{*}C^{(3)}_{\substack {lequ \\  \alpha\beta x1}}\right)\right]^{*}\,.  
\end{align}
where we used the notation $p_{\alpha}=\frac{ m_{\pi^\pm}^2}{   m_{\ell_\alpha}  (m_u + m_d) }$.

\subsection{Neutral current: neutrino propagation}

Regarding neutrino propagation, we need to take into account the interaction of neutrinos with the medium. This is given by the NC interactions between neutrinos and the other fermions~\cite{Campanelli:2002cc,Jenkins:2017jig}:    
    \begin{align}\label{eq:LagNC}
        \mathcal{L}_{\rm WEFT}^{\rm quarks} \supset    - \frac{1}{v^2} &\left\{
    [g_V^{qq} \,\mathbb{1} + \epsilon_V^{qq}]_{\alpha \beta} \left( \bar{q}\gamma^{\mu}q\right)\left(\bar{\nu}_{\alpha}\gamma_{\mu}P_L \nu_{\beta} \right) \right. \nonumber\\
    &\left. + [g_A^{qq} \,\mathbb{1}
    + \epsilon_A^{qq}]_{\alpha \beta} \left( \bar{q}\gamma^{\mu}\gamma^5 q\right)\left(\bar{\nu}_{\alpha}\gamma_{\mu}P_L \nu_{\beta} \right) \right\},\nonumber\\   \mathcal{L}_{\rm WEFT}^{\rm leptons} \supset    - \frac{1}{v^2} &\left\{
    [g_V^{ee} \,\mathbb{1} + \bar{g}_V^{ee} +\epsilon_V^{ee}]_{\alpha \beta} \left( \bar{e}\gamma^{\mu}e\right)\left(\bar{\nu}_{\alpha}\gamma_{\mu}P_L \nu_{\beta} \right) \right. \nonumber\\
    &\left. + [g_A^{ee} \,\mathbb{1}
    + \bar{g}_A^{ee}+\epsilon_A^{ee}]_{\alpha \beta} \left( \bar{e}\gamma^{\mu}\gamma^5 e\right)\left(\bar{\nu}_{\alpha}\gamma_{\mu}P_L \nu_{\beta} \right) \right\}.
\end{align}
Since we are considering neutrino interaction with a medium (Earth), only first generation fermions are to be considered, allowing us to omit some indices. Moreover, $g_{i}^{ff}$ are the SM-coefficients coming from Z-exchange while $\bar{g}_{i}^{\ell\ell}$ are related to W-exchange. Their explicit forms are:
\begin{equation}
 g_V^{ff} = T^3_f - 2 Q_f \sin^2 \theta_W,  
\quad 
g_A^{ff}   = - T^3_f,
\quad
(\bar{g}_V^{\ell\ell})_{\alpha\beta} = -(\bar{g}_A^{\ell\ell})_{\alpha\beta} = \delta_{\alpha e} \delta_{\beta e}\;,   
\end{equation}
where $Q_{f}$ and $T^{3}_{f}$ are the electric charge and weak isospin of fermion $f$, respectively. 
\\

\noindent
The matching to the LEFT is straightforward

\begin{align}
 L_{\substack{\nu q\\ \alpha\beta 11}}^{S,LL} &= -\frac{1}{v^{2}}\left[(g_V^{qq}-g_A^{qq})\delta_{\alpha\beta}  + (\epsilon_V^{qq}-\epsilon_A^{qq})_{\alpha \beta}\right]\,,\\
 L_{\substack{\nu q\\ \alpha\beta 11}}^{S,LR} &= -\frac{1}{v^{2}}\left[(g_V^{qq}+g_A^{qq})\delta_{\alpha\beta}  + (\epsilon_V^{qq}+\epsilon_A^{qq})_{\alpha \beta}\right]\,,
\\
 L_{\substack{\nu e\\ \alpha\beta 11}}^{S,LL} &= -\frac{1}{v^{2}}\left[(g_V^{ee}-g_A^{ee})\delta_{\alpha\beta} + (\bar{g}_V^{ee}-\bar{g}_A^{ee})_{\alpha \beta} + (\epsilon_V^{ee}-\epsilon_A^{ee})_{\alpha \beta}\right]\,,\\
 L_{\substack{\nu e\\ \alpha\beta 11}}^{S,LR} &= -\frac{1}{v^{2}}\left[(g_V^{ee}+g_A^{ee})\delta_{\alpha\beta}  + (\bar{g}_V^{ee}+\bar{g}_A^{ee})_{\alpha \beta} + (\epsilon_V^{ee}+\epsilon_A^{ee})_{\alpha \beta}\right]\,.
\end{align}
\\

\noindent
Finally, the matching to the SMEFT is given by
\begin{align}
   L_{\substack{\nu e\\ prst }}^{S,LL} &= C_{\substack{ ll \\ prst}} + C_{\substack{ ll \\ stpr}} -\frac{\gcw^2}{2 M_W^2} 
\left[W_l\right]_{pt} \left[W_l\right]_{rs}^*  -\frac{\gcZ^2}{M_Z^2} 
\left[Z_\nu \right]_{pr} \left[Z_{e_L} \right]_{st},\\
 L_{\substack{\nu u\\ prst }}^{S,LL} &=
C^{(1)}_{\substack{ lq \\ prst}} + C^{(3)}_{\substack{ lq \\ prst}} 
-\frac{\gcZ^2}{ M_Z^2}   \left[Z_\nu \right]_{pr} \left[Z_{u_L} \right]_{st},\\
 L_{\substack{\nu d\\ prst }}^{S,LL} &=  \left(C^{(1)}_{\substack{ lq \\ prxz}} -  C^{(3)}_{\substack{ lq \\ prxz}}
-\frac{\gcZ^2}{ M_Z^2} 
 \left[Z_\nu \right]_{pr} \left[Z_{d_L} \right]_{xz}\right)V_{tz}V_{sx}^{*},
\end{align}

\begin{align}
L_{\substack{\nu e\\ prst }}^{S,LR} &= C_{\substack{ le \\ prst}} 
-\frac{\gcZ^2}{ M_Z^2}  \left[Z_\nu \right]_{pr} \left[Z_{e_R} \right]_{st}, \\
L_{\substack{\nu u\\ prst }}^{S,LR} &= C_{\substack{ lu \\ prst}}  
-\frac{\gcZ^2}{ M_Z^2}   \left[Z_\nu \right]_{pr} \left[Z_{u_R} \right]_{st}, \\
L_{\substack{\nu d\\ prst }}^{S,LR} &=C_{\substack{ ld \\ prst}}  
-\frac{\gcZ^2}{ M_Z^2}   \left[Z_\nu \right]_{pr} \left[Z_{d_R} \right]_{st}, 
\end{align}
where
\begin{align}
[Z_{\nu}]_{pr} &= \left[\delta_{pr}\left(\frac 12\right) - \frac12 v_T^2  C^{(1)}_{\substack {Hl \\  pr}} + \frac12 v_T^2  C^{(3)}_{\substack {Hl \\  pr}} \right] ,
\\
[Z_{e_L}]_{pr} &= \left[\delta_{pr}\left(-\frac 12+\sc^2 \right) - \frac12 v_T^2  C^{(1)}_{\substack {Hl \\  pr}} - \frac12 v_T^2  C^{(3)}_{\substack {Hl \\  pr}} \right],  &
[Z_{e_R}]_{pr} &= \left[\delta_{pr}\left(+\sc^2 \right) - \frac12 v_T^2  C_{\substack {He \\  pr}}  \right], \\
[Z_{u_L}]_{pr} &=  \left[\delta_{pr}\left(\frac 12-\frac 23 \sc^2 \right) - \frac12 v_T^2  C^{(1)}_{\substack {Hq \\  pr}} + \frac12 v_T^2  C^{(3)}_{\substack {Hq \\  pr}} \right],  &
[Z_{u_R}]_{pr} &=  \left[\delta_{pr}\left(-\frac 23 \sc^2 \right) - \frac12 v_T^2  C_{\substack {Hu \\  pr}}  \right],  \\
[Z_{d_L}]_{pr} &=  \left[\delta_{pr}\left(-\frac 12+ \frac 13 \sc^2 \right) - \frac12 v_T^2  C^{(1)}_{\substack {Hq \\  pr}} - \frac12 v_T^2  C^{(3)}_{\substack {Hq \\  pr}}  \right], &
[Z_{d_R}]_{pr} &=  \left[\delta_{pr}\left(+\frac13\sc^2 \right) - \frac12 v_T^2  C_{\substack {Hd \\  pr}}  \right] .
\end{align}
\\

\noindent
The connection to NSI in propagation for Earth based long-baseline experiments is 
\begin{align}\label{eq:NSI_prop}
    \epsilon_{\alpha \beta} = \frac{v^{2}}{2}\Bigg[&C^{(1)}_{\substack {Hl \\  \alpha \beta}}-C^{(3)}_{\substack {Hl \\  \alpha \beta}}+2\delta_{1\alpha}C^{(3)*}_{\substack {Hl \\   \beta 1}}+2\delta_{1\beta}C^{(3)}_{\substack {Hl \\   \alpha 1}}-C_{\substack {le \\  \alpha \beta 11}} - 3C_{\substack {ld \\  \alpha \beta 11}}- 3C_{\substack {lu \\  \alpha \beta 11}} - C_{\substack {ll \\  \alpha \beta 11}} - C_{\substack {ll \\  11\alpha \beta}} \nonumber\\
    &- 3\left(C^{(1)}_{\substack {lq \\  \alpha \beta 11}}+C^{(3)}_{\substack {lq \\  \alpha \beta 11}} + \sum_{x,z}\Big(C^{(1)}_{\substack {lq \\  \alpha \beta xz}}-C^{(3)}_{\substack {lq \\  \alpha \beta xz}}\Big)V_{1,z}V_{1x}^{*}\right) \nonumber\\&-\delta_{\alpha\beta}\left(3C_{\substack {Hd \\  11}}+3C_{\substack {Hu \\  11}}+C_{\substack {He \\  11}}+C^{(1)}_{\substack {Hl \\  11}}+C^{(3)}_{\substack {Hl \\  11}}\right)\nonumber\\
    &-3\delta_{\alpha\beta}\left(C^{(1)}_{\substack {Hq \\  11}}+C^{(3)}_{\substack {Hq \\  11}}+\sum_{x,z}\Big(C^{(1)}_{\substack {Hq \\  xz}}+C^{(3)}_{\substack {Hq \\  xz}}\Big)V_{1,z}V_{1x}^{*}\right)
    \Bigg]\,.
\end{align}

\section{NSI probed by long-baseline experiments}
\label{sec:analitcal_dune}

In long-baseline experiments, the aim is to measure the conversion rate from a muon neutrino (antineutrino) to an electron neutrino (antineutrino), $R_{\mu\beta}$. The beam of muon neutrinos (anti-neutrinos) is generally generated by pion decay. In the particular case of DUNE, it is expected that 92$\%$ (90.4$\%$) of the beam will correspond to  muon neutrinos (antineutrinos), wrong-sign contamination is as large as 7$\%$ (8.6$\%$) and the background corresponding to electron neutrino and antineutrino is 1$\%$ (1$\%$)~\cite{DUNE:2020ypp}. Thus, by allowing NSI interaction at production, we can probe the combination, $\overline{\epsilon}_{\mu\sigma}$, where
\begin{equation}
[ {\cal P}]_{\mu \sigma} = \delta_{\mu \sigma}  +   \bar{\epsilon}_{\mu \sigma}  ,    
\quad\mbox{with}\quad \bar{\epsilon}_{\mu \sigma} = (\epsilon_L)_{\mu \sigma} -  (\epsilon_R)_{\mu \sigma} 
-    p_{\mu}(\epsilon_P)_{\mu \sigma},    
\end{equation}
and $p_{\mu}=\frac{ m_{\pi^\pm}^2 }{   m_{\mu}  (m_u + m_d) } \sim 27$.
This stands for 3 free parameters in the real case. For illustration, consider the connection between $\bar{\epsilon}_{\mu e}$ and the SMEFT operators (for simplicity, we restrict ourselves to real coefficients and consider the CKM matrix to be approximately diagonal):
\begin{align}
 \bar{\epsilon}_{\mu e} = \frac{v^{2}}{2}\left[2C^{(3)}_{\substack {Hl \\  12}}-2C^{(3)}_{\substack {lq \\  12 11}}+p_{\mu}\left(C_{\substack {ledq \\  1211}}-C^{(3)}_{\substack {lequ \\  12 11}}\right)\right]\,.  
\end{align}

\noindent
In order to have values for $\bar{\epsilon}_{\mu \sigma}$ that can be probed at DUNE (roughly of order $10^{-2}$), the WC has to be at least of order $(\unit{TeV})^{-2}$. However, present constraints coming from charged lepton flavour changing processes, in particular $\mu-e$ conversion in nuclei, requires that these WC are several orders of magnitude lower~\cite{Bischer:2019ttk}, precluding any chance that they can be seen in DUNE. Even more importantly, as we already stated in the introduction, if an anomaly is seen in this experiment in non-diagonal channels, it will definitely require new physics to be other than a weakly coupled theory with masses above the electroweak scale. Even though we focused on $\bar{\epsilon}_{\mu e}$, which is more heavily constrained, a similar reasoning applies to $\bar{\epsilon}_{\mu \tau}$. The main message is: it is extremely difficult to generate sizable non-diagonal NSI in production, once a connection to the SMEFT is made. 
\\

\noindent
On the other hand, diagonal NSI at production could, in principle, be generated, with milder constraints (mainly from pion decay measurements). However, as we show below, the inclusion of indirect effects recently pointed out in~\cite{Breso-Pla:2023tnz}, will remove all dependence of NSI at production from the neutrino transition rate. Therefore, even though they could be sizable, they cannot be observed in long-baseline experiments such as DUNE. Once the NSI is only-diagonal, we obtain
\begin{equation}
[ {\cal P} U ]_{\mu l}   [U^\dagger  {\cal P}^\dagger]_{k \mu}
 = (1+ \bar{\epsilon}_{\mu \mu})^{2}U_{\mu l}U^{\dagger}_{k \mu}\,,   
\end{equation}
while $[ {\cal P} {\cal P}^\dagger]_{\mu \mu}=(1+ \bar{\epsilon}_{\mu \mu})^{2}$. Thus, by inspection of 
eq.\eqref{eq:rate:final}, we see that the standard rate in the three-neutrino paradigm is obtained. In summary, it is possible to consider the SM prediction for the production fluxes even if diagonal NSI in production are generated by a particular UV model. 
\\

\noindent
We conclude by considering NSI in propagation. Recall that long-baseline experiments actually probe the combination
\begin{equation}
    \epsilon_{\alpha\beta} = \sum_{f\in\left\{e,u,d\right\}}\left\langle\frac{N_{f}(x)}{N_{e}(x)}\right\rangle\left(\epsilon^{ff}_{V}\right)_{\alpha\beta}\,\sim \left(\epsilon^{ee}_{V}\right)_{\alpha\beta}+ 3\left(\epsilon^{uu}_{V}\right)_{\alpha\beta}+
    3\left(\epsilon^{dd}_{V}\right)_{\alpha\beta}\,.
\end{equation}
As before, we illustrate the role played by non-diagonal effects with $\epsilon_{\mu e}$, which is related to the SMEFT operators below
\begin{align}
    \epsilon_{\mu e} = \frac{v^{2}}{2}\Bigg[&C^{(1)}_{\substack {Hl \\  21}}+C^{(3)}_{\substack {Hl \\  21}}-C_{\substack {le \\  21 11}} - 3C_{\substack {ld \\  21 11}}- 3C_{\substack {lu \\  21 11}} - C_{\substack {ll \\  21 11}} - C_{\substack {ll \\  1121}} -6C^{(1)}_{\substack {lq \\  21 11}} 
    \Bigg]\,.
\end{align}
\noindent
For simplicity, we consider only real coefficients and a diagonal CKM. In order to have sizable NSI (of order $10^{-2}$), we need that the WC are not smaller than roughly one $(\unit{TeV})^{-2}$. However, even if one can avoid the stringent bounds from conversion in nuclei for some the operators above, there are strong bounds from processes such as $\mu\rightarrow eee$~\cite{Bischer:2019ttk}. Therefore, we arrive at the same conclusion before: it is extremely challenging to have sizable non-diagonal NSI in propagation, for weakly coupled theories with masses at the electroweak scale (or above). Nevertheless, diagonal NSI can still be probed, with sizable values generated by well-motivated UV models. In the next subsection, we present all the weakly coupled models that can generate NSI.

\subsection{Connection to UV model}

In~\cite{deBlas:2017xtg}, a general Lagrangian that maps all renormalizable weakly coupled models up to spin 1 was presented. Given eqs.~(\ref{eq:NSI_prod}-\ref{eq:NSI_prop}), it is straightforward (yet tedious) to find the UV models that can generate the WC entering in these equations, as well as, their matching conditions. This task can be automated making use of the implementation of the tree level dictionary~\cite{deBlas:2017xtg} in the \texttt{MatchingDB} format (see \url{https://gitlab.com/jccriado/matchingdb} and Section 3.6 of~\cite{Aebischer:2023irs}). In order to (re)check the calculation, the explicit tree-level matching was performed for all the models relevant for our purposes. Also the results from the dictionary implemented in \texttt{MatchingDB} were cross-checked against the results of~\cite{deBlas:2017xtg}, showing perfect agreement. For completeness, we have collected in Appendix~\ref{ap:notation} the Lagrangian containing all the relevant models to us, organized in terms of the particle spin. Also, the explicit quantum number for the fields is given. In Appendix~\ref{ap:matching} we present the explicit connection between the UV models and the NSI.
\\

\noindent
We collect in table~\ref{tab:UV} the weakly coupled models that can generate diagonal NSI at propagation and/or production, while in table~\ref{tab:UV-op} we show the explicit connection between the WC in the WEFT, the WC in the SMEFT and the UV model that can generate them. In order to make the table more compact, in each line the UV models depicted can generate at least one of the SMEFT WC presented in the same line. The only exception is for the vector boson models that can generate $\epsilon_{V}$, $\epsilon_{A}$, which are all collected in one single block.  

\begin{table}[h!]
    \centering
    \begin{tabular}{|c|c|c|c|}
       \hline
       NSI  &  Scalar & VL-fermion & Vector boson\\
       \hline
       propagation & $\mathcal{S}_{1},\varphi,\Xi_1,\omega_1,\Pi_1, \Pi_7, \zeta$ & $N,E, \Sigma, \Sigma_1$ & $\mathcal{B},\mathcal{W},\mathcal{L}_1,\mathcal{L}_3,\mathcal{U}_2,\mathcal{Q}_1,\mathcal{Q}_5, \mathcal{X}$\\
       \hline
       production & $\varphi,\omega_1, \Pi_7, \zeta$ & $N,E, \Sigma, \Sigma_1$ & $\mathcal{B}_1,\mathcal{W},\mathcal{L}_1,\mathcal{U}_2,\mathcal{Q}_5, \mathcal{X}$\\
       \hline
    \end{tabular}
    \caption{weakly coupled models that can generate diagonal NSI at propagation and/or production.}
    \label{tab:UV}
\end{table}

\begin{table}[h!]
    \centering
    \begin{tabular}{|c|c|c|c|c|}
       \hline
       WC WEFT  &  SMEFT & Scalar UV & VL-fermion UV & Vector boson UV\\
       \hline
       $\epsilon_{L}$  & $\op_{lq}^{(3)}$ & $\omega_1, \zeta$ & - & $\mathcal{W}, \mathcal{U}_2, \mathcal{X}$  \\
       & $\op^{(3)}_{\phi l}, \op^{(3)}_{\phi q}$ & - & $N, E, \Sigma, \Sigma_{1}, U, D, T_{1}, T_{2}$ & $\mathcal{L}_1$ \\
       \hline
       $\epsilon_{R}$  &  $\op_{\phi ud}$ & - & $Q_{1}$ & $\mathcal{B}_1$\\
       \hline
       $\epsilon_{S}$, $\epsilon_{P}$   &  $\op^{(1)}_{lequ}, \op_{ledq}$ & $\varphi, \omega_1, \Pi_7$ & - & $\mathcal{L}_1, \mathcal{U}_2, \mathcal{Q}_5$ \\
       \hline
       $\epsilon_{T}$  &  $\op^{(3)}_{lequ}$ & $\omega_1, \Pi_7$ & - & -  \\
       \hline
       &  $\op_{ll}, \op_{lq}^{(1)}, \op_{lq}^{(3)}$ & $\mathcal{S}_{1}, \Xi_1, \omega_1, \zeta$ & - & 
       \\
     $\epsilon_{V}$, $\epsilon_{A}$   &  $\op^{(1)}_{\phi l}, \op^{(3)}_{\phi l}, \op^{(1)}_{\phi q}, \op^{(3)}_{\phi q}$ & - & $N, E, \Sigma, \Sigma_1, U, D, T_1, T_2$ & 
     $\mathcal{B}, \mathcal{W}, \mathcal{L}_1, \mathcal{L}_3,$ 
     \\
       & $\op_{le}, \op_{lu}, \op_{ld}$ & $\varphi, \Pi_1, \Pi_7$ & - & 
       $\mathcal{U}_2, \mathcal{Q}_1, \mathcal{Q}_5, \mathcal{X}$
       \\
       & $\op_{\phi e}, \op_{\phi u}, \op_{\phi d}$ & - & $\Delta_1, \Delta_3, Q_1, Q_5, Q_7$ & 
       \\
       \hline
    \end{tabular}
    \caption{Relation among WC in eqs.(\ref{eq:Lag_Pion}),(\ref{eq:LagNC}), the SMEFT operators, and the UV completion.}
    \label{tab:UV-op}
\end{table}

\begin{figure}[h!]
    \centering
    \includegraphics[scale=0.45]{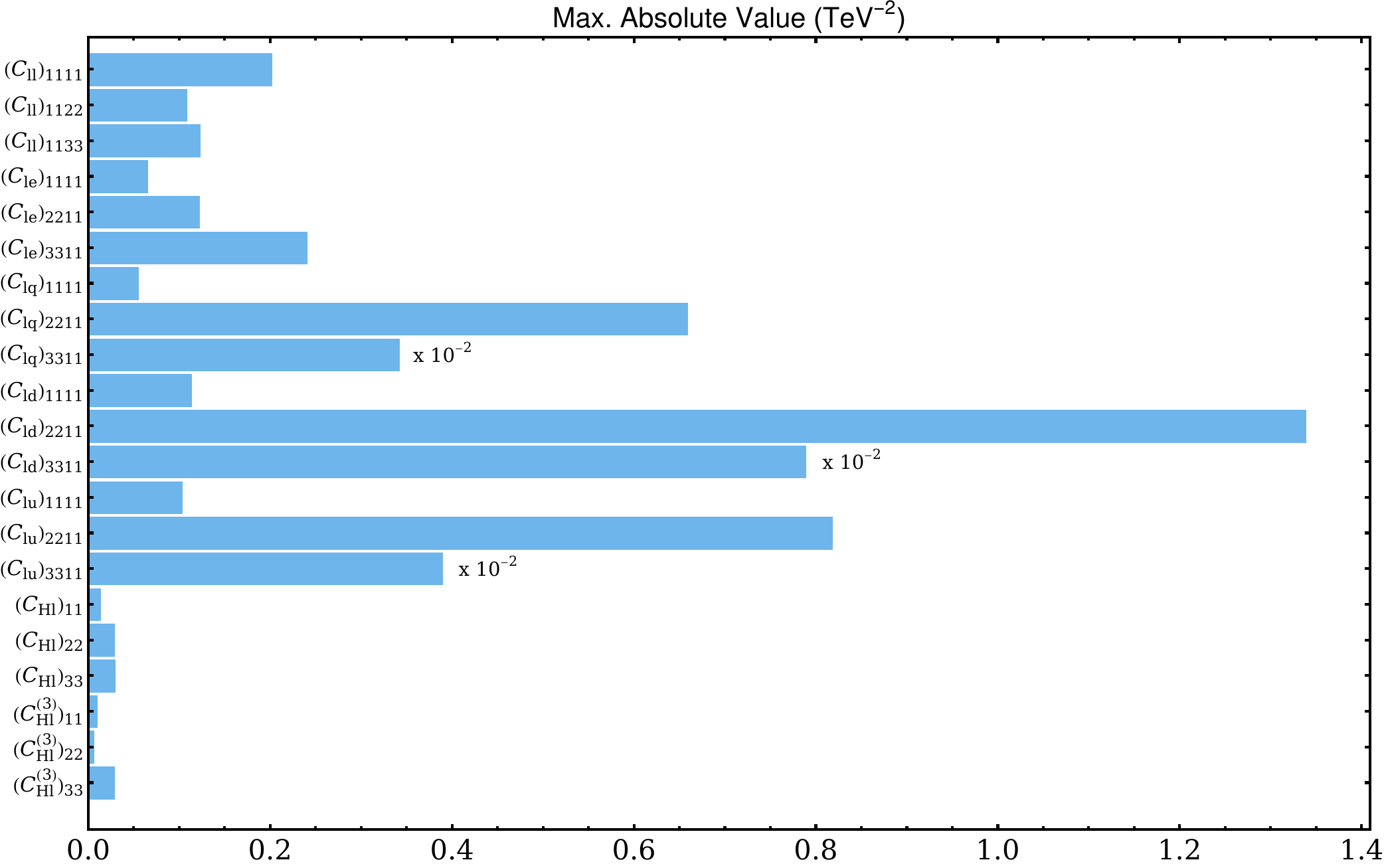}        
    \caption{Maximum absolute value of the WC that affect diagonal NC-NSI at 90$\%$ C.L.~\cite{Falkowski:2017pss,Aebischer:2018iyb}. The bounds were extracted considering only one non-null WC. The bounds for $C_{\substack{ lq \\  33 11}}$, $C_{\substack {ld \\  33 11}}$, and $C_{\substack {lu \\  33 11}}$ were multiplied by $10^{-2}$.}
    \label{fig:bounds}
\end{figure}

\begin{figure}[h!]
    \centering
    \includegraphics[scale=0.45]{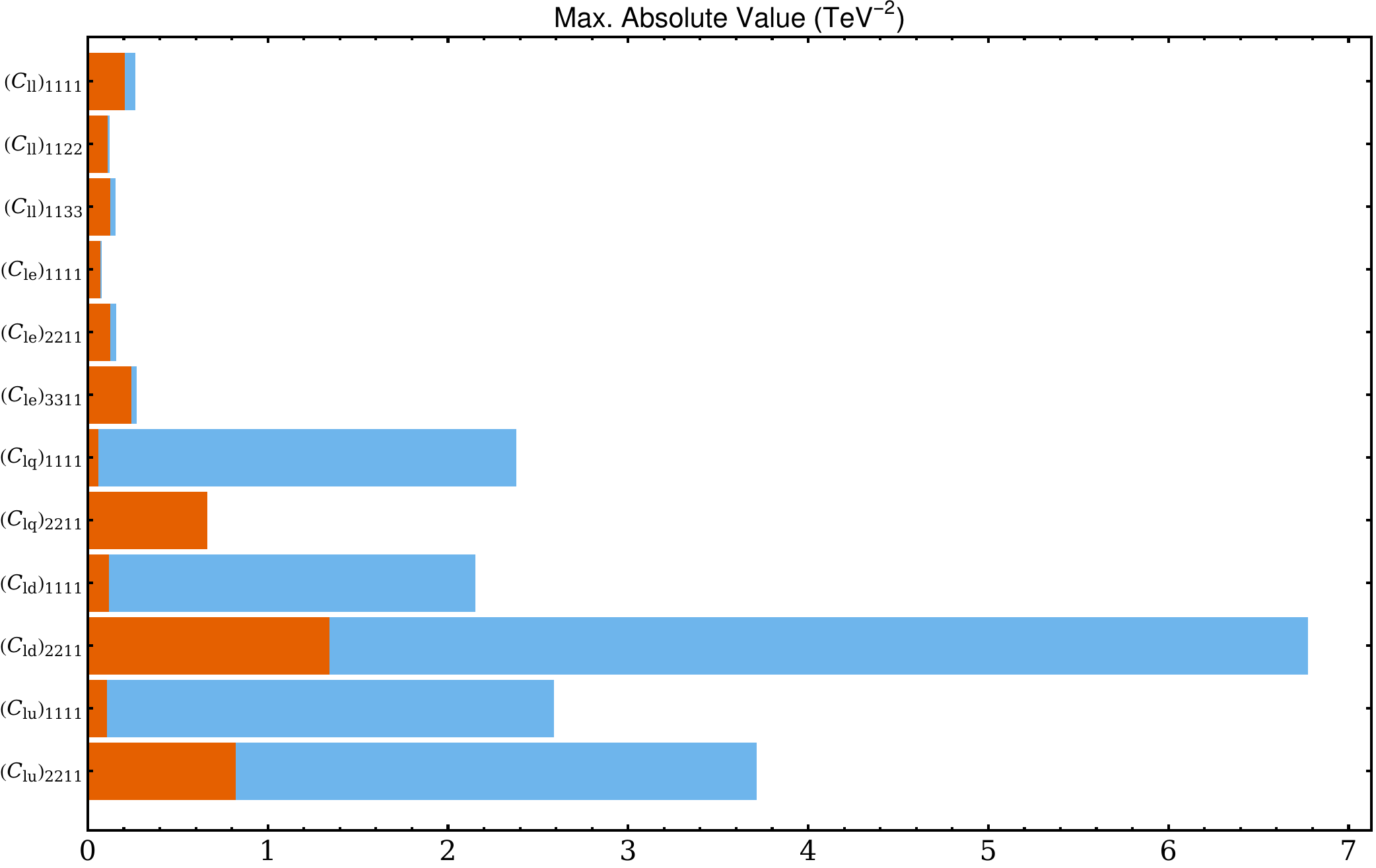}        
    \caption{Maximum absolute value of the WC that affect diagonal NC-NSI at 90$\%$ C.L.~\cite{Falkowski:2017pss}. The blue bounds were extracted considering the $\chi^2$ function given in~\cite{Falkowski:2017pss}, where only the WC affecting NSI are considered to be non-null, and they are marginalized over. For comparison, we repeat in brown the bounds extracted in Fig.~\ref{fig:bounds}, where only one WC is considered non-null at time.
    \label{fig:bounds2}}
\end{figure}

\noindent
In order to have an idea of the impact that each WC can have on the NSI given current bounds, we present in Fig.\ref{fig:bounds} the maximum value that each WC can attain at 90$\%$ C.L. according to the global fit performed in~\cite{Falkowski:2017pss}. Since $C^{(1)}_{\substack {lq \\  33 11}}$, $C_{\substack {ld \\  33 11}}$, $C_{\substack {lu \\  33 11}}$ were not present, we have used \texttt{smelli}~\cite{Aebischer:2018bkb,Straub:2018kue,Aebischer:2018iyb} to obtain the bounds on these coefficients. We have also checked the results for all the other WC's with \texttt{smelli}, and kept the most stringent bounds. We consider only one non-null WC at time, so these bounds are conservative. Given a specific UV model, more than one WC can be induced, with correlations among them. Therefore, the bounds extracted from a global fit analysis may not translate directly to such model. In order to access the impact that simultaneous non-null WC can have, we also consider the situation where WCs related to NSI are non-null at the same time. The 90$\%$ C.L. bounds are given in Fig.\ref{fig:bounds2} following~\cite{Falkowski:2017pss}, where the bound for a given WC is obtained after marginalization of the other WCs. For simplicity, we omit from the marginalization process the WCs that are already large ($C^{(1)}_{\substack {lq \\  33 11}}$, $C_{\substack {ld \\  33 11}}$, $C_{\substack {lu \\  33 11}}$) or suppressed ($C_{\substack {Hl}}$, $C_{\substack {Hl}}^{(3)}$) when only one WC is non-null at time. We are focusing only on diagonal NC-NSI, as already explained, implying that only the operators that generate $\epsilon_{V}$ are relevant. Some comments are in order: 
\begin{itemize}
    \item From Fig.~\ref{fig:bounds}, WC related to quarks and taus can be much larger than $\sim\mathcal{O}(\mathrm{TeV}^{-2})$. They are going to be constrained by future experiments, such as DUNE, implying that it is still possible to have sizable NSI in these channels\footnote{Very recently, an analysis performed at CMS for leptoquarks models coupling the tau and first generation quarks was performed~\cite{CMS:2023bdh}. It can be translated in bounds on $C_{\substack {ld \\  3311}}$ and $C_{\substack {lu \\  3311}}$, which we take into account into our numerical analysis}. They can be generated by models with scalar leptoquarks ($\omega_1$, $\Pi_1$, $\Pi_7$, $\zeta$), neutral isosinglet vector bosons ($\mathcal{B}$), or vector leptoquarks ($\mathcal{U}_2$, $\mathcal{Q}_1$, $\mathcal{Q}_5$, $\mathcal{X}$).

\item When considering one non-null WC at time, with the exception of the tau channels, only $C_{\substack {ld \\  2211}}$ can be of order $(\unit{TeV})^{-2}$, inducing NSI of order $10^{-2}$. It may be generated by models with scalar leptoquarks ($\Pi_1$), neutral isosinglet vector bosons ($\mathcal{B}$), or vector leptoquarks ($\mathcal{Q}_5$). If more than one WC is non-null at the same time, it may be possible to have other coefficients of order $(\unit{TeV})^{-2}$, see Fig.~\ref{fig:bounds2}. Notice however that the looser bounds on the WC related to first generations only appear due to a flat direction in the global fit of~\cite{Falkowski:2017pss}, which is only partially broken in our analysis. For UV specific models, this flat direction may not be present. There is also a correlation between $C_{\substack {ld \\  2211}}$ and $C_{\substack {lu \\  2211}}$, allowing them to be larger than the bounds of Fig.~\ref{fig:bounds}. Notice however that a specific UV model must comply with this correlation in order to the looser bounds of Fig.~\ref{fig:bounds2} to be applicable. In particular, it must allow $C_{\substack {ld \\  2211}}$ and $C_{\substack {lu \\  2211}}$ to be both large at the same time, still avoiding other constraints that the model may be subjected to.

\item In single field models, we generally have $(\epsilon_{V})_{\alpha\beta}\sim \lambda_{\alpha}\lambda_{\beta}$, where $\lambda$ is the generic coupling of the weakly coupled field under consideration. Thus, to avoid flavor-violating NSI, only one of the couplings can be non-zero. To generate sizable NSI in both diagonal directions probed by long-baseline experiments, one needs to consider at least two fields, where one only couples to the muon (chosen from the set $\{\Pi_1\;,\mathcal{B}\;,\mathcal{Q}_5\}$), and the other only couples to the tau (chosen from the set $\{\omega_1\;,\Pi_1\;,\Pi_7\;,\zeta\;,\mathcal{B}\;,\mathcal{U}_2\;,\mathcal{Q}_1\;,\mathcal{Q}_5\;,\mathcal{X}\}$).

\item Models with vector-like fermions are extremely constrained by electroweak precision data, and a sizable NSI cannot be induced. 

\end{itemize}

\noindent
In summary, the most promising combination to induce sizeable diagonal NC-NSI is
\begin{align}
    \epsilon_{e e} - \epsilon_{\mu \mu} = \frac{v^{2}}{2}\left( 3C_{\substack {ld \\  22 11}}
    \right)\;,\quad \quad 
    \epsilon_{\tau \tau} -\epsilon_{\mu \mu} = \frac{v^{2}}{2}\left(- 3C_{\substack {ld \\  33 11}}- 3C_{\substack {lu \\  33 11}}  - 6C^{(1)}_{\substack {lq \\  33 11}}+3C_{\substack {ld \\  22 11}} 
    \right)\;,
\end{align}
\noindent
where we are considering the CKM matrix to be approximately diagonal. 
\\

\noindent
Regarding the connection between sizeable diagonal NSI and TeV mediators, models containing scalar leptoquarks $\omega_1$, $\Pi_{1}$, $\Pi_{7}$, $\zeta$ were considered previously  in~\cite{Babu:2019mfe} (in their notation, fields $\chi$, $\Omega$, $\delta$, $\rho$, respectively)\footnote{In Ref.~\cite{Bischer:2019ttk}, the set of leptoquarks (scalar or vector) that can generate NSI is identified, which agrees with our results. The explicit matching formulas are, however, not provided, which prevents one to conclude how sizable the NSI can actually be in these models.}. To the best of our knowledge, models containing extra vector bosons were considered only for light mediator, for instance in~\cite{Farzan:2017xzy}, or did not take into account bounds from charged-lepton violation processes~\cite{Majhi:2022wyp}. Finally, leptoquarks may also affect neutrino experiments other than long-baseline ones~\cite{Berezinsky:1985yw,Robinett:1987ym,Dorsner:2016wpm}. See, for instance,~\cite{Becirevic:2018uab,Huang:2021mki,Calabrese:2022mnp,Kirk:2023fin,DeRomeri:2023cjt,Schwemberger:2023hee}, for recent updates.

\section{Numerical results}
\label{sec:numerical}

\subsection{Details of the simulation}
\label{sec:details}

To perform our simulation of long-baseline experiments, we will use the GLoBES package~\cite{Huber:2004ka,Huber:2007ji}, together with MonteCUBES~\cite{Blennow:2009pk} for efficient sampling. In this work, we are interested in the DUNE experiment for concreteness, however the same approach could be applied to any other existing (or future) long-baseline experiment. In order to ensure that our results can be reproducible, we used the public GLoBES files supplemented by the DUNE collaboration~\cite{DUNE:2021cuw} which are based on the Technical Design Report configuration~\cite{DUNE:2020jqi}. We will be assuming that the standard three-neutrino paradigm corresponds to our true hypothesis in our simulation, which renders $\Delta\chi^2=\chi^{2}_{\rm BSM} - \chi^{2}_{\rm SM}$. Moreover, we will consider the mixing angles,  $\delta_{\rm CP}$ and the mass squared difference given by the NuFIT\footnote{NuFIT 5.2 (2022), www.nu-fit.org} collaboration~\cite{Esteban:2020cvm}, which we reproduce in table~\ref{tab:nufit}.

\begin{table}[th]
    \centering
    \begin{tabular}{ccc}
        \hline\hline\\[-2.5mm]
         & Normal Ordering
     & Inverted Ordering \\[1mm]
      \hline\hline\\[-2.5mm]
      $\sin^{2}\theta_{12}$ & $0.303^{+0.012}_{-0.011}$ & $0.303^{+0.012}_{-0.011}$\\[1mm]
      $\sin^{2}\theta_{13}$ & $0.02203^{+0.00056}_{-0.00059}$ & $0.02219^{+0.00060}_{-0.00057}$\\[1mm]
      $\sin^{2}\theta_{23}$ & $0.572^{+0.018}_{-0.023}$ & $0.578^{+0.016}_{-0.021}$\\[1mm]
      $\delta_{\rm CP}$/\textdegree & $197^{+42}_{-25}$ & $286^{+27}_{-32}$\\[1mm]
      $\frac{\Delta m^{2}_{21}}{10^{-5} \unit{eV}^{2}}$ & $7.41^{+0.21}_{-0.20}$ & $7.41^{+0.21}_{-0.20}$\\[1mm]
      $\frac{\Delta m^{2}_{31}}{10^{-3} \unit{eV}^{2}}$  & $2.511^{+0.028}_{-0.027}$ & $-2.498^{+0.032}_{-0.025}$\\[1mm]
       \hline\hline 
    \end{tabular}
      \caption{Best-fit values for three-flavor oscillation parameters taken from \cite{Esteban:2020cvm}. 
      }
      \label{tab:nufit}
  \end{table}

\vspace{0.5cm}
\noindent
Since long-baseline experiments are not sensitive to the solar mixing parameters ($\theta_{12}$ and $\Delta m_{21}^{2}$), we will fix them to their best-fit values. For simplicity, we will also only consider normal ordering. The detailed results for inverted ordering will be different but the general conclusions will be similar. When performing our fits, we will consider two scenarios. The first one, denoted by optimistic, will consider that NSI parameters are small enough to not affect the extraction of the remaining parameters $\theta_{13}$, $\theta_{23}$, $\delta_{\rm CP}$, $\Delta m_{31}^{2}$. Thus, they can be measured with good precision by DUNE as well other future experiments, showing no deviation among them. This case corresponds to fixing all oscillation parameters to their best-fit values, which would allow DUNE to set stringent bounds on BSM signals. In the second scenario, denoted by conservative, the remaining parameters will be varied at 1$\sigma$ level when computing our $\Delta\chi^2$. In both scenarios, we will adopt a $5\%$ uncertainty on the average matter density, whose value of $\rho=2.848 \;\unit{g}/\unit{cm}^{3}$ is chosen according to the PREM profile~\cite{Dziewonski:1981xy}. Finally, we checked our own implementation for NSI in GLoBES against previous results in the literature, for instance~\cite{Chatterjee:2021wac}, finding compatible results.

\subsection{Numerical results and current bounds}

\begin{figure}[h!]
    \centering
    \includegraphics[scale=0.6]{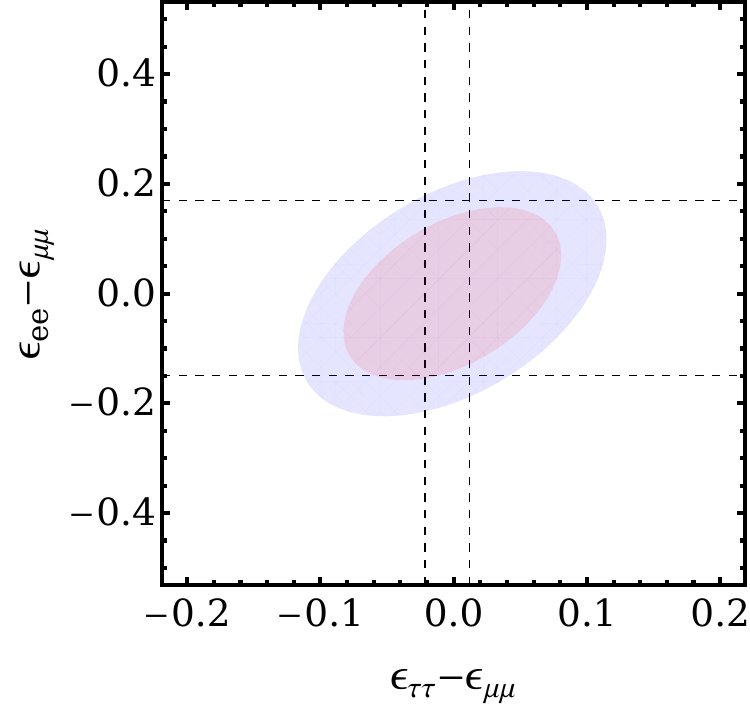}
    \includegraphics[scale=0.6]{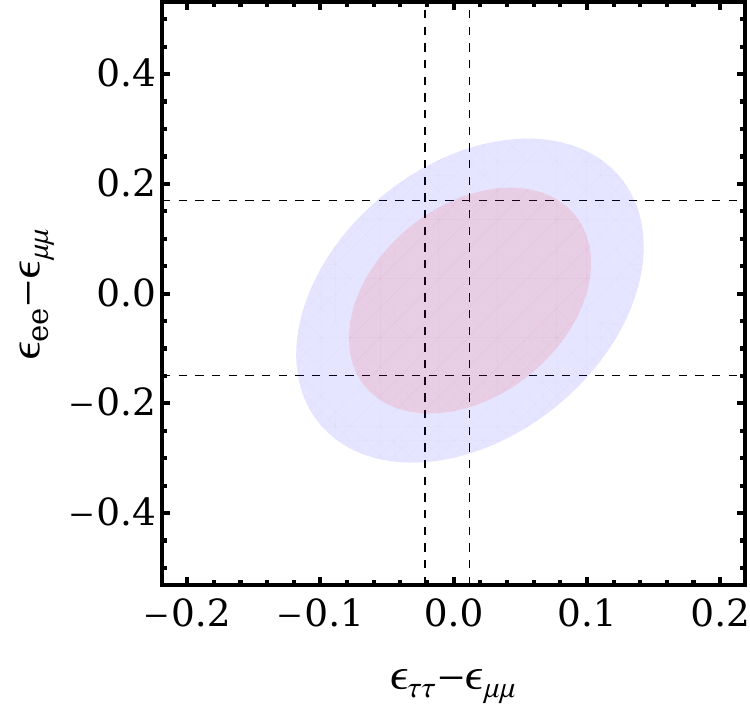}
    \caption{Expected sensitivity on diagonal NC-NSI by DUNE, where the red (blue) region is for  68$\%$ (90$\%$) C.L. The lines represent the current bounds at 90$\%$ C.L. extracted from the most recent global fit to oscillation experiments \cite{Coloma:2023ixt}.The left (right) plot is for the optimistic (conservative) scenario.\label{fig:NC-opt}}
\end{figure}

In this section we present the results of our simulation, following the methodology described in the last section. As pointed out before, inspired by weakly coupled theoretical models, only diagonal NSI in propagation can be probed, which renders two free parameters. In Fig.\ref{fig:NC-opt} we show the 90$\%$C.L. limits on diagonal NC-NSI, considering the two scenarios: optimistic (left) and conservative (right). We also present the most-up-to-date constraints~\cite{Coloma:2023ixt} as dashed lines in the plots. These bounds should be interpreted with care, since we are strictly setting non-diagonal NSI to zero, while in~\cite{Coloma:2023ixt} they were marginalized over. Moreover, since we only focused on long-baseline experiments in our work, we opted to show the 90$\%$ C.L. bounds obtained for the difference of diagonal NSI when only oscillation data is included. In this case, the bounds at 90$\%$ C.L. for $\epsilon_{ee}-\epsilon_{\mu\mu}$ are composed of three disjoint regions, and we only show the one that comprises the hypothesis of null NSI (the true hypothesis in our simulation as explained in the last section). Finally, these bounds were extracted by considering that NC NSI couples simultaneously to quarks and electrons, which may not be realized in a particular UV model. Actually, in light of the discussion presented in section~\ref{sec:analitcal_dune}, sizable NSI are expected from couplings with quarks only. In this case (NC NSI with protons, for instance), the 90$\%$ C.L. allowed region that encompasses the null NSI hypothesis is larger, $-0.21<\epsilon_{ee}-\epsilon_{\mu\mu}<1.0$, $-0.015<\epsilon_{\tau\tau}-\epsilon_{\mu\mu}<0.048$.

\noindent
As can be seen in Fig.\ref{fig:NC-opt}, DUNE will not provide a better sensitivity in the $\epsilon_{\tau\tau}-\epsilon_{\mu\mu}$ direction even if we only consider NC NSI with protons. In this scenario, for the $\epsilon_{ee}-\epsilon_{\mu\mu}$ direction, DUNE has the potential to improve the bound, in particular for positive values of $\epsilon_{ee}-\epsilon_{\mu\mu}$. As a by-product, we also provide a semi-analytical formula for $\Delta\chi^2$ in the optimistic scenario, in terms of $\epsilon_{ee}-\epsilon_{\mu\mu}$, and $\epsilon_{\tau\tau}-\epsilon_{\mu\mu}$
\begin{align}\label{eq:chi2}
   \Delta\chi^2 |_{\text{opt}}= & 115.1 \left(\epsilon _{\text{ee}}-\epsilon _{\mu \mu }\right){}^2-197.1 \left(\epsilon _{\text{ee}}-\epsilon _{\mu \mu }\right)
   \left(\epsilon _{\tau \tau }-\epsilon _{\mu \mu }\right)-0.1577 \left(\epsilon _{\text{ee}}-\epsilon _{\mu \mu }\right)\nonumber\\
   &+429.4
   \left(\epsilon _{\tau \tau }-\epsilon _{\mu \mu }\right){}^2+0.8418 \left(\epsilon _{\tau \tau }-\epsilon _{\mu \mu
   }\right)+0.001713.
\end{align}

\begin{figure}[h!]
    \centering
    \includegraphics[scale=0.6]{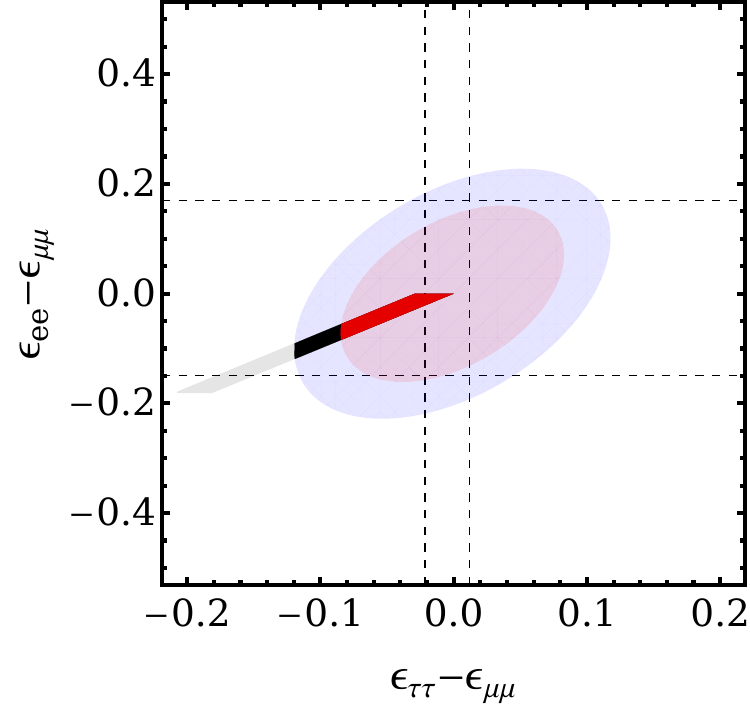}
    \includegraphics[scale=0.6]{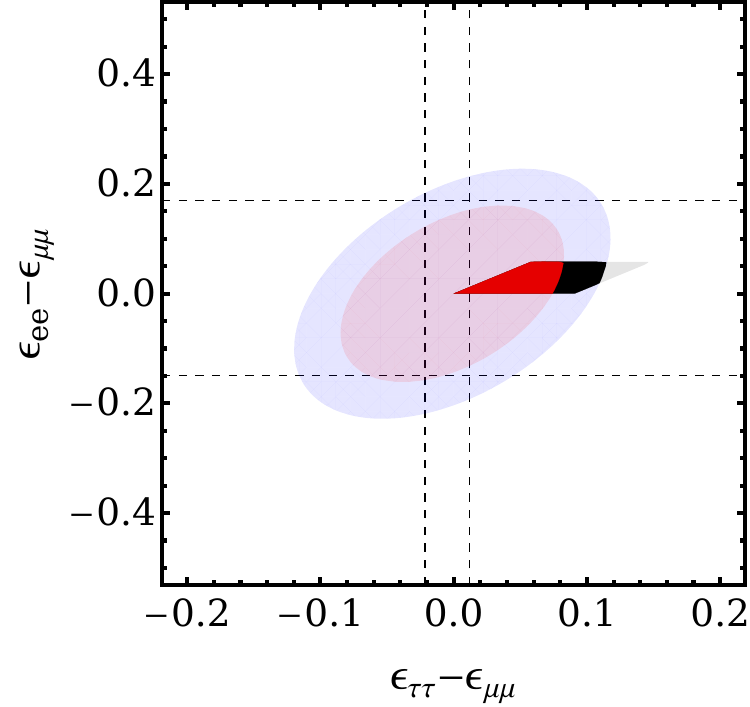}
    \caption{Expected sensitivity on diagonal NC-NSI by DUNE, for some illustrative models containing leptoquarks. The left (right) plot is for the scalar (vector) case, where the red (black) region is for  68$\%$ (90$\%$) C.L. The gray area denote the presently allowed region given other constraints. For illustration we only consider the optimistic scenario.\label{fig:UV}}
\end{figure}

\noindent
By employing eq.\eqref{eq:chi2} and the matching conditions of eq.\eqref{eq:NSI_prop}, it is possible to obtain the allowed region for any weakly coupled model of table~\ref{tab:UV}. For illustration, we consider two examples: models S and V. In model S, we have two scalar leptoquarks, $\Pi_1$, $\omega_1$, where they only couple to first-generation down-type quarks and second/third-generation leptons respectively ($\big(y_{\Pi_1}\big)_{21},\big(y_{\omega_1}^{ql}\big)_{13}\neq0$, see eq.\eqref{eq:lag_scalar} ). A similar model was studied by~\cite{Babu:2019mfe} in connection to radiative neutrino mass generation models. The diagonal NSI induced at tree-level by this model are
\begin{align}
    \epsilon_{e e} - \epsilon_{\mu \mu} = \frac{3v^{2}}{2}\left( -\frac{\left|\big(y_{\Pi_1}\big)_{21}\right|^{2}}{2M^2_{\Pi_1}}
    \right)\;,\quad\quad
    \epsilon_{\tau \tau} -\epsilon_{\mu \mu} = \frac{3v^{2}}{2}\left(  - \frac{\left|\big(y_{\omega_1}^{ql}\big)_{13}\right|^{2}}{2M^2_{\omega_1}}-\frac{\left|\big(y_{\Pi_1}\big)_{21}\right|^{2}}{2M^2_{\Pi_1}}
    \right)\;.
\end{align}
It is immediate to observe that only negative values can be attained, and for non-null values of $y_{\Pi_1}$ both directions have an absolute lower bound. We choose as benchmark for the masses $M_{\Pi_1}=1250\;\unit{GeV}$, $M_{\omega_1}=1000\;\unit{GeV}$ in accordance to the constraints derived in~\cite{Babu:2019mfe}. For the Yukawas, we varied $\big(y_{\Pi_1}\big)_{21}$ in the range allowed by perturbative unitarity $y<\sqrt{2\pi}$, while $\big(y_{\omega_1}^{ql}\big)_{13}$ is bounded to be lower than 0.8 in accordance with the  new result from CMS~\cite{CMS:2023bdh}. The region allowed by present constraints can be seen as the gray area in the left plot of Fig.~\ref{fig:UV}, while in red (black) we show the 68$\%$ (90$\%$) C.L. allowed region from DUNE for the model S.
\\

\noindent
For model V, we consider vector leptoquarks $\mathcal{Q}_5$, and $\mathcal{U}_2$, where they only couples to first-generation down-type quarks and second/third-generation leptons respectively ($\big(g_{\mathcal{Q}_5}^{dl}\big)_{12},\big(g_{\mathcal{U}_2}^{lq}\big)_{31}\neq0$, see eq.\eqref{eq:lag_vector}). Clearly it is a simplified model, whose UV completion is beyond the scope of our work. In this case, the diagonal NSI are given by
\begin{align}
    \epsilon_{e e} - \epsilon_{\mu \mu} = \frac{3v^{2}}{2}\left( \frac{\left|\big(g_{\mathcal{Q}_5}^{dl}\big)_{12}\right|^{2}}{M^2_{\mathcal{Q}_5}}
    \right)\;,\quad\quad
    \epsilon_{\tau \tau} -\epsilon_{\mu \mu} = \frac{3v^{2}}{2}\left(  \frac{|\big(g_{\mathcal{U}_2}^{lq}\big)_{31}|^{2}}{M^2_{\mathcal{U}_2}}+\frac{\left|\big(g_{\mathcal{Q}_5}^{dl}\big)_{12}\right|^{2}}{M^2_{\mathcal{Q}_5}}
    \right)\;.
\end{align}
Contrarily to model S, the NSI can only be positive, while, similarly to before, there is an lower bound for their values if $g_{\mathcal{Q}_5}^{dl}\neq0$. Using as benchmark for the masses the values $M_{\mathcal{Q}_5}=1250\;\unit{GeV}$, $M_{\mathcal{U}_2}=1000\;\unit{GeV}$, and allowing the gauge couplings to be as large as unit, we obtain the right plot of Fig.\ref{fig:UV}. The colours follow the same pattern used for model S. Notice that in both models there is a correlation between $\epsilon_{ee}-\epsilon_{\mu\mu}$ and $\epsilon_{\tau\tau}-\epsilon_{\mu\mu}$. Thus, it is not possible to have large values for the former avoiding the stringent bounds on the latter. We could, nevertheless, devise a mixed model where the diagonal NSI would be given by
\begin{align}
    \epsilon_{e e} - \epsilon_{\mu \mu} &= \frac{3v^{2}}{2}\left( -\frac{\left|\big(y_{\Pi_1}\big)_{21}\right|^{2}}{2M^2_{\Pi_1}}
+\frac{\left|\big(g_{\mathcal{Q}_5}^{dl}\big)_{12}\right|^{2}}{M^2_{\mathcal{Q}_5}}
    \right)\;,\\
    \epsilon_{\tau \tau} -\epsilon_{\mu \mu} &= \frac{3v^{2}}{2}\left(  - \frac{\left|\big(y_{\omega_1}^{ql}\big)_{13}\right|^{2}}{2M^2_{\omega_1}}-\frac{\left|\big(y_{\Pi_1}\big)_{21}\right|^{2}}{2M^2_{\Pi_1}}
    + \frac{|\big(g_{\mathcal{U}_2}^{lq}\big)_{31}|^{2}}{M^2_{\mathcal{U}_2}}+\frac{\left|\big(g_{\mathcal{Q}_5}^{dl}\big)_{12}\right|^{2}}{M^2_{\mathcal{Q}_5}}
    \right)\;.
\end{align}
\noindent
In this case, cancellations between the different coefficients can happen in the $\epsilon_{\tau \tau} -\epsilon_{\mu \mu}$ direction, allowing $\epsilon_{ee} -\epsilon_{\mu \mu}$ to be still large. In particular, for $\big(y_{\omega_1}^{ql}\big)_{13}/M_{\omega_1}\sim\sqrt{2}\big(g_{\mathcal{Q}_5}^{dl}\big)_{12}/M_{\mathcal{Q}_5}$ and $\big(y_{\Pi_1}\big)_{21}/M_{\Pi_1}\sim\sqrt{2}\big(g_{\mathcal{U}_2}^{lq}\big)_{31}/M_{\mathcal{U}_2}$, we obtain that $\epsilon_{\tau \tau} -\epsilon_{\mu \mu}\sim 0$, while $|\epsilon_{ee} -\epsilon_{\mu \mu}|$ can be of order $10^{-1}$.
\\

\noindent
Similar analysis can be performed for an arbitrary number of fields collected in table~\ref{tab:UV} in a straightforward way. This feature will be particularly important after data is collected from the DUNE experiment. At that stage, only the $\chi^{2}$ function given in eq.\eqref{eq:chi2} will have to be updated, and, for instance, exclusion analyses regarding the weakly coupled models of table~\ref{tab:UV} will follow automatically. Even more interestingly is the case where the DUNE experiment reports an anomaly in any of the NSI diagonal directions. In this scenario, all weakly coupled models that could potentially explain the anomaly can be easily identified, and their predictions can be straightforwardly obtained.

\vspace{0.4cm}
\noindent
Finally, since weakly coupled models with mediators at TeV scale (or higher) can only induce diagonal NSI, one can wonder what kind of model is needed to explain an anomaly seen at DUNE in any of the non-diagonal NSI directions. This can be achieved by allowing some of the BSM fields to be light (below weak scale), replacing the SMEFT by another effective theory which contains these new light degrees of freedom. Explicit models containing light charged scalars (masses around 100 GeV) were advocated in~\cite{Forero:2016ghr,Dey:2018yht}, for instance. For these particular models, $\epsilon_{e\tau}$ could be larger than $10^{-2}$, complying with CLFV constraints. Another possibility is to consider models with a light ($\mathcal{O}(\unit{MeV})$ or below) neutral gauge boson ($Z'$), as in~\cite{Farzan:2015hkd,Farzan:2016wym,Farzan:2019xor} for instance. The non-diagonal NSI generated by these models can be of order $10^{-1}$. In order to circumvent CFLV in these models, either the light mediator field~\cite{Forero:2016ghr} or the neutrino field in the $\rm{SU}(2)$ doublet~\cite{Farzan:2016wym} are allowed to mix with other BSM fields, after the electroweak symmetry breaking. In all cases, only NSI in propagation are induced. For NSI in production only, to the best of our knowledge, no UV model has been envisaged. Nevertheless, if a light sterile neutrino is present, it will induce NSI at production, and propagation simultaneously~\cite{Blennow:2016jkn}.
\\

\noindent
Since the main purpose of this work is to provide a complete list of weakly coupled models relevant for long-baseline experiments, we opted to consider only the SMEFT as underlying EFT at the weak scale. This is motivated by the fact that the particles present in the SMEFT framework are the only ones already experimentally observed. Therefore, considering  lighter degrees of freedom necessarily implies an assumption about the nature of still unseen particles, not allowing to draw a complete picture. Nevertheless, if the SMEFT is replaced by another EFT with at least one BSM light degree of freedom, the tree-level matching can be performed, and allowed regions in parameters space can be drawn in similarity to the red and black regions in Fig.~\ref{fig:UV}. To this end, it is necessary to obtain the $\chi^{2}$ function when all NSI at propagation and production are present simultaneously, allowing one, for instance, to obtain the expected sensitivity on a given pair of NSI by DUNE, in similarity to the red and blue regions in Fig.~\ref{fig:UV}. In other words, 
with a given EFT with light BSM degrees of freedom in mind, one can ask which are the best-possible bounds expected by DUNE on NSI, for both propagation and production. A comprehensive analysis was performed in~\cite{Blennow:2016etl}, showing the appearance of degenerate solution for $\theta_{23}$ and $\delta_{\rm CP}$, as well as for the NSI (see also~\cite{Coloma:2015kiu} for an analysis considering NSI at propagation only). Nevertheless, this reference is not only previous to the 2020 update in the three-neutrino paradigm parameters~\cite{Esteban:2020cvm}, but it does not consider the updated experimental setup planned by the DUNE collaboration~\cite{DUNE:2020ypp,DUNE:2020jqi,DUNE:2021cuw}
neither the indirect effects included recently in the treatment of NSI in production~\cite{Breso-Pla:2023tnz}.  Therefore, it is justified to perform an updated analysis, which we discuss on the remaining of this section.
\\

\begin{figure}[h!]
    \centering
    \includegraphics[scale=0.3]{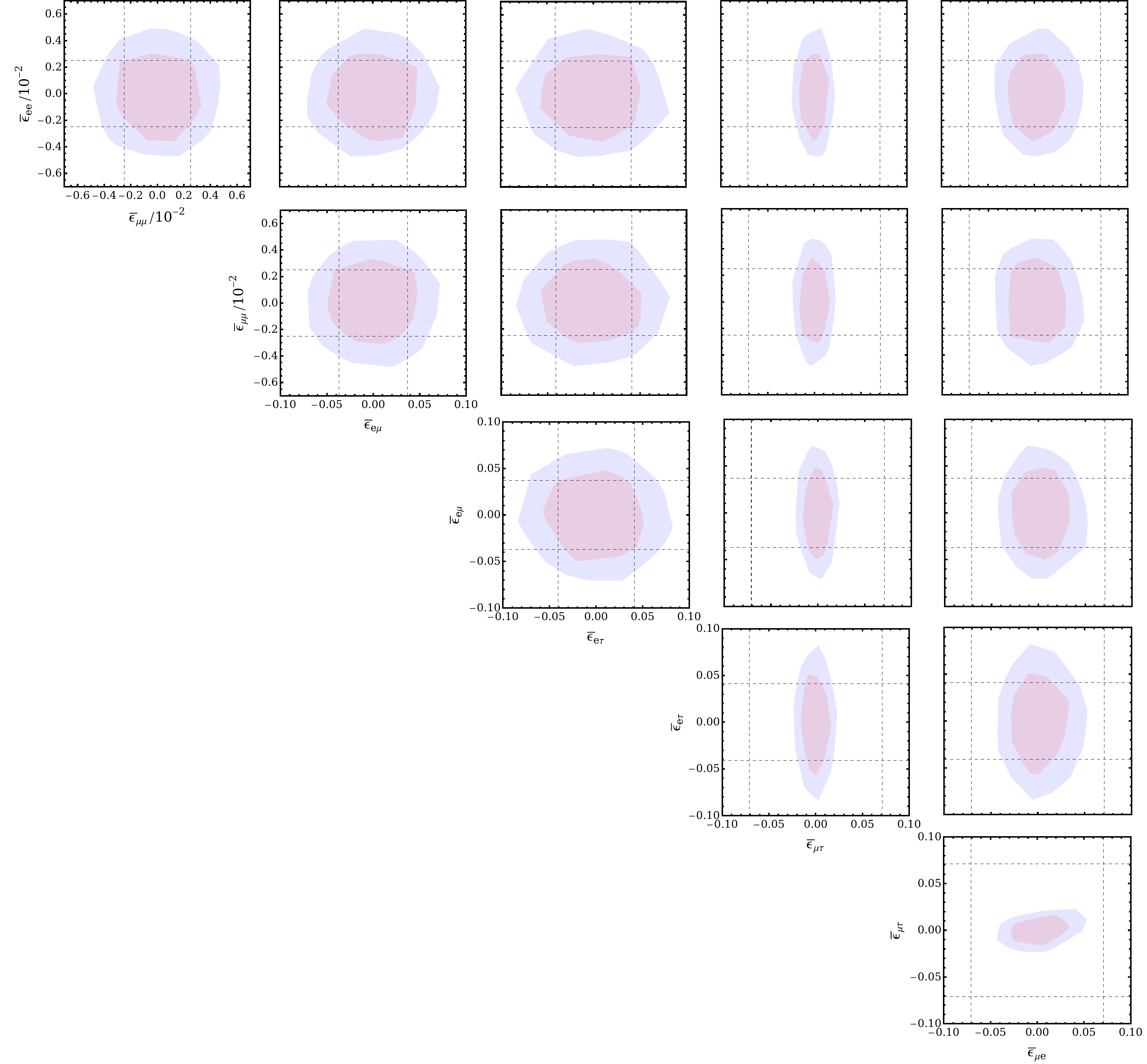}
    \caption{Expected sensitivity on CC-NSI by DUNE for the optimistic scenario, where the red (blue) region is for  68$\%$ (90$\%$) C.L. The lines represent the current bounds at 90$\%$ C.L. coming from $\Gamma_{\pi\to e\nu}/\Gamma_{\pi\to\mu\nu}$.\label{fig:all_prod_op}}
\end{figure}

\begin{figure}[h!]
    \centering
    \includegraphics[scale=0.3]{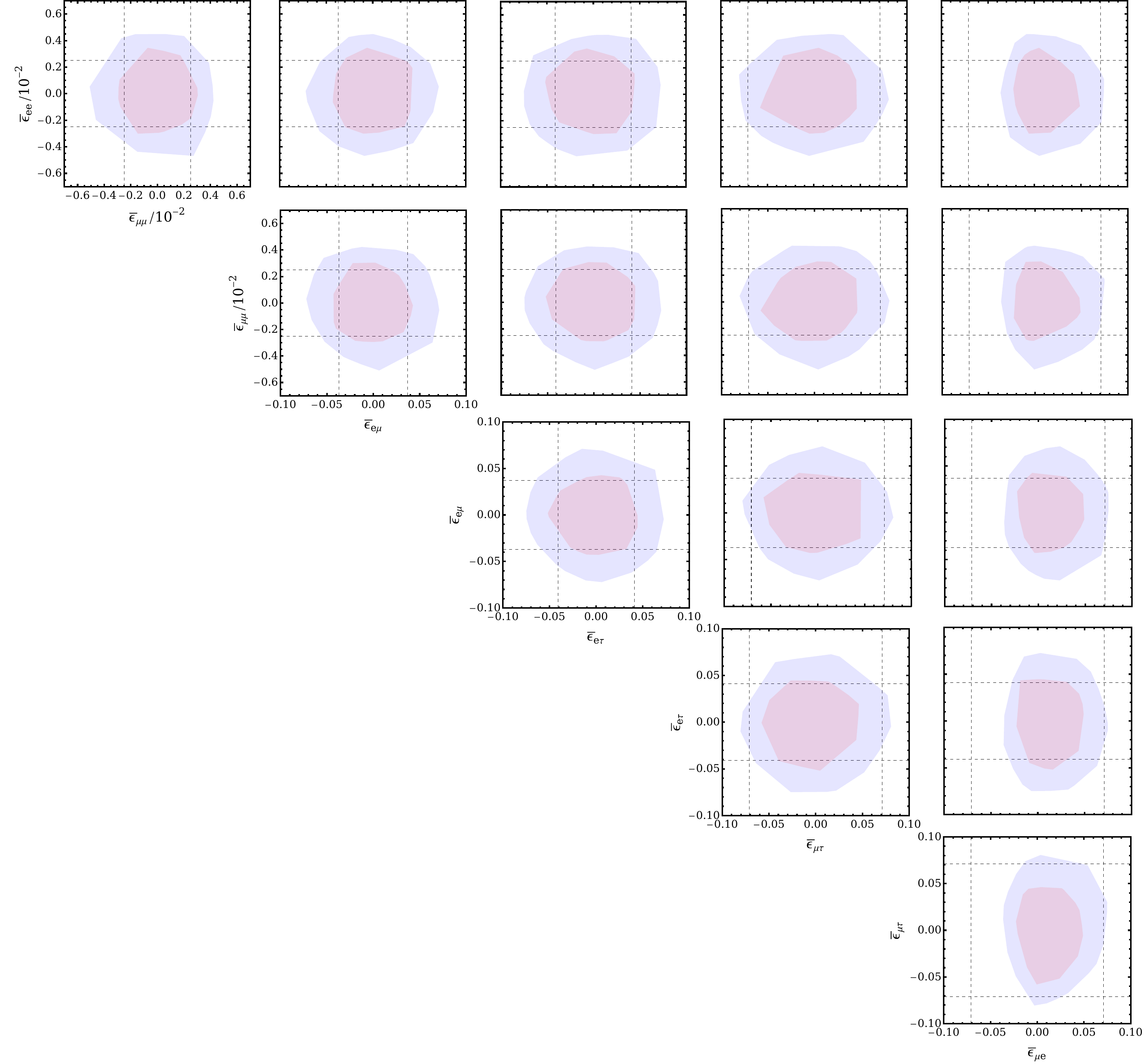}
    \caption{Expected sensitivity on CC-NSI by DUNE for the conservative scenario, where the red (blue) region is for  68$\%$ (90$\%$) C.L. The lines represent the current bounds at 90$\%$ C.L. coming from $\Gamma_{\pi\to e\nu}/\Gamma_{\pi\to\mu\nu}$.\label{fig:all_prod_con}}
\end{figure}

\noindent
As explained in Section~\ref{sec:details}, we will consider two scenarios when performing the global fit. The first (optimistic) corresponds in practice to fix all the three-neutrino standard paradigm parameters to their best-fit values. In the second (conservative) the three-neutrino standard paradigm parameters are varied at 1$\sigma$ level, with the exception of the solar parameters which are kept fixed at their best-fit values. Regarding the NSI parameters, we have now 11 free parameters, five from propagation and six from production (we restrict ourselves to real coefficients). As already explained, if only diagonal CC-NSI are included, they will cancel in the neutrino conversion rate. This constitutes a flat direction in our plot. In order to break this direction, we could add data from other experiments that aim to measure the rate $\Gamma_{\pi\to e\nu}/\Gamma_{\pi\to\mu\nu}$, for instance. This approach, however, requires that the likelihood is obtained independently from GLoBES, with dedicated routines. Another possibility is to include priors on all CC-NSI in the GLoBES engine, penalizing values that do not comply with the experimental rate $\Gamma_{\pi\to e\nu}/\Gamma_{\pi\to\mu\nu}$ at 90$\%$ C.L. We have opted for the latter approach, so the simulation to the DUNE experiment already implemented in GLoBES can be used. Since our goal is to obtain the best-case scenario for DUNE, we have adopted the most restrictive possibility when implementing the constraints. It stands to consider only one non-null CC-NSI at time, in similarity to the approach performed in~\cite{Falkowski:2021bkq}. We will show these bounds as lines in the plot of figs.\ref{fig:all_prod_op}-\ref{fig:all_prod_con}. \\

\begin{figure}[h!]
    \centering
    \includegraphics[scale=0.4]{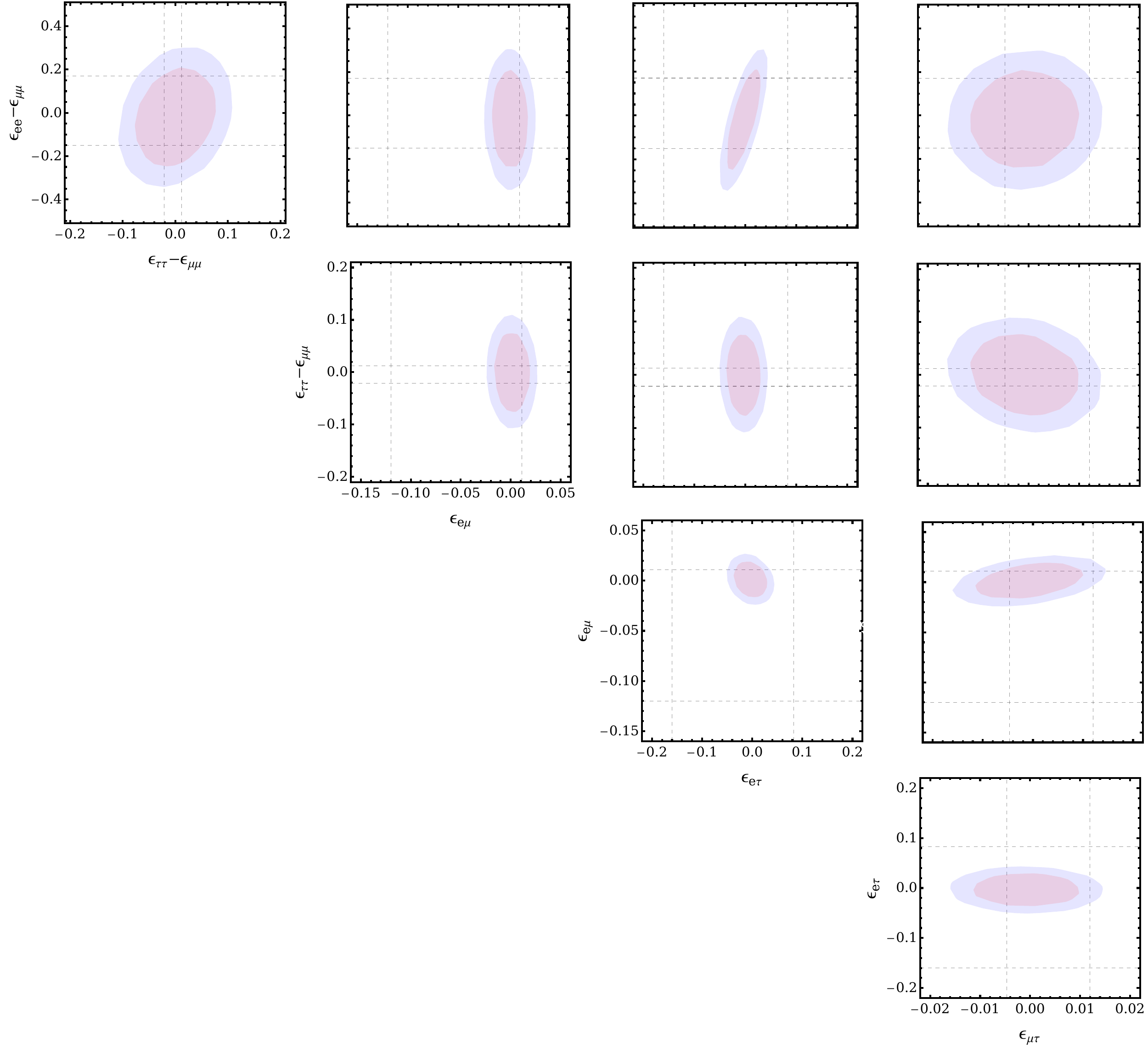}
    \caption{Expected sensitivity on NC-NSI by DUNE for the optimistic scenario, where the red (blue) region is for 68$\%$ (90$\%$) C.L. The lines represent the current bounds at 90$\%$ C.L. extracted from the most recent global fit to oscillation experiments \cite{Coloma:2023ixt}. \label{fig:all_prop_op}}
\end{figure}

\begin{figure}[h!]
    \centering
    \includegraphics[scale=0.4]{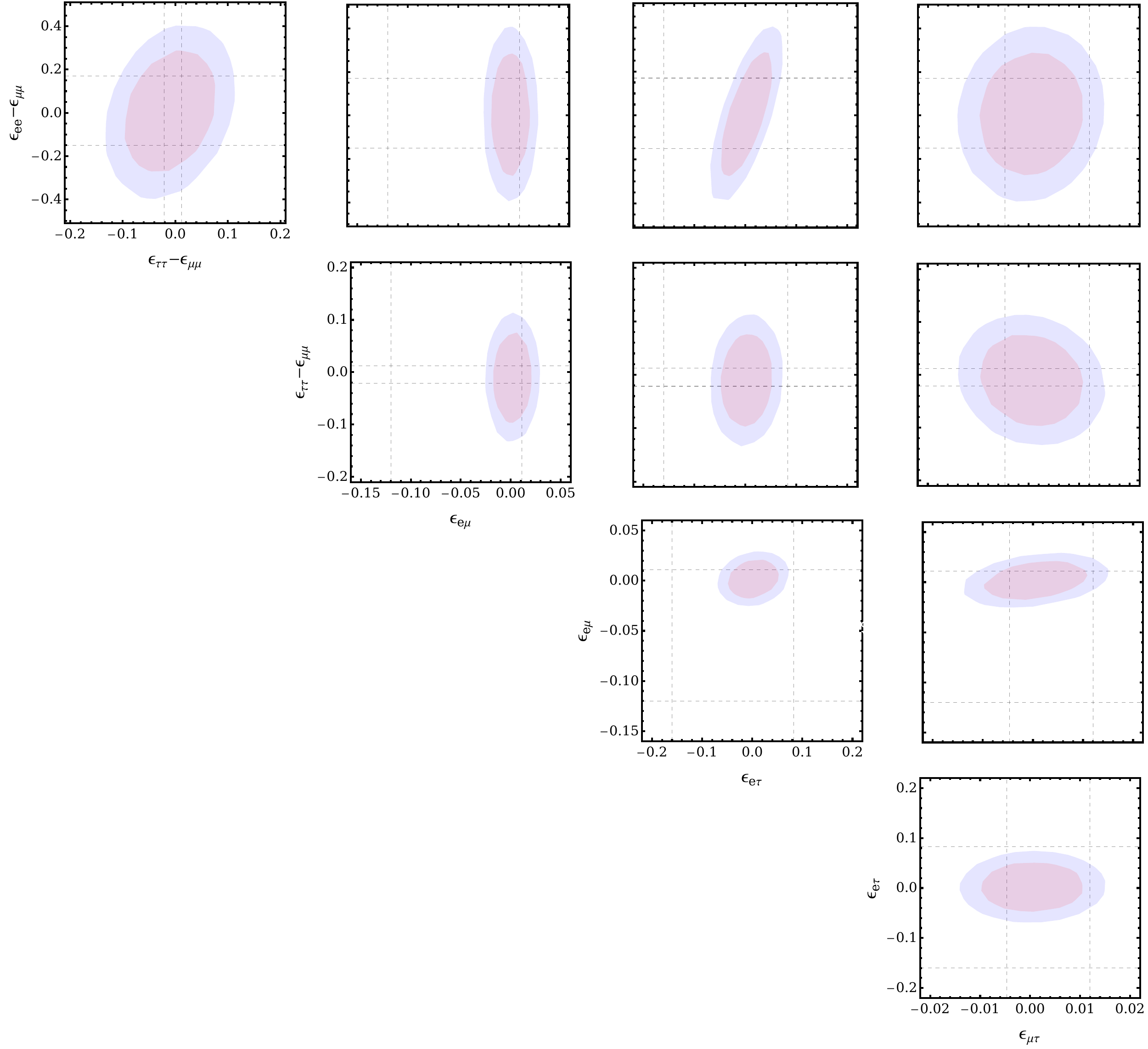}
    \caption{Expected sensitivity on NC-NSI by DUNE for the conservative scenario, where the red (blue) region is for  68$\%$ (90$\%$) C.L. The lines represent the current bounds at 90$\%$ C.L. extracted from the most recent global fit to oscillation experiments \cite{Coloma:2023ixt}. \label{fig:all_prop_con}}
\end{figure}

\noindent
After marginalization of the CC-NSI parameters, we present in figs.\ref{fig:all_prop_op}-\ref{fig:all_prop_con}, the sensitivity for NC-NSI from DUNE, for the optimistic and conservative scenario, respectively. We also show the 90$\%$ C.L. bounds from the most-up-to-date global fit on matter NSI~\cite{Coloma:2023ixt} as black lines. For diagonal NSI, the bounds are obtained by considering data only from oscillation experiments, while for non-diagonal NSI data from other kind of experiments is included as well.
As can be seen in the figure, the sensitivity on the diagonal NSI decreases when compared to our previous analysis of Fig.~\ref{fig:NC-opt}. This is expected since the minimum chi squared in both cases were similar, however, non-diagonal NSI were set to zero before, while they are marginalized over now. Moreover, the most stringent bounds are expected for $\epsilon_{e \mu} $, and $\epsilon_{\mu\tau} $, showing that NSI smaller than $10^{-2}$ cannot be probed by DUNE. Therefore, since in weakly coupled theories the non-diagonal elements are already constrained to be several orders of magnitude smaller~\cite{Bischer:2019ttk}, any possible anomaly seen in these channels will require a new theoretical framework. 
\\

\noindent
Finally, in Fig.\ref{fig:all_prod_op}-\ref{fig:all_prod_con} we present the sensitivity on CC-NSI after marginalization of the NC-NSI parameters, for the optimistic and conservative scenario, respectively.
\noindent
We recall that we have adopted priors coming from the experimental measurement of the pion decay, which are quite stringent. Therefore, for most of the parameters, DUNE, even in the best-case scenario, will not be able to improve the bounds. The only exemption is for the last plot in the optimistic scenario (Fig.\ref{fig:all_prod_op}), considering the parameters $\bar{\epsilon}_{\mu e}$, $\bar{\epsilon}_{\mu \tau}$. In these channels, DUNE could (potentially) report an anomaly. This is reasonable since these parameters will modify the neutrino fluxes for muon neutrino conversion, which is the main channel for neutrino production at DUNE. Notice however, that in the conservative scenario, the sensitivity on these NSI parameters is significantly worse, mainly due to a correlation between $\bar{\epsilon}_{\mu \tau}$ and $\sin^{2}\theta_{23}$.

\clearpage

\section{Summary}
\label{sec:conclusion}

There is at present a very active experimental program in neutrino physics, with many planned experiments in the near future. Their main goal is to measure with precision three ($\theta_{23}$, $\delta_{CP}$, $\Delta m^{2}_{31}$) of the six parameter of the three-neutrino standard paradigm. It cannot be dismissed, however, that the standard scenario is incomplete, since it does not introduce, per se, a mechanism for neutrino mass generation. The simplest proposal, seesaw type I, is realized with very heavy sterile neutrinos, implying that no deviation from the standard paradigm should be seen in future experiments. If, however, the next-generation long-baseline experiments report an anomaly, it may open a window not only to the neutrino mass generation mechanism realized in Nature, but potentially to the presence of new weakly coupled states. 
\\

\noindent
In this work we aimed to identify all the weakly coupled states up to spin 1 that could potentially be probed by long-baseline neutrino experiments, in particular DUNE. This was achieved by employing an effective field theory framework, where the new states are assumed to be heavier than the electroweak scale (SMEFT). We rederived the matching between neutrino non-standard interactions (NSI) coefficients (defined at low-energy) and the WC in the SMEFT at tree-level~\cite{Bischer:2019ttk}, as well as the matching from the SMEFT to the relevant weakly coupled theories. The final formulas are organized in appendix~\ref{ap:matching}, which can be employed not only for long-baseline neutrino experiments, but for all cases that the description by NSI parameters is applicable. In particular, for long-baseline experiments, only a small number of weakly coupled fields can induce sizable NSI parameters, in view of present constrains from collider, electroweak precision data and the non-observance of charged lepton flavor violation. The latter strongly suppress non-diagonal NSI, as noted long ago~\cite{Bergmann:1999pk}. For diagonal NSI, we found that only models containing leptoquarks (scalar or vector) or an extra neutral gauge boson are relevant at tree-level matching. We illustrated the potential from DUNE to unveil some of these models. For ease of the reader the full chain for the matching, as well as the chi squared function in terms of SMEFT WC is provided in \url{https://gitlab.com/alcherchiglia/eft-neutrinos}. Therefore, any combination of fields can be studied, in similarity to the analysis that lead to figure~\ref{fig:UV}. 
\\

\noindent
Finally, if lighter BSM particles (below the electroweak scale) are allowed, the SMEFT should be replaced by another EFT. However, since no new particles has been observed so far, the EFT to be chosen is yet open to discussion. In order to be as general as possible, we performed a global fit to assess DUNE sensitivity to NSI coming from propagation and production simultaneously. Our main aim is to provide model builders with easy access to the strongest bound that DUNE can set on each of the NSI parameters. Our results can be seen in figs.\ref{fig:all_prop_op} and \ref{fig:all_prod_op}, where we assumed that the three-neutrino standard paradigm parameters are fixed to their best-fit values. In this best-case scenario, DUNE could potentially report anomalies in non-diagonal NSI channels, which would necessarily indicate that new physics is light (below the electroweak scale) and, potentially, the gauge group of the SM need to be extended. For completeness, we also present in figs.\ref{fig:all_prop_con} and \ref{fig:all_prod_con} the scenario where the three-neutrino standard paradigm parameters are not fixed, but marginalized over. In this case, the sensitivity on NSI parameters is worsen as expected, however, DUNE can still report anomalies in some of non-diagonal channels.

\section*{Acknowledgments}
We gratefully acknowledge Jorge Blas for valuable comments regarding global fits, Juan Carlos Criado for providing the library of UV models, Pablo Olgoso for his help with \texttt{smelli} and \texttt{flavio}, Pedro Pasquini for his help with GLoBES codes, and Bruno Zamorano García for an enlightening discussion about the DUNE experiment. A.C.~acknowledges support from National Council for Scientific and Technological Development – CNPq through projects 166523\slash2020-8 and 201013\slash2022-3, and FAPESP through Grant 
2022/08954-2. This work has been partially supported by MCIN/AEI (10.13039/501100011033) and ERDF under grants PID2019-106087GB-C22 and PID2022-139466NB-C21 and by the Junta de Andalucía grant FQM 101 and by Consejería de Universidad, Investigación e Innovación, 
Gobierno de España and Unión Europea – NextGenerationEU under grant $\mathrm{AST22\_6.5}$.

\appendix

\section{QFT description for NSI at detection}
\label{ap:detection}

In this appendix we briefly discuss some of the challenges involved in order to include NSI effects at neutrino detection in long-baseline experiments. As in the main text, we are focusing on an QFT description~\cite{Falkowski:2019kfn} for neutrino oscillations, which allows a clear connection to UV models. For a Quantum Mechanical (QM) approach, the introduction of NSI at detection is straightforward~\cite{Ohlsson:2012kf}. As pointed out in~\cite{Falkowski:2019kfn}, however, the matching between QM-NSI and QFT-NSI does not hold in general beyond linear order in the NSI parameters. 
\\

\noindent
In long-baseline experiments, in particular DUNE, neutrino detection relies on the interaction of the neutrino flux with argon targets~\cite{NuSTEC:2017hzk}, where nuclear effects are expected to be substantial. Moreover, DUNE is expected to operate with neutrino fluxes at the few GeV region, where distinct processes can be equally important, namely quasi-elastic scattering (CCQE), resonant meson production (CCRES) and deep inelastic scattering (CCDIS)~\cite{Formaggio:2012cpf}. Given this scenario, there are on-going efforts to better understand the neutrino-nucleus cross-section whose uncertainties are still at the $10-20\%$ level~\cite{Katori:2016yel,Branca:2021vis,DiLodovico:2023jgr}.
\\

\noindent
In any case, neutrino detection is expected to occur through a charged-current process, mediated by the W-boson in the SM. Therefore, when proposing a BSM scenario, it is in principle possible to avoid any (new) complications due to the interaction with the nucleus if the BSM interaction effectively only modifies the neutrino-charged lepton-W boson vertex. In this case, it is straightforward to include the BSM effect both in production and detection in the neutrino event rate defined in eq.\eqref{eq:rate}, since there is a one-by-one correspondence to the QM-NSI approach. However, as the results of our section~\ref{sec:analitcal_dune} show, the Wilson coefficient $C^{(3)}_{Hl}$ (which stands for the deviation from the SM of the interaction vertex neutrino-charged lepton-W boson) is suppressed. Therefore, the BSM effect from this particular vertex is negligible. We should notice that the BSM V-A interaction, parametrized in the LEFT by $\epsilon_{L}$, is matched to other SMEFT operators as well, see eq.~\eqref{eq:LEFT_VLL}. However, we checked with \texttt{smelli}~\cite{Aebischer:2018bkb,Straub:2018kue,Aebischer:2018iyb} that their contribution is also negligible. Therefore effects in neutrino detection are negligible for weakly coupled models effectively described by V-A interaction. For other types of interaction (parametrized by $\epsilon_{R}$, $\epsilon_{S}$, $\epsilon_{P}$, $\epsilon_{T}$), one needs to resort to specific detection processes, whose theoretical description is still lacking.

\section{Lagrangian}
\label{ap:notation}
\newcommand{\hc}{\mathrm{h.c.}}
\newcommand{\ctoprule}{\toprule[0.5mm]}
\newcommand{\cbottomrule}{\bottomrule[0.5mm]}
\newcommand{\cmrule}{\midrule[0.25mm]}
\newcommand{\crowcolor}{\rowcolor[rgb]{0.9,0.9,0.9}}

For convenience of the reader, we reproduce below the parts of the Lagrangian defined in~\cite{deBlas:2017xtg} relevant for our work. We also present the representation under the SM gauge group for each of the fields. The complete set of conventions used can be found in the original reference.

\subsection{New Scalars}

In table~\ref{t:scalars} we collect the new scalars that may contribute to NSI while in eq.\eqref{eq:lag_scalar} we present their interacting Lagrangian.


\begin{table}[h!]
  \begin{center}
    {\small
      \begin{tabular}{lccccccccc}
        \ctoprule
        \crowcolor
        Name &
        ${\cal S}_1$ &
        $\varphi$ &
        $\Xi_1$ &
      ${\omega}_{1}$ &
        $\Pi_1$ &
        $\Pi_7$ &
        $\zeta$ &
      \\
        Irrep &
        $\left(1,1\right)_1$ &
        $\left(1,2\right)_{\frac 12}$ &
        $\left(1,3\right)_1$ &
      $\left(3,1\right)_{-\frac 13}$ &
        $\left(3,2\right)_{\frac 16}$ &
        $\left(3,2\right)_{\frac 76}$ &
        $\left(3,3\right)_{-\frac 13}$
      \\[1.3mm]
        \cbottomrule
      \end{tabular}
    }
    \caption{New scalars and their irrepresentation under the gauge group of the SM $(\rm{SU}(3),\rm{SU}(2))_{\rm{U}(1)}$.}
    \label{t:scalars}
  \end{center}
\end{table}

%
\begin{align}
  -\mathcal{L}_{\mathrm{S}} = & \;
  + \left\{
    (y_{\mathcal{S}_1})_{rij}
    \mathcal{S}_{1r}^\dagger \bar{l}_{Li} i\sigma_2 l^c_{Lj}
    + \hc
  \right\}
  \nonumber \\ &
  + \left\{
    (y^e_{\varphi})_{rij} \varphi_r^\dagger \bar{e}_{Ri} l_{Lj}
    + (y^d_{\varphi})_{rij} \varphi_r^\dagger \bar{d}_{Ri} q_{Lj}
    + (y^u_{\varphi})_{rij} \varphi_r^\dagger i\sigma_2 \bar{q}^T_{Li} u_{Rj}
  \right.
  \nonumber \\ &
  \quad
  \left.
    + (\lambda_{\varphi})_r
    \left(\varphi^\dagger_r \phi \right) \left(\phi^\dagger \phi\right)
    + \hc
  \right\}
  \nonumber \\ &
  + \frac{1}{2} (\lambda_{\Xi_1})_{rs}
  \left(\Xi^{a\dagger}_{1r} \Xi^a_{1s}\right)
  \left(\phi^\dagger\phi\right)
  + \frac{1}{2} (\lambda'_{\Xi_1})_{rs}
  f_{abc} \left(\Xi_{1r}^{a\dagger} \Xi_{1s}^b\right)
  \left(\phi^\dagger \sigma^c \phi\right)
  \nonumber \\ &
  +\left\{
    (y_{\Xi_1})_{rij}
    \Xi^{a\dagger}_{1r} \bar{l}_{Li} \sigma^a i\sigma_2 l^c_{Lj}
    + (\kappa_{\Xi_1})_r
    \Xi_{1r}^{a\dagger} \left(\tilde{\phi}^\dagger \sigma^a \phi\right)
    + \hc
  \right\}
  \nonumber \\ &
  + \left\{
    (y^{ql}_{\omega_1})_{rij}
    \omega_{1r}^\dagger \bar{q}^c_{Li} i\sigma_2 l_{Lj}
    +(y^{qq}_{\omega_1})_{rij}
    \omega_{1r}^{A\dagger} \epsilon_{ABC} \bar{q}^B_{Li} i\sigma_2 q^{c\,C}_{Lj}
  \right.
  \nonumber \\ &
  \quad
  \left.
    +(y^{eu}_{\omega_1})_{rij}
    \omega_{1r}^\dagger \bar{e}^c_{Ri} u_{Rj}
    + (y^{du}_{\omega_1})_{rij}
    \omega_{1r}^{A\dagger} \epsilon_{ABC} \bar{d}^B_{Ri} u^{c\,C}_{Rj}
    + \hc
  \right\}
  \nonumber \\ &
  + \left\{
    (y_{\Pi_1})_{rij} \Pi_{1r}^\dagger i\sigma_2 \bar{l}^T_{Li} d_{Rj}
    + \hc
  \right\}
  \nonumber \\ &
  + \left\{
    (y^{lu}_{\Pi_7})_{rij} \Pi_{7r}^\dagger i\sigma_2 \bar{l}^T_{Li} u_{Rj}
    + (y^{eq}_{\Pi_7})_{rij} \Pi_{7r}^\dagger \bar{e}_{Ri} q_{Lj}
    + \hc
  \right\}
  \nonumber \\ &
  + \left\{
    (y^{ql}_{\zeta})_{rij}
    \zeta_r^{a\dagger} \bar{q}^c_{Li} i\sigma_2 \sigma^a l_{Lj}
    + (y^{qq}_{\zeta})_{rij} \zeta^{a\dagger}_r \epsilon_{ABC}
    \bar{q}^B_{Li}\sigma^a i\sigma_2 q^{c\,C}_{Lj}
    + \hc
  \right\}.
  \label{eq:lag_scalar}
\end{align}

\subsection{New Fermions}

In table~\ref{t:fermions} we collect the new vector-like fermions that may contribute to NSI while in eq.\eqref{eq:lag_fermion} we present their interacting Lagrangian.

%

\begin{table}[h]
  \begin{center}
    {\small
      \begin{tabular}{lccccccc}
        \ctoprule
        \crowcolor
        Name &
        $N$ & $E$ & $\Delta_1$ & $\Delta_3$ & $\Sigma$ & $\Sigma_1$ & \\
        Irrep &
        $\left(1, 1\right)_0$ &
        $\left(1, 1\right)_{-1}$ &
        $\left(1, 2\right)_{-\frac{1}{2}}$ &
        $\left(1, 2\right)_{-\frac{3}{2}}$ &
        $\left(1, 3\right)_0$ &
        $\left(1, 3\right)_{-1}$ & \\[1.3mm]
        \cbottomrule
        &&&&&&&\\[-0.4cm]
        \ctoprule
        \crowcolor
        Name &
        $U$ & $D$ & $Q_1$ & $Q_5$ & $Q_7$ & $T_1$ & $T_2$ \\
        Irrep &
        $\left(3, 1\right)_{\frac{2}{3}}$ &
        $\left(3, 1\right)_{-\frac{1}{3}}$ &
        $\left(3, 2\right)_{\frac{1}{6}}$ &
        $\left(3, 2\right)_{-\frac{5}{6}}$ &
        $\left(3, 2\right)_{\frac{7}{6}}$ &
        $\left(3, 3\right)_{-\frac{1}{3}}$ &
        $\left(3, 3\right)_{\frac{2}{3}}$ \\[1.3mm]
        \cbottomrule
      \end{tabular}
    }
    \caption{New vector-like fermions and their irrepresentation under the gauge group of the SM $(\rm{SU}(3),\rm{SU}(2))_{\rm{U}(1)}$.}
    \label{t:fermions}
  \end{center}
\end{table}

\begin{align}
  -\mathcal{L}_{\mathrm{fermions}} = & \;
  (\lambda_N)_{ri} \bar{N}_{Rr} \tilde{\phi}^\dagger l_{Li}
  + (\lambda_E)_{ri} \bar{E}_{Rr} \phi^\dagger l_{Li}
   + (\lambda_{\Delta_1})_{ri} \bar{\Delta}_{1Lr} \phi e_{Ri}
  + (\lambda_{\Delta_3})_{ri} \bar{\Delta}_{3Lr} \tilde{\phi} e_{Ri}
  \nonumber \\ &
  + \frac{1}{2} (\lambda_{\Sigma})_{ri}
  \bar{\Sigma}^a_{Rr} \tilde{\phi}^\dagger \sigma^a l_{Li}
  + \frac{1}{2} (\lambda_{\Sigma_1})_{ri}
  \bar{\Sigma}^a_{1Rr} \phi^\dagger \sigma^a l_{Li}
  +(\lambda_U)_{ri} \bar{U}_{Rr} \tilde{\phi}^\dagger q_{Li}
  + (\lambda_D)_{ri} \bar{D}_{Rr} \phi^\dagger q_{Li}
  \nonumber \\ &
  + (\lambda^u_{Q_1})_{ri} \bar{Q}_{1Lr} \tilde{\phi} u_{Ri}
  + (\lambda^d_{Q_1})_{ri} \bar{Q}_{1Lr} \phi d_{Ri}
  + (\lambda_{Q_5})_{ri} \bar{Q}_{5Lr} \tilde{\phi} d_{Ri}
  + (\lambda_{Q_7})_{ri} \bar{Q}_{7Lr} \phi u_{Rj}
  \nonumber \\ &
  + \frac{1}{2} (\lambda_{T_1})_{ri}
  \bar{T}^a_{1Rr} \phi^\dagger \sigma^a q_{Li}
  + \frac{1}{2} (\lambda_{T_2})_{ri}
  \bar{T}^a_{2Rr} \tilde{\phi}^\dagger \sigma^a q_{Li}+ \hc~.
  \label{eq:lag_fermion}
\end{align}

\subsection{New Vectors}

In table~\ref{t:vectors} we collect the new vector bosons that may contribute to NSI while in eq.\eqref{eq:lag_vector} we present their interacting Lagrangian.

%
\begin{table}[h]
  \begin{center}
    {\small
      \begin{tabular}{lccccccccc} 
        \ctoprule
        \crowcolor
        Name &
        ${\cal B}$ &
        ${\cal B}_1$ &
        ${\cal W}$ &
        ${\cal L}_1$ &
        ${\cal L}_3$ &
        ${\cal U}_2$ &
        ${\cal Q}_1$ &
        ${\cal Q}_5$ &
        ${\cal X}$ 
       \\
        Irrep &
        $\left(1,1\right)_0$ &
        $\left(1,1\right)_1$ &
        $\left(1,3\right)_0$ &
        $\left(1,2\right)_{\frac 12}$ &
        $\left(1,2\right)_{-\frac 32}$ &
        $\left(3,1\right)_{\frac 23}$ &
        $\left(3,2\right)_{\frac 16}$ &
        $\left(3,2\right)_{-\frac 56}$ &
        $\left(3,3\right)_{\frac 23}$ 
       \\[1.3mm]
        \cbottomrule
      \end{tabular}
    }
    \caption{New vector bosons and their irrepresentation under the gauge group of the SM $(\rm{SU}(3),\rm{SU}(2))_{\rm{U}(1)}$.}
    \label{t:vectors}
  \end{center}
\end{table}

\begin{align}
  -\mathcal{L}_{\mathrm{V}} = & \;
  (g^l_{\mathcal{B}})_{rij} \mathcal{B}^\mu_r \bar{l}_{Li} \gamma_\mu l_{Lj}
  + (g^q_{\mathcal{B}})_{rij} \mathcal{B}^\mu_r \bar{q}_{Li} \gamma_\mu q_{Lj}
  + (g^e_{\mathcal{B}})_{rij} \mathcal{B}^\mu_r \bar{e}_{Li} \gamma_\mu e_{Lj}
  \nonumber \\ &
  + (g^d_{\mathcal{B}})_{rij} \mathcal{B}^\mu_r \bar{d}_{Li} \gamma_\mu d_{Lj}
  + (g^u_{\mathcal{B}})_{rij} \mathcal{B}^\mu_r \bar{u}_{Li} \gamma_\mu u_{Lj}
  + \left\{
    (g^\phi_{\mathcal{B}})_r \mathcal{B}^\mu_r \phi^\dagger iD_\mu \phi + \hc
  \right\}
  \nonumber \\ &
  + \left\{
    (g^{du}_{\mathcal{B}_1})_{rij}
    \mathcal{B}^{\mu\dagger}_{1r} \bar{d}_{Ri}\gamma_\mu u_{Rj}
    + (g^\phi_{\mathcal{B}_1})_r
    \mathcal{B}^{\mu\dagger}_{1r} i D_\mu \phi^T i\sigma_2 \phi
    + \hc
  \right\}
  \nonumber \\ &
  + \frac{1}{2} (g^l_{\mathcal{W}})_{rij}
  \mathcal{W}^{\mu a}_r \bar{l}_{Li} \sigma^a \gamma_\mu l_{Lj}
  + \frac{1}{2} (g^q_{\mathcal{W}})_{rij}
  \mathcal{W}^{\mu a}_r \bar{q}_{Li} \sigma^a \gamma_\mu q_{Lj}
  \nonumber \\ &
  + \left\{
    \frac{1}{2} (g^\phi_{\mathcal{W}})_r
    \mathcal{W}^{\mu a}_r \phi^\dagger \sigma^a iD_\mu \phi + \hc
  \right\}
  \nonumber \\ &
  + \left\{
    (\gamma_{\mathcal{L}_1})_r \mathcal{L}_{1r\mu}^{\dagger} D^{\mu} \phi
    + \hc
  \right\}
  \nonumber \\ &
  + i (g^B_{\mathcal{L}_1})_{rs}
  \mathcal{L}_{1r\mu}^{\dagger} \mathcal{L}_{1s\nu} B^{\mu\nu}
  + i (g^W_{\mathcal{L}_1})_{rs}
  \mathcal{L}_{1i\mu}^{\dagger} \sigma^a \mathcal{L}_{1j\nu} W^{a\,\mu\nu}
  \nonumber \\ &
  + i (g^{\tilde{B}}_{\mathcal{L}_1})_{rs}
  \mathcal{L}_{1r\mu}^{\dagger} \mathcal{L}_{1s\nu} \tilde{B}^{\mu\nu} 
  + i (g^{\tilde{W}}_{\mathcal{L}_1})_{rs}
  \mathcal{L}_{1r\mu}^{\dagger}
  \sigma^{a} \mathcal{L}_{1s\nu} \tilde{W}^{a\,\mu\nu}
  \nonumber \\ &
  + (h^{(1)}_{\mathcal{L}_1})_{rs}
  \left(\mathcal{L}_{1r\mu}^{\dagger} \mathcal{L}^{\mu}_{1s}\right)
  \left(\phi^{\dagger} \phi\right)
  + (h^{(2)}_{\mathcal{L}_1})_{rs}
  \left(\mathcal{L}_{1r\mu}^{\dagger} \phi\right)
  \left(\phi^{\dagger} \mathcal{L}^{\mu}_{1s}\right)
  \nonumber \\ &
  + \left\{
    (h^{(3)}_{\mathcal{L}_1})_{rs}
    \left( \mathcal{L}_{1r\mu}^{1\dagger} \phi\right)
    \left(\mathcal{L}^{\dagger\mu}_{1s} \phi\right)
    + \hc
  \right\}
  \nonumber \\ &
  + \left\{
    (g_{\mathcal{L}_3})_{rij}
    \mathcal{L}^{\mu\dagger}_{3r} \bar{e}^c_{Ri} \gamma_\mu l_{Lj}
    + \hc
  \right\}
  \nonumber \\ &
  + \left\{
    (g^{ed}_{\mathcal{U}_2})_{rij}
    \mathcal{U}^{\mu\dagger}_{2r} \bar{e}_{Ri} \gamma_\mu d_{Rj}
    + (g^{lq}_{\mathcal{U}_2})_{rij}
    \mathcal{U}^{\mu\dagger}_{2r} \bar{l}_{Li} \gamma_\mu q_{Lj}
    + \hc
  \right\}
  \nonumber \\ &
  + \left\{
    (g^{ul}_{\mathcal{Q}_1})_{rij}
    \mathcal{Q}^{\mu\dagger}_{1r} \bar{u}^c_{Ri} \gamma_\mu l_{Lj}
    + (g^{dq}_{\mathcal{Q}_1})_{rij}
    \mathcal{Q}^{A\mu\dagger}_{1r}
    \epsilon_{ABC} \bar{d}^B_{Ri} \gamma_\mu i\sigma_2 q^{c\,C}_{Lj}
    + \hc
  \right\}
  \nonumber \\ &
  + \left\{
    (g^{dl}_{\mathcal{Q}_5})_{rij}
    \mathcal{Q}^{\mu\dagger}_{5r} \bar{d}^c_{Ri} \gamma_\mu l_{Lj}
    + (g^{eq}_{\mathcal{Q}_5})_{rij}
    \mathcal{Q}^{\mu\dagger}_{5r}
    \bar{e}^c_{Ri} \gamma_\mu q_{Lj}
  \right.
  \nonumber \\ &
  \quad
  \left.
    + (g^{uq}_{\mathcal{Q}_5})_{rij}
    \mathcal{Q}^{A\mu\dagger}_{5r}
    \epsilon_{ABC} \bar{u}^B_{Ri} \gamma_\mu q^{c\,C}_{Lj}
    + \hc
  \right\}
  \nonumber \\ &
  + \left\{
    \frac{1}{2} (g_{\mathcal{X}})_{rij}
    \mathcal{X}^{a\mu\dagger}_r
    \bar{l}_{Li} \gamma_\mu \sigma^a q_{Lj}
    + \hc
  \right\}
  \nonumber \\ &
 +
  \Big\{\frac{1}{f} \mathcal{L}^{\mu\dagger}_{1r} \bigg[
  (\tilde{\gamma}^{(1)}_{\mathcal{L}_1})_r
  \left(\phi^\dagger D_\mu \phi\right) \phi
  + (\tilde{\gamma}^{(2)}_{\mathcal{L}_1})_r
  \left(D_\mu\phi^\dagger \phi\right) \phi
  + (\tilde{\gamma}^{(3)}_{\mathcal{L}_1})_r
  \left(\phi^\dagger \phi\right) D_\mu \phi
  \nonumber \\ &
  \quad \qquad \qquad
  + (\tilde{\gamma}^B_{\mathcal{L}_1})_r B_{\mu\nu} D^\nu \phi
  + (\tilde{\gamma}^{\tilde{B}}_{\mathcal{L}_1})_r \tilde{B}_{\mu\nu} D^\nu \phi
  \nonumber \\ &
  \quad \qquad \qquad
  + (\tilde{\gamma}^W_{\mathcal{L}_1})_r W^a_{\mu\nu} \sigma^a D^\nu \phi
  + (\tilde{\gamma}^{\tilde{W}}_{\mathcal{L}_1})_r
  \tilde{W}^a_{\mu\nu} \sigma^a D^\nu \phi
  \nonumber \\ &
  \quad \qquad \qquad
  + (\tilde{g}^{eDl}_{\mathcal{L}_1})_{rij} \bar{e}_{Ri} D_\mu l_{Lj}
  + (\tilde{g}^{Del}_{\mathcal{L}_1})_{rij} D_\mu \bar{e}_{Ri} l_{Lj}
  + (\tilde{g}^{dDq}_{\mathcal{L}_1})_{rij} \bar{d}_{Ri} D_\mu q_{Lj}
  \nonumber \\ &
  \quad \qquad \qquad
  + (\tilde{g}^{Ddq}_{\mathcal{L}_1})_{rij} D_\mu \bar{d}_{Ri} q_{Lj}
  + (\tilde{g}^{qDu}_{\mathcal{L}_1})_{rij} i\sigma_2 \bar{q}^T_{Li} D_\mu u_{Rj} 
  + (\tilde{g}^{Dqu}_{\mathcal{L}_1})_{rij} i\sigma_2 D_\mu \bar{q}^T_{Li} u_{Rj}
  \nonumber \\ &
  \quad \qquad \qquad
  + (\tilde{g}^{du}_{\mathcal{L}_1})_{rij}
  \tilde{\phi} \bar{d}_{Ri} \gamma_\mu u_{Rj}
  + (\tilde{g}^e_{\mathcal{L}_1})_{rij} \phi \bar{e}_{Ri} \gamma_\mu e_{Rj}
  + (\tilde{g}^d_{\mathcal{L}_1})_{rij} \phi \bar{d}_{Ri} \gamma_\mu d_{Rj}
  \nonumber \\ &
  \quad \qquad \qquad
  + (\tilde{g}^u_{\mathcal{L}_1})_{rij} \phi \bar{u}_{Ri} \gamma_\mu u_{Rj}
  + (\tilde{g}^{l}_{\mathcal{L}_1})_{rij} \phi \bar{l}_{Ri} \gamma_\mu l_{Lj}
  + (\tilde{g}^{l\prime}_{\mathcal{L}_1})_{rij} \left(\sigma^a \phi\right)
  \left(\bar{l}_{Li} \gamma_\mu \sigma^a l_{Lj}\right)
  \nonumber \\ &
  \quad \qquad \qquad
  + (\tilde{g}^{q}_{\mathcal{L}_1})_{rij} \phi \bar{q}_{Li} \gamma_\mu q_{Lj}
  + (\tilde{g}^{q\prime}_{\mathcal{L}_1})_{rij} \left(\sigma^a \phi\right)
  \left(\bar{q}_{Li} \gamma_\mu \sigma^a q_{Lj}\right)
  \bigg]
  + \hc~\!\Big\}.
  \label{eq:lag_vector}
\end{align}

\section{Explicit matching results}
\label{ap:matching}

We provide in the following explicit matching results for the UV models identified in table \ref{tab:UV-op}.

\clearpage
\subsection{Scalar models}

\renewcommand{\arraystretch}{1.8}
\begin{table}[h!]
    \centering
    \begin{tabular}{|c|c| }
       \hline
       \multicolumn{2}{|c|}{$\varphi$ model}
       \\
       \hline
       WEFT WC  & Matching  \\
       \hline
        $\frac{2V_{k j}^{*}}{v^{2}} (\epsilon_{S}^{kj})^{*}_{\alpha\beta}$ & $-\frac{1}{M_{\varphi}^{2}}\left[(y^{d}_{\varphi})_{jk} + \sum_{x}V_{xj}^{*}(y^{u}_{\varphi})_{xk}\right](y^{e}_{\varphi})^{*}_{\alpha\beta}$ \\
        \hline
        $\frac{2V_{k j}^{*}}{v^{2}} (\epsilon_{P}^{kj})^{*}_{\alpha\beta}$ & $-\frac{1}{M_{\varphi}^{2}}\left[(y^{d}_{\varphi})_{jk} - \sum_{x}V_{xj}^{*}(y^{u}_{\varphi})_{xk}\right](y^{e}_{\varphi})^{*}_{\alpha\beta}$ \\
        \hline
         $\frac{2}{v^{2}} (\epsilon_{V}^{ee})_{\alpha\beta}$ & $\frac{(y^{e}_{\varphi})^{*}_{1\alpha}(y^{e}_{\varphi})_{1\beta}}{2M_{\varphi}^{2}}$ \\
        \hline
         $\frac{2}{v^{2}} (\epsilon_{A}^{ee})_{\alpha\beta}$ & $\frac{(y^{e}_{\varphi})^{*}_{1\alpha}(y^{e}_{\varphi})_{1\beta}}{2M_{\varphi}^{2}}$ \\
        \hline        
    \end{tabular}
    \caption{Matching to $\varphi$ model.}
    \label{tab:my_label}
\end{table}
\renewcommand{\arraystretch}{1.0}

\renewcommand{\arraystretch}{1.8}
\begin{table}[h!]
    \centering
    \begin{tabular}{|c|c|c|}
       \hline
 & $\omega_{1}$ model & $\Pi_7$ model
       \\
       \hline
       WEFT WC  & \multicolumn{2}{c|}{Matching}  \\
       \hline
       $\frac{2V_{k j}^{*}}{v^{2}} (\epsilon_{L}^{kj})^{*}_{\alpha\beta}$ & $\frac{\sum_{x}V_{xj}^{*}(y^{ql}_{\omega_{1}})_{x\beta}^{*}(y^{ql}_{\omega_{1}})_{k\alpha}}{2M_{\omega_{1}}^{2}}$ & -  \\
        \hline
        $\frac{2V_{k j}^{*}}{v^{2}} (\epsilon_{S}^{kj})^{*}_{\alpha\beta}$ & $-\frac{\sum_{x}V_{xj}^{*}(y^{eu}_{\omega_{1}})_{\alpha k}(y^{ql}_{\omega_{1}})_{x\beta}^{*}}{2M_{\omega_{1}}^{2}}$ & $-\frac{\sum_{x}V_{xj}^{*}(y^{eq}_{\Pi_{7}})^{*}_{\alpha x}(y^{lu}_{\Pi_{7}})_{\beta k}}{2M_{\Pi_{7}}^{2}}$  \\
        \hline
        $\frac{2V_{k j}^{*}}{v^{2}} (\epsilon_{P}^{kj})^{*}_{\alpha\beta}$ & $\frac{\sum_{x}V_{xj}^{*}(y^{eu}_{\omega_{1}})_{\alpha k}(y^{ql}_{\omega_{1}})_{x\beta}^{*}}{2M_{\omega_{1}}^{2}}$ & $\frac{\sum_{x}V_{xj}^{*}(y^{eq}_{\Pi_{7}})^{*}_{\alpha x}(y^{lu}_{\Pi_{7}})_{\beta k}}{2M_{\Pi_{7}}^{2}}$ \\
        \hline
         $\frac{2}{v^{2}} (\epsilon_{V}^{uu})_{\alpha\beta}$ & - & $\frac{(y^{lu}_{\Pi_{7}})_{\beta 1}^{*}(y^{lu}_{\Pi_{7}})_{\alpha 1}}{2M_{\Pi_{7}}^{2}}$\\
        \hline
         $\frac{2}{v^{2}} (\epsilon_{A}^{uu})_{\alpha\beta}$ & - & $\frac{(y^{lu}_{\Pi_{7}})_{\beta 1}^{*}(y^{lu}_{\Pi_{7}})_{\alpha 1}}{2M_{\Pi_{7}}^{2}}$ \\
        \hline
         $\frac{2}{v^{2}} (\epsilon_{V}^{dd})_{\alpha\beta}$ & $-\frac{\sum_{x,z}V_{1z}V_{1x}^{*}(y^{ql}_{\omega_{1}})_{x\alpha}^{*}(y^{ql}_{\omega_{1}})_{z\beta}}{2M_{\omega_{1}}^{2}}$ & -\\
        \hline
         $\frac{2}{v^{2}} (\epsilon_{A}^{dd})_{\alpha\beta}$ & $\frac{\sum_{x,z}V_{1z}V_{1x}^{*}(y^{ql}_{\omega_{1}})_{x\alpha}^{*}(y^{ql}_{\omega_{1}})_{z\beta}}{2M_{\omega_{1}}^{2}}$ & - \\
        \hline        
    \end{tabular}
    \caption{Matching to $\omega_{1}$, and $\Pi_7$ models.}
\end{table}
\renewcommand{\arraystretch}{1.0}

\renewcommand{\arraystretch}{1.8}
\begin{table}[h!]
    \centering
    \begin{tabular}{|c|c|c|c|}
       \hline
 & $\mathcal{S}_{1}$ model & $\Xi_1$ model  & $\Pi_{1}$ model
       \\
       \hline
       WEFT WC  & \multicolumn{3}{c|}{Matching}  \\
       \hline
        $\frac{2}{v^{2}} (\epsilon_{V}^{ee})_{\alpha\beta}$ & $-\frac{(y_{\mathcal{S}_{1}})_{\beta 1}^{*}(y_{\mathcal{S}_{1}})_{\alpha 1}}{M_{\mathcal{S}_{1}}^{2}}-\frac{(y_{\mathcal{S}_{1}})_{1\beta }^{*}(y_{\mathcal{S}_{1}})_{1\alpha }}{M_{\mathcal{S}_{1}}^{2}}$ & $-\frac{(y_{\Xi_{1}})_{\beta 1}^{*}(y_{\Xi_{1}})_{\alpha 1}}{M_{\Xi_{1}}^{2}}-\frac{(y_{\Xi_{1}})_{1\beta }^{*}(y_{\Xi_{1}})_{1\alpha }}{M_{\Xi_{1}}^{2}}$ & - \\
        \hline
         $\frac{2}{v^{2}} (\epsilon_{A}^{ee})_{\alpha\beta}$ & $\frac{(y_{\mathcal{S}_{1}})_{\beta 1}^{*}(y_{\mathcal{S}_{1}})_{\alpha 1}}{M_{\mathcal{S}_{1}}^{2}}+\frac{(y_{\mathcal{S}_{1}})_{1\beta }^{*}(y_{\mathcal{S}_{1}})_{1\alpha }}{M_{\mathcal{S}_{1}}^{2}}$ & $\frac{(y_{\Xi_{1}})_{\beta 1}^{*}(y_{\Xi_{1}})_{\alpha 1}}{M_{\Xi_{1}}^{2}}+\frac{(y_{\Xi_{1}})_{1\beta }^{*}(y_{\Xi_{1}})_{1\alpha }}{M_{\Xi_{1}}^{2}}$ & - \\
        \hline
        $\frac{2}{v^{2}} (\epsilon_{V}^{dd})_{\alpha\beta}$ & -& -&   $\frac{(y_{\Pi_{1}})_{\beta 1}^{*}(y_{\Pi_{1}})_{\alpha 1}}{2M_{\Pi_{1}}^{2}}$\\
        \hline
        $\frac{2}{v^{2}} (\epsilon_{A}^{dd})_{\alpha\beta}$ & -& -&  $\frac{(y_{\Pi_{1}})_{\beta 1}^{*}(y_{\Pi_{1}})_{\alpha 1}}{2M_{\Pi_{1}}^{2}}$ \\
        \hline
    \end{tabular}
    \caption{Matching to $\mathcal{S}_{1}$,  $\Xi_1$, and $\Pi_{1}$ models.}
\end{table}
\renewcommand{\arraystretch}{1.0}

\renewcommand{\arraystretch}{1.8}
\begin{table}[h!]
    \centering
    \begin{tabular}{|c|c|}
       \hline
  & $\zeta$ model
       \\
       \hline
       WEFT WC  & Matching  \\
       \hline
        $\frac{2V_{k j}^{*}}{v^{2}} (\epsilon_{L}^{kj})^{*}_{\alpha\beta}$  & $\frac{\sum_{x}V_{xj}^{*}(y^{ql}_{\zeta})_{x \beta}^{*}(y^{ql}_{\zeta})_{k \alpha}}{M_{\zeta}^{2}}$\\
        \hline
        $\frac{2}{v^{2}} (\epsilon_{V}^{dd})_{\alpha\beta}$  & $-\frac{\sum_{x,z}V_{1z}V_{1x}^{*}(y^{ql}_{\zeta})_{x \alpha}^{*}(y^{ql}_{\zeta})_{z \beta}}{2M_{\Xi_{1}}^{2}}$\\
        \hline
         $\frac{2}{v^{2}} (\epsilon_{A}^{dd})_{\alpha\beta}$  & $\frac{\sum_{x,z}V_{1z}V_{1x}^{*}(y^{ql}_{\zeta})_{x \alpha}^{*}(y^{ql}_{\zeta})_{z \beta}}{2M_{\Xi_{1}}^{2}}$\\
        \hline
        $\frac{2}{v^{2}} (\epsilon_{V}^{uu})_{\alpha\beta}$  & $-\frac{(y^{ql}_{\zeta})_{1 \alpha}^{*}(y^{ql}_{\zeta})_{1 \beta}}{M_{\zeta}^{2}}$\\
        \hline
         $\frac{2}{v^{2}} (\epsilon_{A}^{uu})_{\alpha\beta}$  & $\frac{(y^{ql}_{\zeta})_{1 \alpha}^{*}(y^{ql}_{\zeta})_{1 \beta}}{M_{\zeta}^{2}}$ \\
         \hline
    \end{tabular}
    \caption{Matching to $\zeta$ model.}
\end{table}
\renewcommand{\arraystretch}{1.0}

\clearpage

\subsection{Vector-like models}

\renewcommand{\arraystretch}{1.8}
\begin{table}[h!]
    \centering
    \begin{tabular}{|c|c|}
       \hline
  & $N$ model
       \\
       \hline
       WEFT WC  & Matching  \\
       \hline
       $\frac{2V_{k j}^{*}}{v^{2}} (\epsilon_{L}^{kj})^{*}_{\alpha\beta}$  & $-\frac{(\lambda_{N})_{\alpha}^{*} (\lambda_{N})_{\beta}V_{k j}^{*}}{2M_{N}^{2}}$\\
       \hline
       $\frac{2}{v^{2}} (\epsilon_{V}^{ee})_{\alpha\beta}$  & $-\frac{(\lambda_{N})_{1}^{*}\delta_{1\alpha}(\lambda_{N})_{\beta}+(\lambda_{N})_{\alpha}^{*}(\lambda_{N})_{1}\delta_{1\beta}-(1-4s^2)(\lambda_{N})_{\alpha}^{*}(\lambda_{N})_{\beta}}{2M_{N}^{2}}$ \\
       \hline
       $\frac{2}{v^{2}} (\epsilon_{A}^{ee})_{\alpha\beta}$  & $\frac{(\lambda_{N})_{1}^{*}\delta_{1\alpha}(\lambda_{N})_{\beta}+(\lambda_{N})_{\alpha}^{*}(\lambda_{N})_{1}\delta_{1\beta}-(\lambda_{N})_{\alpha}^{*}(\lambda_{N})_{\beta}}{2M_{N}^{2}}$ \\
       \hline
        $\frac{2}{v^{2}} (\epsilon_{V}^{uu})_{\alpha\beta}$  & $\frac{(-3+8s^2)(\lambda_{N})_{\alpha}^{*}(\lambda_{N})_{\beta}}{6M_{N}^{2}}$\\
        \hline
         $\frac{2}{v^{2}} (\epsilon_{A}^{uu})_{\alpha\beta}$  & $\frac{(\lambda_{N})_{\alpha}^{*}(\lambda_{N})_{\beta}}{2M_{N}^{2}}$\\
        \hline
        $\frac{2}{v^{2}} (\epsilon_{V}^{dd})_{\alpha\beta}$  & $\frac{(3-4s^2)(\lambda_{N})_{\alpha}^{*}(\lambda_{N})_{\beta}}{6M_{N}^{2}}$\\
        \hline
         $\frac{2}{v^{2}} (\epsilon_{A}^{dd})_{\alpha\beta}$  & $-\frac{(\lambda_{N})_{\alpha}^{*}(\lambda_{N})_{\beta}}{2M_{N}^{2}}$\\
         \hline
    \end{tabular}
    \caption{Matching to $N$ model.}
\end{table}
\renewcommand{\arraystretch}{1.0}

\renewcommand{\arraystretch}{1.8}
\begin{table}[h!]
    \centering
    \begin{tabular}{|c|c|}
       \hline
  & $E$ model
       \\
       \hline
       WEFT WC  & Matching  \\
       \hline
       $\frac{2V_{k j}^{*}}{v^{2}} (\epsilon_{L}^{kj})^{*}_{\alpha\beta}$  & $-\frac{(\lambda_{E})_{\alpha}^{*} (\lambda_{E})_{\beta}V_{k j}^{*}}{2M_{E}^{2}}$\\
       \hline
       $\frac{2}{v^{2}} (\epsilon_{V}^{ee})_{\alpha\beta}$  & $-\frac{(\lambda_{E})_{1}^{*}\delta_{1\alpha}(\lambda_{E})_{\beta}+(\lambda_{E})_{\alpha}^{*}(\lambda_{E})_{1}\delta_{1\beta}-(\lambda_{E})_{1}^{*}(\lambda_{E})_{1}\delta_{\alpha\beta}}{2M_{E}^{2}}$ \\
       \hline
       $\frac{2}{v^{2}} (\epsilon_{A}^{ee})_{\alpha\beta}$  & $\frac{(\lambda_{E})_{1}^{*}\delta_{1\alpha}(\lambda_{E})_{\beta}+(\lambda_{E})_{\alpha}^{*}(\lambda_{E})_{1}\delta_{1\beta}-(\lambda_{E})_{1}^{*}(\lambda_{E})_{1}\delta_{\alpha\beta}}{2M_{E}^{2}}$  \\
       \hline
    \end{tabular}
    \caption{Matching to $E$ model.}
\end{table}
\renewcommand{\arraystretch}{1.0}

\renewcommand{\arraystretch}{1.8}
\begin{table}[h!]
    \centering
    \begin{tabular}{|c|c|c|}
       \hline
  & $\Delta_1$ model & $\Delta_3$ model
       \\
       \hline
       WEFT WC  & \multicolumn{2}{c|}{Matching}   \\
       \hline
       $\frac{2}{v^{2}} (\epsilon_{V}^{ee})_{\alpha\beta}$  & $-\frac{(\lambda_{\Delta_1})_{1}^{*}(\lambda_{\Delta_1})_{1}\delta_{\alpha\beta}}{2M_{\Delta_1}^{2}}$ & $\frac{(\lambda_{\Delta_3})_{1}^{*}(\lambda_{\Delta_3})_{1}\delta_{\alpha\beta}}{2M_{\Delta_3}^{2}}$  \\
       \hline
       $\frac{2}{v^{2}} (\epsilon_{A}^{ee})_{\alpha\beta}$  & $-\frac{(\lambda_{\Delta_1})_{1}^{*}(\lambda_{\Delta_1})_{1}\delta_{\alpha\beta}}{2M_{\Delta_1}^{2}}$ & $\frac{(\lambda_{\Delta_3})_{1}^{*}(\lambda_{\Delta_3})_{1}\delta_{\alpha\beta}}{2M_{\Delta_3}^{2}}$  \\
       \hline
    \end{tabular}
    \caption{Matching to $\Delta_1$, and $\Delta_3$ models.}
\end{table}
\renewcommand{\arraystretch}{1.0}

\renewcommand{\arraystretch}{1.8}
\begin{table}[h!]
    \centering
    \begin{tabular}{|c|c|}
       \hline
  & $\Sigma$ model
       \\
       \hline
       WEFT WC  & Matching  \\
       \hline
       $\frac{2V_{k j}^{*}}{v^{2}} (\epsilon_{L}^{kj})^{*}_{\alpha\beta}$  & $\frac{(\lambda_{\Sigma})_{\alpha}^{*} (\lambda_{\Sigma})_{\beta}V_{k j}^{*}}{8M_{\Sigma}^{2}}$\\
       \hline
       $\frac{2}{v^{2}} (\epsilon_{V}^{ee})_{\alpha\beta}$  & $\frac{(\lambda_{\Sigma})_{1}^{*}\delta_{1\alpha}(\lambda_{\Sigma})_{\beta}+(\lambda_{\Sigma})_{\alpha}^{*}(\lambda_{\Sigma})_{1}\delta_{1\beta}-2(\lambda_{\Sigma})_{1}^{*}(\lambda_{\Sigma})_{1}\delta_{\alpha\beta}+(1-4s^2)(\lambda_{\Sigma})_{\alpha}^{*}(\lambda_{\Sigma})_{\beta}}{8M_{\Sigma}^{2}}$ \\
       \hline
       $\frac{2}{v^{2}} (\epsilon_{A}^{ee})_{\alpha\beta}$  & $-\frac{(\lambda_{\Sigma})_{1}^{*}\delta_{1\alpha}(\lambda_{\Sigma})_{\beta}+(\lambda_{\Sigma})_{\alpha}^{*}(\lambda_{\Sigma})_{1}\delta_{1\beta}-2(\lambda_{\Sigma})_{1}^{*}(\lambda_{\Sigma})_{1}\delta_{\alpha\beta}+(\lambda_{\Sigma})_{\alpha}^{*}(\lambda_{\Sigma})_{\beta}}{8M_{\Sigma}^{2}}$ \\
       \hline
        $\frac{2}{v^{2}} (\epsilon_{V}^{uu})_{\alpha\beta}$  & $\frac{(-3+8s^2)(\lambda_{\Sigma})_{\alpha}^{*}(\lambda_{\Sigma})_{\beta}}{24M_{\Sigma}^{2}}$\\
        \hline
         $\frac{2}{v^{2}} (\epsilon_{A}^{uu})_{\alpha\beta}$  & $\frac{(\lambda_{\Sigma})_{\alpha}^{*}(\lambda_{\Sigma})_{\beta}}{8M_{\Sigma}^{2}}$\\
        \hline
        $\frac{2}{v^{2}} (\epsilon_{V}^{dd})_{\alpha\beta}$  & $\frac{(3-4s^2)(\lambda_{\Sigma})_{\alpha}^{*}(\lambda_{\Sigma})_{\beta}}{24M_{\Sigma}^{2}}$\\
        \hline
         $\frac{2}{v^{2}} (\epsilon_{A}^{dd})_{\alpha\beta}$  & $-\frac{(\lambda_{\Sigma})_{\alpha}^{*}(\lambda_{\Sigma})_{\beta}}{8M_{\Sigma}^{2}}$\\
         \hline
    \end{tabular}
    \caption{Matching to $\Sigma$ model.}
\end{table}
\renewcommand{\arraystretch}{1.0}

\renewcommand{\arraystretch}{1.8}
\begin{table}[h!]
    \centering
    \begin{tabular}{|c|c|}
       \hline
  & $\Sigma_1$ model
       \\
       \hline
       WEFT WC  & Matching  \\
       \hline
       $\frac{2V_{k j}^{*}}{v^{2}} (\epsilon_{L}^{kj})^{*}_{\alpha\beta}$  & $\frac{(\lambda_{\Sigma_1})_{\alpha}^{*} (\lambda_{\Sigma_1})_{\beta}V_{k j}^{*}}{8M_{\Sigma_1}^{2}}$\\
       \hline
       $\frac{2}{v^{2}} (\epsilon_{V}^{ee})_{\alpha\beta}$  & $\frac{(\lambda_{\Sigma_1})_{1}^{*}\delta_{1\alpha}(\lambda_{\Sigma_1})_{\beta}+(\lambda_{\Sigma_1})_{\alpha}^{*}(\lambda_{\Sigma_1})_{1}\delta_{1\beta}+(\lambda_{\Sigma_1})_{1}^{*}(\lambda_{\Sigma_1})_{1}\delta_{\alpha\beta}-2(1-4s^2)(\lambda_{\Sigma_1})_{\alpha}^{*}(\lambda_{\Sigma_1})_{\beta}}{8M_{\Sigma_1}^{2}}$ \\
       \hline
       $\frac{2}{v^{2}} (\epsilon_{A}^{ee})_{\alpha\beta}$  & $-\frac{(\lambda_{\Sigma_1})_{1}^{*}\delta_{1\alpha}(\lambda_{\Sigma_1})_{\beta}+(\lambda_{\Sigma_1})_{\alpha}^{*}(\lambda_{\Sigma_1})_{1}\delta_{1\beta}+(\lambda_{\Sigma_1})_{1}^{*}(\lambda_{\Sigma_1})_{1}\delta_{\alpha\beta}-2(\lambda_{\Sigma_1})_{\alpha}^{*}(\lambda_{\Sigma_1})_{\beta}}{8M_{\Sigma_1}^{2}}$ \\
       \hline
        $\frac{2}{v^{2}} (\epsilon_{V}^{uu})_{\alpha\beta}$  & $\frac{(3-8s^2)(\lambda_{\Sigma_1})_{\alpha}^{*}(\lambda_{\Sigma_1})_{\beta}}{12M_{\Sigma_1}^{2}}$\\
        \hline
         $\frac{2}{v^{2}} (\epsilon_{A}^{uu})_{\alpha\beta}$  & $-\frac{(\lambda_{\Sigma_1})_{\alpha}^{*}(\lambda_{\Sigma_1})_{\beta}}{4M_{\Sigma_1}^{2}}$\\
        \hline
        $\frac{2}{v^{2}} (\epsilon_{V}^{dd})_{\alpha\beta}$  & $\frac{(-3+4s^2)(\lambda_{\Sigma_1})_{\alpha}^{*}(\lambda_{\Sigma_1})_{\beta}}{12M_{\Sigma_1}^{2}}$\\
        \hline
         $\frac{2}{v^{2}} (\epsilon_{A}^{dd})_{\alpha\beta}$  & $\frac{(\lambda_{\Sigma_1})_{\alpha}^{*}(\lambda_{\Sigma_1})_{\beta}}{4M_{\Sigma_1}^{2}}$\\
         \hline
    \end{tabular}
    \caption{Matching to $\Sigma_1$ model.}
\end{table}
\renewcommand{\arraystretch}{1.0}

\renewcommand{\arraystretch}{1.8}
\begin{table}[h!]
    \centering
    \begin{tabular}{|c|c|c|}
       \hline
   & $U$ model & $D$ model 
       \\
       \hline
       WEFT WC  & \multicolumn{2}{c|}{Matching}   \\
       \hline
       $\frac{2V_{k j}^{*}}{v^{2}} (\epsilon_{L}^{kj})^{*}_{\alpha\beta}$  & $-\frac{\sum_{x}V_{xj}^{*}(\lambda_{U})_{x}^{*} (\lambda_{U})_{k}\delta_{\alpha\beta}}{2M_{U}^{2}}$ & $-\frac{\sum_{x}V_{xj}^{*}(\lambda_{D})_{x}^{*} (\lambda_{D})_{k}\delta_{\alpha\beta}}{2M_{D}^{2}}$  \\
       \hline
       $\frac{2}{v^{2}} (\epsilon_{V}^{dd})_{\alpha\beta}$  & - & $\frac{\sum_{x,z}V_{1x}^{*}V_{1z}(\lambda_{D})_{x}^{*}(\lambda_{D})_{z}\delta_{\alpha\beta}}{2M_{D}^{2}}$  \\
       \hline
       $\frac{2}{v^{2}} (\epsilon_{A}^{dd})_{\alpha\beta}$ & - & $-\frac{\sum_{x,z}V_{1x}^{*}V_{1z}(\lambda_{D})_{x}^{*}(\lambda_{D})_{z}\delta_{\alpha\beta}}{2M_{D}^{2}}$   \\
       \hline
       $\frac{2}{v^{2}} (\epsilon_{V}^{uu})_{\alpha\beta}$  & $-\frac{(\lambda_{U})_{1}^{*}(\lambda_{U})_{1}\delta_{\alpha\beta}}{2M_{U}^{2}}$  & -  \\
       \hline
       $\frac{2}{v^{2}} (\epsilon_{A}^{uu})_{\alpha\beta}$ & $\frac{(\lambda_{U})_{1}^{*}(\lambda_{U})_{1}\delta_{\alpha\beta}}{2M_{U}^{2}}$  &-  \\
    \hline
    \end{tabular}
    \caption{Matching to $U$, and $D$ models.}
\end{table}
\renewcommand{\arraystretch}{1.0}

\renewcommand{\arraystretch}{1.8}
\begin{table}[h!]
    \centering
    \begin{tabular}{|c|c|c|c|}
       \hline
   & $Q_1$ model & $Q_5$ model & $Q_7$ model
       \\
       \hline
       WEFT WC  & \multicolumn{3}{c|}{Matching}   \\
       \hline
       $\frac{2V_{k j}^{*}}{v^{2}} (\epsilon_{R}^{kj})^{*}_{\alpha\beta}$  & $-\frac{(\lambda_{Q_1})_{j}^{*} (\lambda_{Q_1})_{k}\delta_{\alpha\beta}}{M_{Q_1}^{2}}$ & - & - \\
       \hline
       $\frac{2}{v^{2}} (\epsilon_{V}^{dd})_{\alpha\beta}$  & $-\frac{(\lambda_{Q_1})_{1}^{*}(\lambda_{Q_1})_{1}\delta_{\alpha\beta}}{2M_{Q_1}^{2}}$ & $\frac{(\lambda_{Q_5})_{1}^{*}(\lambda_{Q_5})_{1}\delta_{\alpha\beta}}{2M_{Q_5}^{2}}$ & - \\
       \hline
       $\frac{2}{v^{2}} (\epsilon_{A}^{dd})_{\alpha\beta}$ & $-\frac{(\lambda_{Q_1})_{1}^{*}(\lambda_{Q_1})_{1}\delta_{\alpha\beta}}{2M_{Q_1}^{2}}$ & $\frac{(\lambda_{Q_5})_{1}^{*}(\lambda_{Q_5})_{1}\delta_{\alpha\beta}}{2M_{Q_5}^{2}}$ & -  \\
       \hline
       $\frac{2}{v^{2}} (\epsilon_{V}^{uu})_{\alpha\beta}$  & $\frac{(\lambda_{Q_1})_{1}^{*}(\lambda_{Q_1})_{1}\delta_{\alpha\beta}}{2M_{Q_1}^{2}}$& - & $-\frac{(\lambda_{Q_7})_{1}^{*}(\lambda_{Q_7})_{1}\delta_{\alpha\beta}}{2M_{Q_7}^{2}}$  \\
       \hline
       $\frac{2}{v^{2}} (\epsilon_{A}^{uu})_{\alpha\beta}$  & $\frac{(\lambda_{Q_1})_{1}^{*}(\lambda_{Q_1})_{1}\delta_{\alpha\beta}}{2M_{Q_1}^{2}}$ & - & $-\frac{(\lambda_{Q_7})_{1}^{*}(\lambda_{Q_7})_{1}\delta_{\alpha\beta}}{2M_{Q_7}^{2}}$  \\
    \hline
    \end{tabular}
    \caption{Matching to $Q_1$, $Q_5$ and $Q_7$ models.}
\end{table}
\renewcommand{\arraystretch}{1.0}

\renewcommand{\arraystretch}{1.8}
\begin{table}[h!]
    \centering
    \begin{tabular}{|c|c|c|}
       \hline
   & $T_1$ model & $T_2$ model 
       \\
       \hline
       WEFT WC  & \multicolumn{2}{c|}{Matching}   \\
       \hline
       $\frac{2V_{k j}^{*}}{v^{2}} (\epsilon_{L}^{kj})^{*}_{\alpha\beta}$  & $\frac{\sum_{x}V_{xj}^{*}(\lambda_{T_1})_{x}^{*} (\lambda_{T_1})_{k}\delta_{\alpha\beta}}{8M_{T_1}^{2}}$ & $\frac{\sum_{x}V_{xj}^{*}(\lambda_{T_2})_{x}^{*} (\lambda_{T_2})_{k}\delta_{\alpha\beta}}{8M_{T_2}^{2}}$  \\
       \hline
       $\frac{2}{v^{2}} (\epsilon_{V}^{dd})_{\alpha\beta}$  & $\frac{\sum_{x,z}V_{1x}^{*}V_{1z}(\lambda_{T_1})_{x}^{*}(\lambda_{T_1})_{z}\delta_{\alpha\beta}}{8M_{T_1}^{2}}$ & $-\frac{\sum_{x,z}V_{1x}^{*}V_{1z}(\lambda_{T_2})_{x}^{*}(\lambda_{T_2})_{z}\delta_{\alpha\beta}}{4M_{T_2}^{2}}$  \\
       \hline
       $\frac{2}{v^{2}} (\epsilon_{A}^{dd})_{\alpha\beta}$ & $-\frac{\sum_{x,z}V_{1x}^{*}V_{1z}(\lambda_{T_1})_{x}^{*}(\lambda_{T_1})_{z}\delta_{\alpha\beta}}{8M_{T_1}^{2}}$ & $\frac{\sum_{x,z}V_{1x}^{*}V_{1z}(\lambda_{T_2})_{x}^{*}(\lambda_{T_2})_{z}\delta_{\alpha\beta}}{4M_{T_2}^{2}}$   \\
       \hline
       $\frac{2}{v^{2}} (\epsilon_{V}^{uu})_{\alpha\beta}$  & $\frac{(\lambda_{T_1})_{1}^{*}(\lambda_{T_1})_{1}\delta_{\alpha\beta}}{4M_{T_1}^{2}}$  & $-\frac{(\lambda_{T_2})_{1}^{*}(\lambda_{T_2})_{1}\delta_{\alpha\beta}}{8M_{T_2}^{2}}$  \\
       \hline
       $\frac{2}{v^{2}} (\epsilon_{A}^{uu})_{\alpha\beta}$ & $-\frac{(\lambda_{T_1})_{1}^{*}(\lambda_{T_1})_{1}\delta_{\alpha\beta}}{4M_{T_1}^{2}}$  & $\frac{(\lambda_{T_2})_{1}^{*}(\lambda_{T_2})_{1}\delta_{\alpha\beta}}{8M_{T_2}^{2}}$  \\
    \hline
    \end{tabular}
    \caption{Matching to $T_1$, and $T_2$ models.}
\end{table}
\renewcommand{\arraystretch}{1.0}

\clearpage

\subsection{Gauge boson models}

\renewcommand{\arraystretch}{1.8}
\begin{table}[h!]
    \centering
    \begin{tabular}{|c|c|}
       \hline
  & $\mathcal{B}$ model
       \\
       \hline
       WEFT WC  & Matching  \\
       \hline
       $\frac{2}{v^{2}} (\epsilon_{V}^{ee})_{\alpha\beta}$  & $\frac{(g^{l}_{\mathcal{B}})_{\alpha\beta}\left((g^{e}_{\mathcal{B}})_{11}+(g^{l}_{\mathcal{B}})_{11}\right)}{M_{\mathcal{B}}^{2}}+\frac{g^{\phi}_{\mathcal{B}}\left[\left((g^{e}_{\mathcal{B}})_{11}+(g^{l}_{\mathcal{B}})_{11}\right)\delta_{\alpha\beta}-(1-4s^{2})(g^{l}_{\mathcal{B}})_{\alpha\beta}\right]}{2M_{\mathcal{B}}^{2}}$ \\
       \hline
       $\frac{2}{v^{2}} (\epsilon_{A}^{ee})_{\alpha\beta}$  & $\frac{(g^{l}_{\mathcal{B}})_{\alpha\beta}\left((g^{e}_{\mathcal{B}})_{11}-(g^{l}_{\mathcal{B}})_{11}\right)}{M_{\mathcal{B}}^{2}}+\frac{g^{\phi}_{\mathcal{B}}\left[\left((g^{e}_{\mathcal{B}})_{11}-(g^{l}_{\mathcal{B}})_{11}\right)\delta_{\alpha\beta}+(g^{l}_{\mathcal{B}})_{\alpha\beta}\right]}{2M_{\mathcal{B}}^{2}}$ \\
       \hline
        $\frac{2}{v^{2}} (\epsilon_{V}^{uu})_{\alpha\beta}$  & $\frac{(g^{l}_{\mathcal{B}})_{\alpha\beta}\left((g^{u}_{\mathcal{B}})_{11}+(g^{q}_{\mathcal{B}})_{11}\right)}{M_{\mathcal{B}}^{2}}+\frac{g^{\phi}_{\mathcal{B}}\left[3\left((g^{u}_{\mathcal{B}})_{11}+(g^{q}_{\mathcal{B}})_{11}\right)\delta_{\alpha\beta}+(3-8s^{2})(g^{l}_{\mathcal{B}})_{\alpha\beta}\right]}{6M_{\mathcal{B}}^{2}}$\\
        \hline
         $\frac{2}{v^{2}} (\epsilon_{A}^{uu})_{\alpha\beta}$  & $\frac{(g^{l}_{\mathcal{B}})_{\alpha\beta}\left((g^{u}_{\mathcal{B}})_{11}-(g^{q}_{\mathcal{B}})_{11}\right)}{M_{\mathcal{B}}^{2}}+\frac{g^{\phi}_{\mathcal{B}}\left[3\left((g^{u}_{\mathcal{B}})_{11}-(g^{q}_{\mathcal{B}})_{11}\right)\delta_{\alpha\beta}-(g^{l}_{\mathcal{B}})_{\alpha\beta}\right]}{2M_{\mathcal{B}}^{2}}$\\
        \hline
        $\frac{2}{v^{2}} (\epsilon_{V}^{dd})_{\alpha\beta}$  & $\frac{(g^{l}_{\mathcal{B}})_{\alpha\beta}\left((g^{d}_{\mathcal{B}})_{11}+\sum_{xz}V_{1z}V_{1x}^{*}(g^{q}_{\mathcal{B}})_{xz}\right)}{M_{\mathcal{B}}^{2}}+\frac{g^{\phi}_{\mathcal{B}}\left[3\left((g^{d}_{\mathcal{B}})_{11}+\sum_{xz}V_{1z}V_{1x}^{*}(g^{q}_{\mathcal{B}})_{xz}\right)\delta_{\alpha\beta}-(3-4s^{2})(g^{l}_{\mathcal{B}})_{\alpha\beta}\right]}{6M_{\mathcal{B}}^{2}}$\\
        \hline
         $\frac{2}{v^{2}} (\epsilon_{A}^{dd})_{\alpha\beta}$  & $\frac{(g^{l}_{\mathcal{B}})_{\alpha\beta}\left((g^{d}_{\mathcal{B}})_{11}-\sum_{xz}V_{1z}V_{1x}^{*}(g^{q}_{\mathcal{B}})_{xz}\right)}{M_{\mathcal{B}}^{2}}+\frac{g^{\phi}_{\mathcal{B}}\left[3\left((g^{d}_{\mathcal{B}})_{11}-\sum_{xz}V_{1z}V_{1x}^{*}(g^{q}_{\mathcal{B}})_{xz}\right)\delta_{\alpha\beta}+(g^{l}_{\mathcal{B}})_{\alpha\beta}\right]}{2M_{\mathcal{B}}^{2}}$\\
         \hline
    \end{tabular}
    \caption{Matching to $\mathcal{B}$ model.}
\end{table}
\renewcommand{\arraystretch}{1.0}

\renewcommand{\arraystretch}{1.8}
\begin{table}[h!]
    \centering
    \begin{tabular}{|c|c|}
       \hline
  & $\mathcal{W}$ model
       \\
       \hline
       WEFT WC  & Matching  \\
       \hline
       $\frac{2V_{k j}^{*}}{v^{2}} (\epsilon_{L}^{kj})^{*}_{\alpha\beta}$  & $\frac{(g^{l}_{\mathcal{W}})_{\beta\alpha }\sum_{x}V_{xj}^{*}(g^{q}_{\mathcal{W}})^{*}_{xk}}{2M_{\mathcal{W}}^{2}}-\frac{g^{\phi}_{\mathcal{W}}\left[\sum_{x}V_{xj}^{*}(g^{q}_{\mathcal{W}})_{kx}^{*}\delta_{\alpha\beta}+(g^{l}_{\mathcal{W}})_{\alpha\beta}V_{kj}^{*}\right]}{4M_{\mathcal{W}}^{2}}$\\
       \hline
       $\frac{2}{v^{2}} (\epsilon_{V}^{ee})_{\alpha\beta}$  & $\frac{(g^{l}_{\mathcal{W}})_{\alpha 1}(g^{l}_{\mathcal{W}})_{1\beta}}{2M_{\mathcal{W}}^{2}}+\frac{g^{\phi}_{\mathcal{W}}\left[(g^{l}_{\mathcal{W}})_{11}\delta_{\alpha\beta}-2(g^{l}_{\mathcal{W}})_{\beta 1}^{*}\delta_{1\alpha}-2(g^{l}_{\mathcal{W}})_{\alpha 1}\delta_{1\beta}+(1-4s^{2})(g^{l}_{\mathcal{W}})_{\alpha\beta}\right]}{8M_{\mathcal{W}}^{2}}$ \\
       \hline
       $\frac{2}{v^{2}} (\epsilon_{A}^{ee})_{\alpha\beta}$  & $-\frac{(g^{l}_{\mathcal{W}})_{\alpha 1}(g^{l}_{\mathcal{W}})_{1\beta}}{2M_{\mathcal{W}}^{2}}-\frac{g^{\phi}_{\mathcal{W}}\left[(g^{l}_{\mathcal{W}})_{11}\delta_{\alpha\beta}-2(g^{l}_{\mathcal{W}})_{\beta 1}^{*}\delta_{1\alpha}-2(g^{l}_{\mathcal{W}})_{\alpha 1}\delta_{1\beta}+(g^{l}_{\mathcal{W}})_{\alpha\beta}\right]}{8M_{\mathcal{W}}^{2}}$ \\
       \hline
        $\frac{2}{v^{2}} (\epsilon_{V}^{uu})_{\alpha\beta}$  & $\frac{(g^{l}_{\mathcal{W}})_{\alpha \beta}(g^{q}_{\mathcal{W}})_{11}}{4M_{\mathcal{W}}^{2}}-\frac{g^{\phi}_{\mathcal{W}}\left[3(g^{q}_{\mathcal{W}})_{11}\delta_{\alpha\beta}+(3-8s^{2})(g^{l}_{\mathcal{W}})_{\alpha\beta}\right]}{24M_{\mathcal{W}}^{2}}$\\
        \hline
         $\frac{2}{v^{2}} (\epsilon_{A}^{uu})_{\alpha\beta}$  & $-\frac{(g^{l}_{\mathcal{W}})_{\alpha \beta}(g^{q}_{\mathcal{W}})_{11}}{4M_{\mathcal{W}}^{2}}+\frac{g^{\phi}_{\mathcal{W}}\left[(g^{q}_{\mathcal{W}})_{11}\delta_{\alpha\beta}+(g^{l}_{\mathcal{W}})_{\alpha\beta}\right]}{8M_{\mathcal{W}}^{2}}$\\
        \hline
        $\frac{2}{v^{2}} (\epsilon_{V}^{dd})_{\alpha\beta}$  & $-\frac{(g^{l}_{\mathcal{W}})_{\alpha\beta}\sum_{xz}V_{1z}V_{1x}^{*}(g^{q}_{\mathcal{W}})_{xz}}{4M_{\mathcal{W}}^{2}}+\frac{g^{\phi}_{\mathcal{W}}\left[3\sum_{xz}V_{1z}V_{1x}^{*}(g^{q}_{\mathcal{W}})_{xz}\delta_{\alpha\beta}+(3-4s^{2})(g^{l}_{\mathcal{W}})_{\alpha\beta}\right]}{24M_{\mathcal{W}}^{2}}$\\
        \hline
         $\frac{2}{v^{2}} (\epsilon_{A}^{dd})_{\alpha\beta}$  & $\frac{(g^{l}_{\mathcal{W}})_{\alpha\beta}\sum_{xz}V_{1z}V_{1x}^{*}(g^{q}_{\mathcal{W}})_{xz}}{4M_{\mathcal{W}}^{2}}-\frac{g^{\phi}_{\mathcal{W}}\left[3\sum_{xz}V_{1z}V_{1x}^{*}(g^{q}_{\mathcal{W}})_{xz}\delta_{\alpha\beta}+(g^{l}_{\mathcal{W}})_{\alpha\beta}\right]}{8M_{\mathcal{W}}^{2}}$\\
         \hline
    \end{tabular}
    \caption{Matching to $\mathcal{W}$ model.}
\end{table}
\renewcommand{\arraystretch}{1.0}

\renewcommand{\arraystretch}{1.8}
\begin{table}[h!]
    \centering
    \begin{tabular}{|c|c|c|c|}
       \hline
   & $\mathcal{L}_1$ model & $\mathcal{L}_3$ model &
$\mathcal{B}_1$ model       \\
       \hline
       WEFT WC  & \multicolumn{3}{c|}{Matching}   \\
       \hline
       $\frac{2V_{k j}^{*}}{v^{2}} (\epsilon_{R}^{kj})^{*}_{\alpha\beta}$  & $\frac{1}{f}\frac{\sum_{x}\gamma_{\mathcal{L}_{1}}(\tilde{y}_{u})^{*}_{k x}(\tilde{g}^{dDq}_{\mathcal{L}_1})_{j x}\delta_{\alpha\beta}}{2M_{\mathcal{L}_1}^{2}}$ & - & $\frac{g_{\mathcal{B}_1}^{\phi}(g_{\mathcal{B}_1}^{du})_{jk}\delta_{\alpha\beta}}{M_{\mathcal{B}_1}^{2}}$   \\
       \hline
       $\frac{2V_{k j}^{*}}{v^{2}} (\epsilon_{P}^{kj})^{*}_{\alpha\beta}$  & $\frac{1}{f}\frac{\gamma_{\mathcal{L}_{1}}(\tilde{g}^{eDl}_{\mathcal{L}_1})_{\alpha\beta}\left[\sum_{x}V_{xj}(\tilde{y}_{u})^{*}_{k x}+(\tilde{y}_{d})^{*}_{j k}\right]}{2M_{\mathcal{L}_1}^{2}}$ & - & -\\
       \hline
       $\frac{2V_{k j}^{*}}{v^{2}} (\epsilon_{S}^{kj})^{*}_{\alpha\beta}$  & $-\frac{1}{f}\frac{\gamma_{\mathcal{L}_{1}}(\tilde{g}^{eDl}_{\mathcal{L}_1})_{\alpha\beta}\left[\sum_{x}V_{xj}(\tilde{y}_{u})^{*}_{k x}-(\tilde{y}_{d})^{*}_{j k}\right]}{2M_{\mathcal{L}_1}^{2}}$ & - & - \\
       \hline
       $\frac{2}{v^{2}} (\epsilon_{V}^{ee})_{\alpha\beta}$  & $-\frac{1}{f}\frac{\gamma_{\mathcal{L}_{1}}(\tilde{y}_{e})_{1\beta}(\tilde{g}^{eDL}_{\mathcal{L}_1})_{1\alpha}^{*}}{4M_{\mathcal{L}_1}^{2}}$  & $-\frac{(g_{\mathcal{L}_3})_{1\beta}(g_{\mathcal{L}_3})_{1\alpha}^{*}}{M_{\mathcal{L}_3}^{2}}$ & - \\
       \hline
       $\frac{2}{v^{2}} (\epsilon_{A}^{uu})_{\alpha\beta}$ & $-\frac{1}{f}\frac{\gamma_{\mathcal{L}_{1}}(\tilde{y}_{e})_{1\beta}(\tilde{g}^{eDL}_{\mathcal{L}_1})_{1\alpha}^{*}}{4M_{\mathcal{L}_1}^{2}}$  & $-\frac{(g_{\mathcal{L}_3})_{1\beta}(g_{\mathcal{L}_3})_{1\alpha}^{*}}{M_{\mathcal{L}_3}^{2}}$ & - \\
    \hline
    \end{tabular}
    \caption{Matching to $\mathcal{L}_1$, $\mathcal{L}_3$, and $\mathcal{B}_1$ models.}
\end{table}
\renewcommand{\arraystretch}{1.0}

\renewcommand{\arraystretch}{1.8}
\begin{table}[h!]
    \centering
    \begin{tabular}{|c|c|c|c|c|}
       \hline
   & $\mathcal{Q}_1$ model & $\mathcal{Q}_5$ model & $\mathcal{U}_{2}$ model & $\mathcal{X}$ model
       \\
       \hline
       WEFT WC  & \multicolumn{4}{c|}{Matching}   \\
       \hline
       $\frac{2V_{k j}^{*}}{v^{2}} (\epsilon_{L}^{kj})^{*}_{\alpha\beta}$  & - & - &$\frac{\sum_{x}V_{xj}^{*}(g^{lq}_{\mathcal{U}_2})^{*}_{\alpha x}(g^{lq}_{\mathcal{U}_2})_{\beta k}}{M_{\mathcal{U}_2}^{2}}$ & $-\frac{\sum_{x}V_{xj}^{*}(g_{\mathcal{X}})_{\alpha x}^{*}(g_{\mathcal{X}})_{\beta k}}{4M_{\mathcal{X}}^{2}}$   \\
       \hline
       $\frac{2V_{k j}^{*}}{v^{2}} (\epsilon_{P}^{kj})^{*}_{\alpha\beta}$  & - & $\frac{2(g^{eq}_{\mathcal{Q}_5})_{\alpha k}(g^{dl}_{\mathcal{Q}_5})_{j\beta}^{*}}{M_{\mathcal{Q}_5}^{2}}$ & $-\frac{2(g^{ed}_{\mathcal{U}_2})_{\alpha j}^{*}(g^{lq}_{\mathcal{U}_2})_{\beta k}}{M_{\mathcal{U}_2}^{2}}$ & - \\
       \hline
       $\frac{2V_{k j}^{*}}{v^{2}} (\epsilon_{S}^{kj})^{*}_{\alpha\beta}$  & - & $\frac{2(g^{eq}_{\mathcal{Q}_5})_{\alpha k}(g^{dl}_{\mathcal{Q}_5})_{j\beta}^{*}}{M_{\mathcal{Q}_5}^{2}}$ & $-\frac{2(g^{ed}_{\mathcal{U}_2})_{\alpha j}^{*}(g^{lq}_{\mathcal{U}_2})_{\beta k}}{M_{\mathcal{U}_2}^{2}}$ & -\\
       \hline
       $\frac{2}{v^{2}} (\epsilon_{V}^{dd})_{\alpha\beta}$  & - & $-\frac{(g^{dl}_{\mathcal{Q}_5})_{1\alpha}^{*}(g^{dl}_{\mathcal{Q}_5})_{1\beta}}{M_{\mathcal{Q}_5}^{2}}$  & - & $\frac{\sum_{xz}V_{1x}V_{1z}^{*}(g_{\mathcal{X}})_{\alpha x}(g_{\mathcal{X}})_{\beta z}^{*}}{2M_{\mathcal{X}}^{2}}$ \\
       \hline
       $\frac{2}{v^{2}} (\epsilon_{A}^{dd})_{\alpha\beta}$ & - & $-\frac{(g^{dl}_{\mathcal{Q}_5})_{1\alpha}^{*}(g^{dl}_{\mathcal{Q}_5})_{1\beta}}{M_{\mathcal{Q}_5}^{2}}$ & - & $-\frac{\sum_{xz}V_{1x}V_{1z}^{*}(g_{\mathcal{X}})_{\alpha x}(g_{\mathcal{X}})_{\beta z}^{*}}{2M_{\mathcal{X}}^{2}}$   \\
       \hline
       $\frac{2}{v^{2}} (\epsilon_{V}^{uu})_{\alpha\beta}$  & $-\frac{(g^{ul}_{\mathcal{Q}_1})_{1\alpha}^{*}(g^{ul}_{\mathcal{Q}_1})_{1\beta}}{M_{\mathcal{Q}_1}^{2}}$  & -  & $\frac{(g^{lq}_{\mathcal{U}_2})_{\alpha 1}(g^{lq}_{\mathcal{U}_2})_{\beta 1}^{*}}{M_{\mathcal{U}_2}^{2}}$ & $\frac{(g_{\mathcal{X}})_{\alpha 1}(g_{\mathcal{X}})_{\beta 1}^{*}}{4M_{\mathcal{X}}^{2}}$ \\
       \hline
       $\frac{2}{v^{2}} (\epsilon_{A}^{uu})_{\alpha\beta}$ & $-\frac{(g^{ul}_{\mathcal{Q}_1})_{1\alpha}^{*}(g^{ul}_{\mathcal{Q}_1})_{1\beta}}{M_{\mathcal{Q}_1}^{2}}$  & - & $-\frac{(g^{lq}_{\mathcal{U}_2})_{\alpha 1}(g^{lq}_{\mathcal{U}_2})_{\beta 1}^{*}}{M_{\mathcal{U}_2}^{2}}$ & $-\frac{(g_{\mathcal{X}})_{\alpha 1}(g_{\mathcal{X}})_{\beta 1}^{*}}{4M_{\mathcal{X}}^{2}}$\\
    \hline
    \end{tabular}
    \caption{Matching to $\mathcal{Q}_1$, $\mathcal{Q}_5$, $\mathcal{U}_{2}$, and $\mathcal{X}$ models.}
\end{table}
\renewcommand{\arraystretch}{1.0}

 \clearpage

\bibliographystyle{JHEP}
\bibliography{main_v3}

\providecommand{\href}[2]{#2}\begingroup\raggedright\begin{thebibliography}{10}

\bibitem{Farzan:2017xzy}
Y.~Farzan and M.~Tortola, \emph{{Neutrino oscillations and Non-Standard
  Interactions}}, \href{https://doi.org/10.3389/fphy.2018.00010}{\emph{Front.
  in Phys.} {\bfseries 6} (2018) 10}
  [\href{https://arxiv.org/abs/1710.09360}{{\ttfamily 1710.09360}}].

\bibitem{Proceedings:2019qno}
\emph{{Neutrino Non-Standard Interactions: A Status Report}}, vol.~2, 2019.
\newblock 10.21468/SciPostPhysProc.2.001.

\bibitem{Arguelles:2022tki}
C.A.~Arg\"uelles et~al., \emph{{Snowmass white paper: beyond the standard model
  effects on neutrino flavor: Submitted to the proceedings of the US community
  study on the future of particle physics (Snowmass 2021)}},
  \href{https://doi.org/10.1140/epjc/s10052-022-11049-7}{\emph{Eur. Phys. J. C}
  {\bfseries 83} (2023) 15} [\href{https://arxiv.org/abs/2203.10811}{{\ttfamily
  2203.10811}}].

\bibitem{Wolfenstein:1977ue}
L.~Wolfenstein, \emph{{Neutrino Oscillations in Matter}},
  \href{https://doi.org/10.1103/PhysRevD.17.2369}{\emph{Phys. Rev. D}
  {\bfseries 17} (1978) 2369}.

\bibitem{Mikheyev:1985zog}
S.P.~Mikheyev and A.Y.~Smirnov, \emph{{Resonance Amplification of Oscillations
  in Matter and Spectroscopy of Solar Neutrinos}}, {\emph{Sov. J. Nucl. Phys.}
  {\bfseries 42} (1985) 913}.

\bibitem{Gourlay:2022odf}
S.~Gourlay et~al., \emph{{Snowmass'21 Accelerator Frontier Report}},
  \href{https://arxiv.org/abs/2209.14136}{{\ttfamily 2209.14136}}.

\bibitem{Artuso:2022ouk}
M.~Artuso et~al., \emph{{Report of the Frontier For Rare Processes and
  Precision Measurements}},  \href{https://arxiv.org/abs/2210.04765}{{\ttfamily
  2210.04765}}.

\bibitem{deBlas:2017xtg}
J.~de~Blas, J.C.~Criado, M.~Perez-Victoria and J.~Santiago, \emph{{Effective
  description of general extensions of the Standard Model: the complete
  tree-level dictionary}},
  \href{https://doi.org/10.1007/JHEP03(2018)109}{\emph{JHEP} {\bfseries 03}
  (2018) 109} [\href{https://arxiv.org/abs/1711.10391}{{\ttfamily
  1711.10391}}].

\bibitem{Gavela:2008ra}
M.B.~Gavela, D.~Hernandez, T.~Ota and W.~Winter, \emph{{Large gauge invariant
  non-standard neutrino interactions}},
  \href{https://doi.org/10.1103/PhysRevD.79.013007}{\emph{Phys. Rev. D}
  {\bfseries 79} (2009) 013007}
  [\href{https://arxiv.org/abs/0809.3451}{{\ttfamily 0809.3451}}].

\bibitem{Li:2023cwy}
X.-X.~Li, Z.~Ren and J.-H.~Yu, \emph{{A complete tree-level dictionary between
  simplified BSM models and SMEFT (d $\leq$ 7) operators}},
  \href{https://arxiv.org/abs/2307.10380}{{\ttfamily 2307.10380}}.

\bibitem{Guedes:2023azv}
G.~Guedes, P.~Olgoso and J.~Santiago, \emph{{Towards the one loop IR/UV
  dictionary in the SMEFT: one loop generated operators from new scalars and
  fermions}},  \href{https://arxiv.org/abs/2303.16965}{{\ttfamily 2303.16965}}.

\bibitem{Bergmann:1999pk}
S.~Bergmann, Y.~Grossman and D.M.~Pierce, \emph{{Can lepton flavor violating
  interactions explain the atmospheric neutrino problem?}},
  \href{https://doi.org/10.1103/PhysRevD.61.053005}{\emph{Phys. Rev. D}
  {\bfseries 61} (2000) 053005}
  [\href{https://arxiv.org/abs/hep-ph/9909390}{{\ttfamily hep-ph/9909390}}].

\bibitem{Antusch:2008tz}
S.~Antusch, J.P.~Baumann and E.~Fernandez-Martinez, \emph{{Non-Standard
  Neutrino Interactions with Matter from Physics Beyond the Standard Model}},
  \href{https://doi.org/10.1016/j.nuclphysb.2008.11.018}{\emph{Nucl. Phys. B}
  {\bfseries 810} (2009) 369}
  [\href{https://arxiv.org/abs/0807.1003}{{\ttfamily 0807.1003}}].

\bibitem{Davidson:2019iqh}
S.~Davidson and M.~Gorbahn, \emph{{Charged lepton flavor change and nonstandard
  neutrino interactions}},
  \href{https://doi.org/10.1103/PhysRevD.101.015010}{\emph{Phys. Rev. D}
  {\bfseries 101} (2020) 015010}
  [\href{https://arxiv.org/abs/1909.07406}{{\ttfamily 1909.07406}}].

\bibitem{Bischer:2019ttk}
I.~Bischer and W.~Rodejohann, \emph{{General neutrino interactions from an
  effective field theory perspective}},
  \href{https://doi.org/10.1016/j.nuclphysb.2019.114746}{\emph{Nucl. Phys. B}
  {\bfseries 947} (2019) 114746}
  [\href{https://arxiv.org/abs/1905.08699}{{\ttfamily 1905.08699}}].

\bibitem{Falkowski:2021bkq}
A.~Falkowski, M.~Gonz\'alez-Alonso, J.~Kopp, Y.~Soreq and Z.~Tabrizi,
  \emph{{EFT at FASER\ensuremath{\nu}}},
  \href{https://doi.org/10.1007/JHEP10(2021)086}{\emph{JHEP} {\bfseries 10}
  (2021) 086} [\href{https://arxiv.org/abs/2105.12136}{{\ttfamily
  2105.12136}}].

\bibitem{Meloni:2009cg}
D.~Meloni, T.~Ohlsson, W.~Winter and H.~Zhang, \emph{{Non-standard interactions
  versus non-unitary lepton flavor mixing at a neutrino factory}},
  \href{https://doi.org/10.1007/JHEP04(2010)041}{\emph{JHEP} {\bfseries 04}
  (2010) 041} [\href{https://arxiv.org/abs/0912.2735}{{\ttfamily 0912.2735}}].

\bibitem{Altmannshofer:2018xyo}
W.~Altmannshofer, M.~Tammaro and J.~Zupan, \emph{{Non-standard neutrino
  interactions and low energy experiments}},
  \href{https://doi.org/10.1007/JHEP11(2021)113}{\emph{JHEP} {\bfseries 09}
  (2019) 083} [\href{https://arxiv.org/abs/1812.02778}{{\ttfamily
  1812.02778}}].

\bibitem{Falkowski:2019kfn}
A.~Falkowski, M.~Gonz\'alez-Alonso and Z.~Tabrizi, \emph{{Consistent QFT
  description of non-standard neutrino interactions}},
  \href{https://doi.org/10.1007/JHEP11(2020)048}{\emph{JHEP} {\bfseries 11}
  (2020) 048} [\href{https://arxiv.org/abs/1910.02971}{{\ttfamily
  1910.02971}}].

\bibitem{Falkowski:2019xoe}
A.~Falkowski, M.~Gonz\'alez-Alonso and Z.~Tabrizi, \emph{{Reactor neutrino
  oscillations as constraints on Effective Field Theory}},
  \href{https://doi.org/10.1007/JHEP05(2019)173}{\emph{JHEP} {\bfseries 05}
  (2019) 173} [\href{https://arxiv.org/abs/1901.04553}{{\ttfamily
  1901.04553}}].

\bibitem{Babu:2019mfe}
K.S.~Babu, P.S.B.~Dev, S.~Jana and A.~Thapa, \emph{{Non-Standard Interactions
  in Radiative Neutrino Mass Models}},
  \href{https://doi.org/10.1007/JHEP03(2020)006}{\emph{JHEP} {\bfseries 03}
  (2020) 006} [\href{https://arxiv.org/abs/1907.09498}{{\ttfamily
  1907.09498}}].

\bibitem{Terol-Calvo:2019vck}
J.~Terol-Calvo, M.~T\'ortola and A.~Vicente, \emph{{High-energy constraints
  from low-energy neutrino nonstandard interactions}},
  \href{https://doi.org/10.1103/PhysRevD.101.095010}{\emph{Phys. Rev. D}
  {\bfseries 101} (2020) 095010}
  [\href{https://arxiv.org/abs/1912.09131}{{\ttfamily 1912.09131}}].

\bibitem{Babu:2020nna}
K.S.~Babu, D.~Gon\c{c}alves, S.~Jana and P.A.N.~Machado, \emph{{Neutrino
  Non-Standard Interactions: Complementarity Between LHC and Oscillation
  Experiments}},
  \href{https://doi.org/10.1016/j.physletb.2021.136131}{\emph{Phys. Lett. B}
  {\bfseries 815} (2021) 136131}
  [\href{https://arxiv.org/abs/2003.03383}{{\ttfamily 2003.03383}}].

\bibitem{Du:2020dwr}
Y.~Du, H.-L.~Li, J.~Tang, S.~Vihonen and J.-H.~Yu, \emph{{Non-standard
  interactions in SMEFT confronted with terrestrial neutrino experiments}},
  \href{https://doi.org/10.1007/JHEP03(2021)019}{\emph{JHEP} {\bfseries 03}
  (2021) 019} [\href{https://arxiv.org/abs/2011.14292}{{\ttfamily
  2011.14292}}].

\bibitem{Du:2021rdg}
Y.~Du, H.-L.~Li, J.~Tang, S.~Vihonen and J.-H.~Yu, \emph{{Exploring SMEFT
  induced nonstandard interactions: From COHERENT to neutrino oscillations}},
  \href{https://doi.org/10.1103/PhysRevD.105.075022}{\emph{Phys. Rev. D}
  {\bfseries 105} (2022) 075022}
  [\href{https://arxiv.org/abs/2106.15800}{{\ttfamily 2106.15800}}].

\bibitem{Breso-Pla:2023tnz}
V.~Bres\'o-Pla, A.~Falkowski, M.~Gonz\'alez-Alonso and K.~Mons\'alvez-Pozo,
  \emph{{EFT analysis of New Physics at COHERENT}},
  \href{https://doi.org/10.1007/JHEP05(2023)074}{\emph{JHEP} {\bfseries 05}
  (2023) 074} [\href{https://arxiv.org/abs/2301.07036}{{\ttfamily
  2301.07036}}].

\bibitem{Coloma:2023ixt}
P.~Coloma, M.C.~Gonzalez-Garcia, M.~Maltoni, J.a.P.~Pinheiro and S.~Urrea,
  \emph{{Global constraints on non-standard neutrino interactions with quarks
  and electrons}}, \href{https://doi.org/10.1007/JHEP08(2023)032}{\emph{JHEP}
  {\bfseries 08} (2023) 032}
  [\href{https://arxiv.org/abs/2305.07698}{{\ttfamily 2305.07698}}].

\bibitem{Giunti:1993se}
C.~Giunti, C.W.~Kim, J.A.~Lee and U.W.~Lee, \emph{{On the treatment of neutrino
  oscillations without resort to weak eigenstates}},
  \href{https://doi.org/10.1103/PhysRevD.48.4310}{\emph{Phys. Rev. D}
  {\bfseries 48} (1993) 4310}
  [\href{https://arxiv.org/abs/hep-ph/9305276}{{\ttfamily hep-ph/9305276}}].

\bibitem{Akhmedov:2010ms}
E.K.~Akhmedov and J.~Kopp, \emph{{Neutrino Oscillations: Quantum Mechanics vs.
  Quantum Field Theory}},
  \href{https://doi.org/10.1007/JHEP04(2010)008}{\emph{JHEP} {\bfseries 04}
  (2010) 008} [\href{https://arxiv.org/abs/1001.4815}{{\ttfamily 1001.4815}}].

\bibitem{Kobach:2017osm}
A.~Kobach, A.V.~Manohar and J.~McGreevy, \emph{{Neutrino Oscillation
  Measurements Computed in Quantum Field Theory}},
  \href{https://doi.org/10.1016/j.physletb.2018.06.021}{\emph{Phys. Lett. B}
  {\bfseries 783} (2018) 59}
  [\href{https://arxiv.org/abs/1711.07491}{{\ttfamily 1711.07491}}].

\bibitem{NuSTEC:2017hzk}
{\scshape NuSTEC} collaboration, \emph{{NuSTEC White Paper: Status and
  challenges of neutrino\textendash{}nucleus scattering}},
  \href{https://doi.org/10.1016/j.ppnp.2018.01.006}{\emph{Prog. Part. Nucl.
  Phys.} {\bfseries 100} (2018) 1}
  [\href{https://arxiv.org/abs/1706.03621}{{\ttfamily 1706.03621}}].

\bibitem{Lisi:1997yc}
E.~Lisi and D.~Montanino, \emph{{Earth regeneration effect in solar neutrino
  oscillations: An Analytic approach}},
  \href{https://doi.org/10.1103/PhysRevD.56.1792}{\emph{Phys. Rev. D}
  {\bfseries 56} (1997) 1792}
  [\href{https://arxiv.org/abs/hep-ph/9702343}{{\ttfamily hep-ph/9702343}}].

\bibitem{Cirigliano:2009wk}
V.~Cirigliano, J.~Jenkins and M.~Gonzalez-Alonso, \emph{{Semileptonic decays of
  light quarks beyond the Standard Model}},
  \href{https://doi.org/10.1016/j.nuclphysb.2009.12.020}{\emph{Nucl. Phys. B}
  {\bfseries 830} (2010) 95} [\href{https://arxiv.org/abs/0908.1754}{{\ttfamily
  0908.1754}}].

\bibitem{Jenkins:2017jig}
E.E.~Jenkins, A.V.~Manohar and P.~Stoffer, \emph{{Low-Energy Effective Field
  Theory below the Electroweak Scale: Operators and Matching}},
  \href{https://doi.org/10.1007/JHEP03(2018)016}{\emph{JHEP} {\bfseries 03}
  (2018) 016} [\href{https://arxiv.org/abs/1709.04486}{{\ttfamily
  1709.04486}}].

\bibitem{Campanelli:2002cc}
M.~Campanelli and A.~Romanino, \emph{{Effects of new physics in neutrino
  oscillations in matter}},
  \href{https://doi.org/10.1103/PhysRevD.66.113001}{\emph{Phys. Rev. D}
  {\bfseries 66} (2002) 113001}
  [\href{https://arxiv.org/abs/hep-ph/0207350}{{\ttfamily hep-ph/0207350}}].

\bibitem{DUNE:2020ypp}
{\scshape DUNE} collaboration, \emph{{Deep Underground Neutrino Experiment
  (DUNE), Far Detector Technical Design Report, Volume II: DUNE Physics}},
  \href{https://arxiv.org/abs/2002.03005}{{\ttfamily 2002.03005}}.

\bibitem{Aebischer:2023irs}
J.~Aebischer et~al., \emph{{Computing Tools for Effective Field Theories}},  7,
  2023 [\href{https://arxiv.org/abs/2307.08745}{{\ttfamily 2307.08745}}].

\bibitem{Falkowski:2017pss}
A.~Falkowski, M.~Gonz\'alez-Alonso and K.~Mimouni, \emph{{Compilation of
  low-energy constraints on 4-fermion operators in the SMEFT}},
  \href{https://doi.org/10.1007/JHEP08(2017)123}{\emph{JHEP} {\bfseries 08}
  (2017) 123} [\href{https://arxiv.org/abs/1706.03783}{{\ttfamily
  1706.03783}}].

\bibitem{Aebischer:2018bkb}
J.~Aebischer, J.~Kumar and D.M.~Straub, \emph{{Wilson: a Python package for the
  running and matching of Wilson coefficients above and below the electroweak
  scale}}, \href{https://doi.org/10.1140/epjc/s10052-018-6492-7}{\emph{Eur.
  Phys. J. C} {\bfseries 78} (2018) 1026}
  [\href{https://arxiv.org/abs/1804.05033}{{\ttfamily 1804.05033}}].

\bibitem{Straub:2018kue}
D.M.~Straub, \emph{{flavio: a Python package for flavour and precision
  phenomenology in the Standard Model and beyond}},
  \href{https://arxiv.org/abs/1810.08132}{{\ttfamily 1810.08132}}.

\bibitem{Aebischer:2018iyb}
J.~Aebischer, J.~Kumar, P.~Stangl and D.M.~Straub, \emph{{A Global Likelihood
  for Precision Constraints and Flavour Anomalies}},
  \href{https://doi.org/10.1140/epjc/s10052-019-6977-z}{\emph{Eur. Phys. J. C}
  {\bfseries 79} (2019) 509}
  [\href{https://arxiv.org/abs/1810.07698}{{\ttfamily 1810.07698}}].

\bibitem{CMS:2023bdh}
{\scshape CMS} collaboration, \emph{{Search for scalar leptoquarks produced in
  lepton-quark collisions and coupled to $\tau$ leptons}},
  \href{https://arxiv.org/abs/2308.06143}{{\ttfamily 2308.06143}}.

\bibitem{Majhi:2022wyp}
R.~Majhi, D.K.~Singha, K.N.~Deepthi and R.~Mohanta, \emph{{Vector leptoquark
  $U_3$ and CP violation at T2K, NOvA experiments}},
  \href{https://doi.org/10.1140/epjc/s10052-022-10900-1}{\emph{Eur. Phys. J. C}
  {\bfseries 82} (2022) 919}
  [\href{https://arxiv.org/abs/2205.04269}{{\ttfamily 2205.04269}}].

\bibitem{Berezinsky:1985yw}
V.S.~Berezinsky, \emph{{ON GENERATION OF 'LIGHT' LEPTOQUARKS AND SQUARKS IN
  DEEP WATER DETECTORS. (IN RUSSIAN)}}, {\emph{Yad. Fiz.} {\bfseries 41} (1985)
  393}.

\bibitem{Robinett:1987ym}
R.W.~Robinett, \emph{{The Production of Leptoquarks in Ultrahigh-energy
  Neutrino Interactions}},
  \href{https://doi.org/10.1103/PhysRevD.37.84}{\emph{Phys. Rev. D} {\bfseries
  37} (1988) 84}.

\bibitem{Dorsner:2016wpm}
I.~Dor\v{s}ner, S.~Fajfer, A.~Greljo, J.F.~Kamenik and N.~Ko\v{s}nik,
  \emph{{Physics of leptoquarks in precision experiments and at particle
  colliders}}, \href{https://doi.org/10.1016/j.physrep.2016.06.001}{\emph{Phys.
  Rept.} {\bfseries 641} (2016) 1}
  [\href{https://arxiv.org/abs/1603.04993}{{\ttfamily 1603.04993}}].

\bibitem{Becirevic:2018uab}
D.~Be\v{c}irevi\'c, B.~Panes, O.~Sumensari and R.~Zukanovich~Funchal,
  \emph{{Seeking leptoquarks in IceCube}},
  \href{https://doi.org/10.1007/JHEP06(2018)032}{\emph{JHEP} {\bfseries 06}
  (2018) 032} [\href{https://arxiv.org/abs/1803.10112}{{\ttfamily
  1803.10112}}].

\bibitem{Huang:2021mki}
G.-y.~Huang, S.~Jana, M.~Lindner and W.~Rodejohann, \emph{{Probing new physics
  at future tau neutrino telescopes}},
  \href{https://doi.org/10.1088/1475-7516/2022/02/038}{\emph{JCAP} {\bfseries
  02} (2022) 038} [\href{https://arxiv.org/abs/2112.09476}{{\ttfamily
  2112.09476}}].

\bibitem{Calabrese:2022mnp}
R.~Calabrese, J.~Gunn, G.~Miele, S.~Morisi, S.~Roy and P.~Santorelli,
  \emph{{Constraining scalar leptoquarks using COHERENT data}},
  \href{https://doi.org/10.1103/PhysRevD.107.055039}{\emph{Phys. Rev. D}
  {\bfseries 107} (2023) 055039}
  [\href{https://arxiv.org/abs/2212.11210}{{\ttfamily 2212.11210}}].

\bibitem{Kirk:2023fin}
M.~Kirk, S.~Okawa and K.~Wu, \emph{{A $\nu$ window onto leptoquarks?}},
  \href{https://arxiv.org/abs/2307.11152}{{\ttfamily 2307.11152}}.

\bibitem{DeRomeri:2023cjt}
V.~De~Romeri, V.M.~Lozano and G.~Sanchez~Garcia, \emph{{A neutrino window to
  scalar leptoquarks: from low energy to colliders}},
  \href{https://arxiv.org/abs/2307.13790}{{\ttfamily 2307.13790}}.

\bibitem{Schwemberger:2023hee}
T.~Schwemberger, V.~Takhistov and T.-T.~Yu, \emph{{Hunting Nonstandard Neutrino
  Interactions and Leptoquarks in Dark Matter Experiments}},
  \href{https://arxiv.org/abs/2307.15736}{{\ttfamily 2307.15736}}.

\bibitem{Huber:2004ka}
P.~Huber, M.~Lindner and W.~Winter, \emph{{Simulation of long-baseline neutrino
  oscillation experiments with GLoBES (General Long Baseline Experiment
  Simulator)}}, \href{https://doi.org/10.1016/j.cpc.2005.01.003}{\emph{Comput.
  Phys. Commun.} {\bfseries 167} (2005) 195}
  [\href{https://arxiv.org/abs/hep-ph/0407333}{{\ttfamily hep-ph/0407333}}].

\bibitem{Huber:2007ji}
P.~Huber, J.~Kopp, M.~Lindner, M.~Rolinec and W.~Winter, \emph{{New features in
  the simulation of neutrino oscillation experiments with GLoBES 3.0: General
  Long Baseline Experiment Simulator}},
  \href{https://doi.org/10.1016/j.cpc.2007.05.004}{\emph{Comput. Phys. Commun.}
  {\bfseries 177} (2007) 432}
  [\href{https://arxiv.org/abs/hep-ph/0701187}{{\ttfamily hep-ph/0701187}}].

\bibitem{Blennow:2009pk}
M.~Blennow and E.~Fernandez-Martinez, \emph{{Neutrino oscillation parameter
  sampling with MonteCUBES}},
  \href{https://doi.org/10.1016/j.cpc.2009.09.014}{\emph{Comput. Phys. Commun.}
  {\bfseries 181} (2010) 227}
  [\href{https://arxiv.org/abs/0903.3985}{{\ttfamily 0903.3985}}].

\bibitem{DUNE:2021cuw}
{\scshape DUNE} collaboration, \emph{{Experiment Simulation Configurations
  Approximating DUNE TDR}},  \href{https://arxiv.org/abs/2103.04797}{{\ttfamily
  2103.04797}}.

\bibitem{DUNE:2020jqi}
{\scshape DUNE} collaboration, \emph{{Long-baseline neutrino oscillation
  physics potential of the DUNE experiment}},
  \href{https://doi.org/10.1140/epjc/s10052-020-08456-z}{\emph{Eur. Phys. J. C}
  {\bfseries 80} (2020) 978}
  [\href{https://arxiv.org/abs/2006.16043}{{\ttfamily 2006.16043}}].

\bibitem{Esteban:2020cvm}
I.~Esteban, M.C.~Gonzalez-Garcia, M.~Maltoni, T.~Schwetz and A.~Zhou,
  \emph{{The fate of hints: updated global analysis of three-flavor neutrino
  oscillations}}, \href{https://doi.org/10.1007/JHEP09(2020)178}{\emph{JHEP}
  {\bfseries 09} (2020) 178}
  [\href{https://arxiv.org/abs/2007.14792}{{\ttfamily 2007.14792}}].

\bibitem{Dziewonski:1981xy}
A.M.~Dziewonski and D.L.~Anderson, \emph{{Preliminary reference earth model}},
  \href{https://doi.org/10.1016/0031-9201(81)90046-7}{\emph{Phys. Earth Planet.
  Interiors} {\bfseries 25} (1981) 297}.

\bibitem{Chatterjee:2021wac}
S.S.~Chatterjee, P.S.B.~Dev and P.A.N.~Machado, \emph{{Impact of improved
  energy resolution on DUNE sensitivity to neutrino non-standard
  interactions}}, \href{https://doi.org/10.1007/JHEP08(2021)163}{\emph{JHEP}
  {\bfseries 08} (2021) 163}
  [\href{https://arxiv.org/abs/2106.04597}{{\ttfamily 2106.04597}}].

\bibitem{Forero:2016ghr}
D.V.~Forero and W.-C.~Huang, \emph{{Sizable NSI from the SU(2)$_{L}$ scalar
  doublet-singlet mixing and the implications in DUNE}},
  \href{https://doi.org/10.1007/JHEP03(2017)018}{\emph{JHEP} {\bfseries 03}
  (2017) 018} [\href{https://arxiv.org/abs/1608.04719}{{\ttfamily
  1608.04719}}].

\bibitem{Dey:2018yht}
U.K.~Dey, N.~Nath and S.~Sadhukhan, \emph{{Non-Standard Neutrino Interactions
  in a Modified $\nu$2HDM}},
  \href{https://doi.org/10.1103/PhysRevD.98.055004}{\emph{Phys. Rev. D}
  {\bfseries 98} (2018) 055004}
  [\href{https://arxiv.org/abs/1804.05808}{{\ttfamily 1804.05808}}].

\bibitem{Farzan:2015hkd}
Y.~Farzan and I.M.~Shoemaker, \emph{{Lepton Flavor Violating Non-Standard
  Interactions via Light Mediators}},
  \href{https://doi.org/10.1007/JHEP07(2016)033}{\emph{JHEP} {\bfseries 07}
  (2016) 033} [\href{https://arxiv.org/abs/1512.09147}{{\ttfamily
  1512.09147}}].

\bibitem{Farzan:2016wym}
Y.~Farzan and J.~Heeck, \emph{{Neutrinophilic nonstandard interactions}},
  \href{https://doi.org/10.1103/PhysRevD.94.053010}{\emph{Phys. Rev. D}
  {\bfseries 94} (2016) 053010}
  [\href{https://arxiv.org/abs/1607.07616}{{\ttfamily 1607.07616}}].

\bibitem{Farzan:2019xor}
Y.~Farzan, \emph{{A model for lepton flavor violating non-standard neutrino
  interactions}},
  \href{https://doi.org/10.1016/j.physletb.2020.135349}{\emph{Phys. Lett. B}
  {\bfseries 803} (2020) 135349}
  [\href{https://arxiv.org/abs/1912.09408}{{\ttfamily 1912.09408}}].

\bibitem{Blennow:2016jkn}
M.~Blennow, P.~Coloma, E.~Fernandez-Martinez, J.~Hernandez-Garcia and
  J.~Lopez-Pavon, \emph{{Non-Unitarity, sterile neutrinos, and Non-Standard
  neutrino Interactions}},
  \href{https://doi.org/10.1007/JHEP04(2017)153}{\emph{JHEP} {\bfseries 04}
  (2017) 153} [\href{https://arxiv.org/abs/1609.08637}{{\ttfamily
  1609.08637}}].

\bibitem{Blennow:2016etl}
M.~Blennow, S.~Choubey, T.~Ohlsson, D.~Pramanik and S.K.~Raut, \emph{{A
  combined study of source, detector and matter non-standard neutrino
  interactions at DUNE}},
  \href{https://doi.org/10.1007/JHEP08(2016)090}{\emph{JHEP} {\bfseries 08}
  (2016) 090} [\href{https://arxiv.org/abs/1606.08851}{{\ttfamily
  1606.08851}}].

\bibitem{Coloma:2015kiu}
P.~Coloma, \emph{{Non-Standard Interactions in propagation at the Deep
  Underground Neutrino Experiment}},
  \href{https://doi.org/10.1007/JHEP03(2016)016}{\emph{JHEP} {\bfseries 03}
  (2016) 016} [\href{https://arxiv.org/abs/1511.06357}{{\ttfamily
  1511.06357}}].

\bibitem{Ohlsson:2012kf}
T.~Ohlsson, \emph{{Status of non-standard neutrino interactions}},
  \href{https://doi.org/10.1088/0034-4885/76/4/044201}{\emph{Rept. Prog. Phys.}
  {\bfseries 76} (2013) 044201}
  [\href{https://arxiv.org/abs/1209.2710}{{\ttfamily 1209.2710}}].

\bibitem{Formaggio:2012cpf}
J.A.~Formaggio and G.P.~Zeller, \emph{{From eV to EeV: Neutrino Cross Sections
  Across Energy Scales}},
  \href{https://doi.org/10.1103/RevModPhys.84.1307}{\emph{Rev. Mod. Phys.}
  {\bfseries 84} (2012) 1307}
  [\href{https://arxiv.org/abs/1305.7513}{{\ttfamily 1305.7513}}].

\bibitem{Katori:2016yel}
T.~Katori and M.~Martini, \emph{{Neutrino\textendash{}nucleus cross sections
  for oscillation experiments}},
  \href{https://doi.org/10.1088/1361-6471/aa8bf7}{\emph{J. Phys. G} {\bfseries
  45} (2018) 013001} [\href{https://arxiv.org/abs/1611.07770}{{\ttfamily
  1611.07770}}].

\bibitem{Branca:2021vis}
A.~Branca, G.~Brunetti, A.~Longhin, M.~Martini, F.~Pupilli and F.~Terranova,
  \emph{{A New Generation of Neutrino Cross Section Experiments: Challenges and
  Opportunities}}, \href{https://doi.org/10.3390/sym13091625}{\emph{Symmetry}
  {\bfseries 13} (2021) 1625}
  [\href{https://arxiv.org/abs/2108.12212}{{\ttfamily 2108.12212}}].

\bibitem{DiLodovico:2023jgr}
F.~Di~Lodovico, R.B.~Patterson, M.~Shiozawa and E.~Worcester,
  \emph{{Experimental Considerations in Long-Baseline Neutrino Oscillation
  Measurements}},
  \href{https://doi.org/10.1146/annurev-nucl-102020-101615}{\emph{Ann. Rev.
  Nucl. Part. Sci.} {\bfseries 73} (2023) 69}.

\end{thebibliography}\endgroup

\end{document}